\documentclass[aps, amsmath,amssymb, superscriptaddress, 12pt]{revtex4-1}

\usepackage{float}
\usepackage{graphicx}
\usepackage{xcolor,color,soul}

\definecolor{fb}{rgb}{1, 0, 0}
\definecolor{da}{rgb}{0, 1, 0}
\definecolor{og}{rgb}{0, 0, 1}

\newcommand{\BSCCO}{Bi$_2$Sr$_2$CaCu$_2$O$_{8+\delta}$}

\usepackage{lineno}

\usepackage{graphicx}% Include figure files
\usepackage{dcolumn}% Align table columns on decimal point
\usepackage{bm}% bold math
\usepackage{relsize}% for large math symbols
\usepackage{xcolor,color}
\usepackage{ulem}
\usepackage{hyperref}
\hypersetup{colorlinks=true,urlcolor=blue,citecolor=blue}

\usepackage[
]{changes}
\definechangesauthor[name=Antoine, color=purple]{AG}

\begin{document}

%\title{Dynamical Coexistence of Superconductivity and Pseudogap in Bi-based cuprates}
% OR
%\title{Light-induced Asymmetric Pseudogap in Cuprates}
%OR
\title{Light-induced Asymmetric Pseudogap below T$_\text{c}$ in cuprates}

\author{D. Armanno}
\affiliation{Advanced Laser Light Source, Institut National de la Recherche Scientifique, Varennes QC J3X 1P7 Canada}
\affiliation{Department of Physics, Center for the Physics of Materials, McGill University, 3600 rue Université, Montréal, Québec H3A 2T8, Canada}
\affiliation{Department of Chemistry, McGill University, 801 rue Sherbrooke Ouest, Montréal, Québec H3A 0B8, Canada}
\author{O. Gingras}
\email[]{ogingras@flatironinstitute.org}
\affiliation{Center for Computational Quantum Physics, Flatiron Institute, 162 Fifth Avenue, New York, New York 10010, USA}
\affiliation{Université Paris-Saclay, CNRS, CEA, Institut de physique théorique, 91191, Gif-sur-Yvette, France}
\author{F. Goto}
\author{J.-M. Parent}
\author{A. Longa}
\author{A. Jabed}
\author{B. Frimpong}
\affiliation{Advanced Laser Light Source, Institut National de la Recherche Scientifique, Varennes QC J3X 1P7 Canada}
\author{R.\,D.\,Zhong}
\affiliation{Condensed Matter Physics and Materials Science, Brookhaven National Laboratory, Upton, NY 11973, USA}
\author{J.\,Schneeloch}
\affiliation{Condensed Matter Physics and Materials Science, Brookhaven National Laboratory, Upton, NY 11973, USA}
\affiliation{Department of Physics $\&$ Astronomy, Stony Brook University, Stony Brook, NY 11795-3800, USA}
\author{G.\,D.\,Gu}
\affiliation{Condensed Matter Physics and Materials Science, Brookhaven National Laboratory, Upton, NY 11973, USA}
\author{G. Jargot}
\author{H. Ibrahim}
\author{F. L\'{e}gar\'{e}}
\affiliation{Advanced Laser Light Source, Institut National de la Recherche Scientifique, Varennes QC J3X 1P7 Canada}
%\author{S. Beaulieu}
%\affiliation{Université de Bordeaux-CNRS-CEA,CELIA,UMR5107,F33405 Talence,France}
\author{B.J. Siwick}
\affiliation{Department of Physics, Center for the Physics of Materials, McGill University, 3600 rue Université, Montréal, Québec H3A 2T8, Canada}
\affiliation{Department of Chemistry, McGill University, 801 rue Sherbrooke Ouest, Montréal, Québec H3A 0B8, Canada}
\author{N. Gauthier}
\affiliation{Advanced Laser Light Source, Institut National de la Recherche Scientifique, Varennes QC J3X 1P7 Canada}
\author{A. Georges}
\affiliation{Coll\`{e}ge de France, 11 place Marcelin Berthelot, F-75005 Paris, France}
\affiliation{Center for Computational Quantum Physics, Flatiron Institute, 162 Fifth Avenue, New York, New York 10010, USA}
\affiliation{CPHT, CNRS, Ecole Polytechnique, IP Paris, F-91128 Palaiseau, France}
\affiliation{Department of Quantum Matter Physics, University of Geneva, CH-1211 Geneva 4, Switzerland}
\author{A.J. Millis}
\affiliation{Center for Computational Quantum Physics, Flatiron Institute, 162 Fifth Avenue, New York, New York 10010, USA}
\affiliation{Department of Physics, Columbia University, 538 West 120th Street, New York, New York 10027, USA}
\author{F. Boschini}
\email[]{fabio.boschini@inrs.ca}
\affiliation{Advanced Laser Light Source, Institut National de la Recherche Scientifique, Varennes QC J3X 1P7 Canada}
\affiliation{Quantum Matter Institute, University of British Columbia, Vancouver, BC V6T 1Z4, Canada}

\maketitle

\textbf{
   To this day, high-temperature cuprate superconductors remain an unparalleled platform for studying the competition and coexistence of emergent, static and dynamic, quantum phases of matter exhibiting  high transition temperature non-s-wave superconductivity, non-Fermi liquid transport and a still enigmatic pseudogap regime. However, how superconductivity emerges alongside and competes with the pseudogap regime remains an open question.
   Here, we present a high-resolution, time- and angle-resolved photoemission study of the near-antinodal region of optimally-doped \BSCCO. 
   %\deleted{Using mid-infrared light, \fb{we disrupt superconductivity and 
   %\deleted{use time and angle resolved photoemission to} probe the changes in density of states.}} 
   For a sufficiently high excitation fluence, we disrupt superconductivity and drive a transient change from a symmetric superconducting-like to an asymmetric pseudogap-like density of states, for electronic temperatures well below the equilibrium superconducting critical temperature. Conversely, when the superconductivity is fully restored, the pseudogap is suppressed, as signaled by a fully   
   particle-hole symmetric density of states. A unique aspect of our experiments  is that the pseudogap coexists with superconducting features at intermediate times or at intermediate fluence.  Our findings challenge the paradigm that superconductivity emerges by establishing phase coherence in  the pseudogap. Instead, our experimental results, supported by phenomenological theory, demonstrate that the two states compete, and that the low-temperature ground state of the cuprates originates from a competition between superconducting and pseudogap states.} %\sout{rather than from a simple sum of the two.}
  % \AG{The last part of the sentence above is TBD. I am not sure what is meant by 'a simple sum of the two'. In any (reasonable) theory study I know of, SC and PG are introduced by having two distinct self-energy in 
  % a Nambu matrix form, so that at the end the gap adds basically in quadrature.} 

   %we disrupt \da{the particle-hole symmetric} superconductivity and, for a high enough excitation fluence, we reveal a particle-hole asymmetric, pseudogap-like, density of states (DOS) \da{well below the equilibrium superconducting critical temperature}. \sout{We measure the electronic temperature in the non-equilibrium state and show that the pseudogap-like DOS persists for electronic temperatures well below the equilibrium superconducting critical temperature.} 
   
   %Our results reveal an intrinsic pseudogap DOS that is fundamentally different from the superconducting DOS, strongly suggesting that the two phenomena are at least independent and compete. In the superconducting state the pseudogap contribution to the DOS is weak, while suppressing superconductivity allows pseudogap physics to emerge and alter the DOS, thus providing a new framework for understanding the complex interplay between intertwined phases in correlated electron systems.}

\vspace{5mm}
\large\noindent\textbf{Introduction}
\vspace{2mm}
\normalsize

%\fb{General thoughts: the intro should be improved. 
%Competition between SC and pseudogap is well known: when decreasing the temperature (i) the AN gap increases from T* to Tc, and then decrease below Tc, and (ii) integrated AN spectral weight has a non-monotonic behavior, (iii) + several other evidences
%However, no definitive proof since pseudogap is intrinsically p-h asymmetric and looking at peak position and spectral weight could be misleading.
%In particular when SC correlations are present (preformed pairs) analysis of the coherent peak or spectral weight is misleading. The only way to really evaluate directly the competition is to evaluate the p-h asymmetry level.}

%\fb{OUR FINDINGS
%Light-induced phase transition into a non-thermal state where we expose SC and pseudogap coexistence and intertwining
%(1) long range SC kills pseudogap, but peak does not move: is the order parameter of pseudogap still there?
%(2) fluctuating SC coexist with pseudogap: i) minimum at 0 but asymmetric (ii) peak shift back}

\begin{figure}[H]
    \centering
    \includegraphics[width=\textwidth]{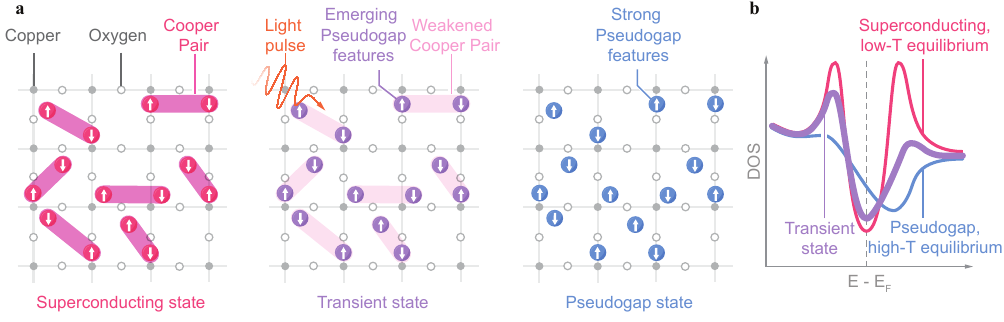}
    \caption{\textbf{Emergence of the light-induced pseudogap state.}
    \textbf{a} Pictorial sketch of the equilibrium superconducting (left) and pseudogap (right) states in cuprates. These two states correspond, respectively, to the particle-hole symmetric (red) and asymmetric (blue) density of states (DOS) displayed in \textbf{b}. 
    %transient particle-hole asymmetric state. Left: below T$_c$, long-range SC order is established. Middle: 
    Light excitation breaks pairs while only slightly increasing the electronic temperature, thereby prompting the emergence of a light-induced pseudogap with a particle-hole asymmetric DOS (panel a, middle, and thick violet curve in b). }
    %Right: above T$_c$ the intrinsically asymmetric Pseudogap order is established. A schematic DOS for the equilibrium SC (pink), transient asymmetric state (violet) and equilibrium PG (blue) are displayed in the right-most panel.}
    \label{Fig0}
\end{figure}

Since the discovery of unconventional high-temperature superconductivity (SC) in layered copper oxides ``cuprates'', one of the long-standing challenges has been to understand the underlying normal state from which high critical temperature (T$_c$) SC order emerges \cite{keimer2015quantum}.
In particular, upon hole doping, a linear-in-temperature resistivity 
which can exceed the Mott-Ioffe-Regel limit is reported \cite{phillips2022stranger}, in concert with the appearance of a mysterious pseudogap regime characterized by  a partial suppression of the spectral weight near the Fermi level E$_\text{F}$, that remains evident 
%above the superconducting critical temperature (T$_c$) 
over  wide ranges of doping and temperature \cite{vishik2018photoemission,kordyuk2015pseudogap,yang2008emergence,chen2019incoherent,hashimoto2015direct,he2011single}.
Although both SC and pseudogap states lead to gaps in the electronic spectrum that are largest where the nominal Fermi surface touches the Brillouin boundary (antinodal region), they have qualitatively different spectral fingerprints. The SC gap exhibits point nodes and an intrinsically particle-hole symmetric DOS featuring coherent Bogoliubov peaks of equal intensity in both the occupied and unoccupied electronic states \cite{yang2008emergence,matsui2003bcs} (Fig.\,\ref{Fig0}a, left, and the corresponding red curve in b). In contrast, key manifestations of the pseudogap state are the appearance of disconnected Fermi arcs \cite{lee2007abrupt,kondo2013formation}, in concert with a spectral gap not centered at E$_\text{F}$ \cite{yang2008emergence,hashimoto2010particle,yang2011reconstructed} and a shift of Fermi momentum k$_\text{F}$ from the normal state position \cite{hashimoto2010particle}, thus implying particle-hole asymmetry, as depicted in Fig.\,\ref{Fig0}a, right, and the corresponding blue curves in b.

The competition and/or coexistence of the SC and pseudogap states have been widely investigated using various experimental techniques over the past decades \cite{hufner2008two,guyard2008breakpoint,sakai2013raman,kondo2009competition,huecker2014competing}. In particular, high-resolution angle-resolved photoemission spectroscopy (ARPES) \cite{sobota2021angle,damascelli2003angle} has been instrumental in identifying that the pseudogap is not a precursor to superconductivity but instead competes with the SC phase by contending for spectral weight \cite{kondo2009competition,kondo2011disentangling,hashimoto2015direct,kuspert2022pseudogap,pushp2009extending}.
However, these studies have primarily focused on mapping the evolution of the equilibrium antinodal spectral gap amplitude and the in-gap spectral weight as a function of temperature. Because they address mainly the occupied states, these experimental approaches are not well suited to the investigation of the (cluster-) dynamical mean field prediction that the intrinsically particle-hole asymmetric pseudogap evolves into symmetric coherent peaks when entering the SC phase \cite{gull2013superconductivity,gull2015quasiparticle,civelli2008,sakai2013raman}. These studies were also not able to address the low temperature properties of the pseudogap regime in the absence of superconductivity, i.e. when long-range SC order is quenched for temperatures below T$_c$.  
Here, by using ultrafast laser light excitations to disrupt the SC order \cite{giannetti2016ultrafast}, in combination with a new experimental strategy to access the antinodal region of cuprates with low-photon energy light, we present a high-resolution time-resolved ARPES (TR-ARPES) \cite{boschini2024time} investigation of the interplay between superconductivity and the pseudogap and its effect on the electronic DOS in the near-antinodal region of optimally-doped \BSCCO\, with T$_\text{c} \sim$91\,K (Bi2212-OP91).
%we provide the first evidence of a low-temperature light-induced pseudogaped state in optimally-doped \BSCCO (Bi2212) with T$_c \sim$91\, K via time- and angle-resolved photoemission spectroscopy (TR-ARPES) \cite{boschini2024time}. 
Remarkably, we find the appearance of a light-induced particle-hole asymmetric pseudogap when the SC order parameter is suppressed even at electronic temperatures below the equilibrium T$_\text{c}$, as displayed in the middle panel of Fig.\,\ref{Fig0}a, and the corresponding violet curve in b. By accessing the transient evolution of the coherent Bogoliubov peaks in the unoccupied near-antinodal states, we provide compelling evidence of the dynamic coexistence between SC correlations and the pseudogap state on ultrafast timescales. Moreover, we follow the antinodal spectrum evolution from particle-hole asymmetric to symmetric within a few picoseconds, thus offering direct tracking of the reestablishment of the competition between SC and pseudogap in equilibrium conditions. 
These findings are supported by calculations on an equilibrium phenomenological theory, which %\fb{Although this theory does not describe the ultrafast emergence of the light-induced pseudogap state within the first 0.5\,ps, it} 
reproduce the experimental data only when the interplay between pseudogap and SC is considered.

\vspace{5mm}
\large\noindent\textbf{TR-ARPES at the antinode of Bi-based cuprates}
\vspace{2mm}
\normalsize

%In particular, we selectively enhance pair-breaking scattering events via mid-infrared light excitation (300\,meV photon energy, well below the charge transfer gap of Bi2212 \cite{giannetti2016ultrafast,baldini2020electron,cilento2018dynamics}) while maintaining the electronic temperature below T$_c$, as illustrated in Fig.\,\ref{Fig1}a, left. This experimental strategy is validated by recent ultrafast spectroscopy works reporting, upon light excitation, a transient state similar to what is achieved at high magnetic fields \cite{wandel2022enhanced,jang2022characterization}, but without the actual need for an external magnetic field, which is not easy to implement in ARPES experiments \cite{huang2023angle} (certainly not with a strength high enough to disrupt the SC order).

\begin{figure}[H]
    \centering
    \includegraphics[width=0.55\textwidth]{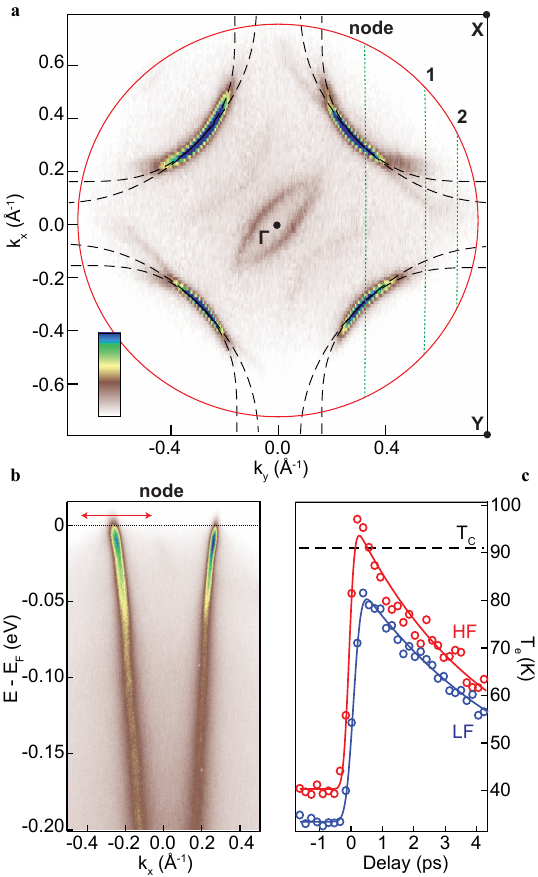}
    \caption{\textbf{Fermi surface mapping and tracking of the transient electronic temperature.}
    \textbf{a} Fermi surface mapping of Bi2212-OP91 ([-15;15]\,meV integration window). The red circle shows the maximum momentum reachable with a 6 eV photon probe, corresponding to the edges of the photoemission cone. The green dashed lines indicate the three momentum cuts examined in this work.
    \textbf{b} ARPES map along the nodal direction. 
    %The node is crossed along the $\Gamma$--Y direction (negative momenta). 
    \textbf{c} Transient evolution of the electronic temperature T$_e(\tau)$, extracted by fitting the momentum-integrated nodal energy distribution curves (EDCs, red arrow in panel c) via Fermi-Dirac fit, for two excitation regimes: low fluence (LF, blue) and high fluence (HF, red).
    }
    \label{Fig1}
\end{figure}

Figure\,\ref{Fig1}a displays the Fermi surface mapping of Bi2212-OP91, as well as the three momentum cuts investigated in this work, acquired with low-photon energy (6 eV) probe pulses that enable a high momentum and energy resolution. Most notably, our low-photon-energy Fermi mapping simultaneously covers all four quadrants of the first Brillouin zone, a long-sought-after aspiration in the TR-ARPES community previously believed to be achievable only at high photon energies \cite{na2023advancing,cilento2016advancing}.
The access to such a large range in momentum space with 6 eV photons is enabled by biasing the sample \cite{gauthier2021expanding} in concert with the use of a hemispherical electron analyzer with deflector technology and $\pm$30$^o$ maximum acceptance angle \cite{longa2024time}. This novel experimental strategy enabled the acquisition of all three momentum cuts of Fig.\,\ref{Fig1}a, namely the node, the end of the Fermi arc (cut 1), and the near antinodal region (cut 2) (additionally, a near-nodal cut is shown in the SI, Fig.\,S2), without spatially moving and realigning the sample, and in a single iterative pump-probe acquisition. This assures that the pump and probe interaction geometry with the sample remained unchanged for all three cuts, which addresses the common challenge of reliably comparing different TR-ARPES scans, thereby making our data analysis and findings robust.

The ARPES map of Fig.\,\ref{Fig1}b crosses the node along the $\Gamma$-Y direction (k$_\text{x}<$0) and near-nodal states along $\Gamma$-X (k$_\text{x}>$0). In order to avoid any contributions from superstructure replicas, we focus our analysis along the $\Gamma$-Y direction \cite{king2011structural}. 
The Bi2212-OP91 sample is photoexcited with mid-infrared light (300\,meV photon energy, well below the charge transfer gap of Bi2212 \cite{giannetti2016ultrafast,baldini2020electron}) in two excitation regimes, namely low fluence (LF, $\sim$74\,$\mu$J/cm$^2$) and high fluence (HF, $\sim$138\,$\mu$J/cm$^2$).
Light-induced non-thermal carriers in cuprates rapidly thermalize via electron-electron and electron-boson scattering within $\sim$100\,fs \cite{perfetti2007ultrafast,giannetti2016ultrafast}. Because the electronic spectrum in the nodal direction is gapless both in the SC and PG regimes, we may fit the momentum-integrated nodal ARPES energy distribution curves (EDC; momentum integration range indicated by the red arrow in Fig.\,\ref{Fig1}b) to a Fermi-Dirac  distribution function. After approximately 200-300\,fs, an effective quasi-equilibrium electronic distribution can be well fit along the nodal direction, yielding estimates of the electronic temperature (see Fig.\,S5). 
%Moreover, although the n  odal direction is gapless at the Fermi level (E$_\text{F}$) in the SC phase, a pseudogap may be present in the unoccupied states [Refs] (Fig.\,\ref{FIG_fit_Node_SI}).
%Consequently, 
%in the [-20,20]\,meV energy range 
Figure\,\ref{Fig1}c displays the resulting T$_e(\tau)$ for both LF and HF regimes. 

%The transient electronic temperatures T$_e(\tau)$ for two incident fluences ($\sim$74\,$\mu$J/cm$^2$ low fluence, LF, and $\sim$138\,$\mu$J/cm$^2$ high fluence, HF) are shown in Fig.\,\ref{Fig1}c, and they are extracted by fitting the momentum-integrated nodal energy distribution curves (EDCs, momentum-integration range indicated by the red arrow in Fig.\,\ref{Fig1}b) with a Fermi-Dirac function. Light-induced non-thermal carriers in cuprates thermalize via electron-electron and electron-boson scattering within $\sim$100\,fs [Refs]. After approximately 100-200\,fs (in this specific case, when pump and probe beams are not temporally overlapped), an effective quasi-equilibrium electronic distribution can be fit along the nodal direction (see Fig.\,\ref{FIG_fit_Node_SI}). Moreover, although the nodal direction is gapless at the Fermi level E$_\text{F}$ in the SC phase, a pseudogap may be present in the unoccupied states [Refs] (Fig.\,\ref{FIG_fit_Node_SI}). Consequently, T$_e(\tau)$ is extracted by fitting the momentum-integrated nodal EDCs in the [-20,20]\,meV energy range and solely for time delays $\tau >$200\,fs.  

\vspace{5mm}
\large\noindent\textbf{Ultrafast filling of the near-antinodal superconducting gap}
\vspace{2mm}
\normalsize

\begin{figure}[H]
    \centering
    \includegraphics[scale=1.22]{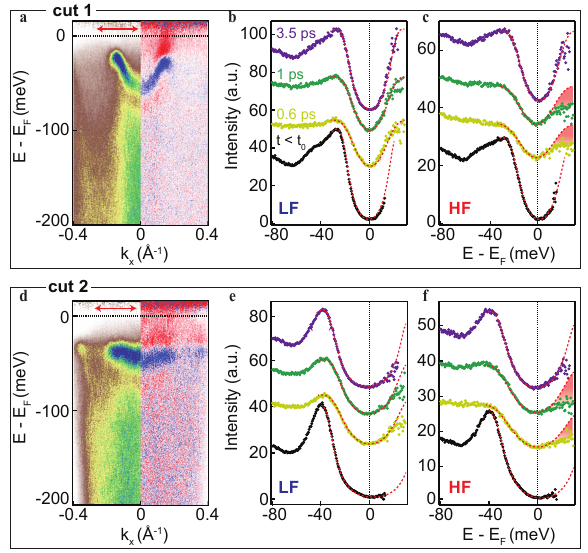}
    \caption{\textbf{Transient evolution of the off-nodal and near-antinodal superconducting gap.}
    \textbf{a} and \textbf{d} $\tau <$0 (left) and differential (right, obtained by subtracting the $\tau <$0 map from its counterpart at $\tau =$1\,ps) ARPES maps along cuts 1 and 2 defined in Fig.\,\ref{Fig1}a. The red arrow indicates the integration region for the $\text{TDOS}(\omega)$ extraction.
    \textbf{b} and \textbf{e} transient evolution of the TDOS for cuts 1 and 2, respectively, in the LF case at various pump-probe delays. The EDCs are plotted up to the experimental limit dictated by $4.5\times$\,$k_\text{B}$T$_e(\tau)$. The superimposed Dynes fit (dashed lines) captures well the evolution of the DOS from the equilibrium SC DOS through a transient intermediate state back to the SC state. Under low fluence conditions (panels b, e) the DOS remains particle-hole symmetric and well fit by the standard Dynes form (dashed lines) indicating a slightly disrupted superconducting state.
    \textbf{c} and \textbf{f} transient evolution of the TDOS for cuts 1 and 2, respectively, in the HF case at various pump-probe delays. In this case, we report a transient pseudogap-like asymmetric TDOS spectrum, as highlighted by 0.6\,ps and 1\,ps pump-probe delays. For these delays, the TDOS is fundamentally different from the SC DOS, exhibiting a marked particle-hole asymmetry that is inconsistent with the Dynes fit, as emphasized by the red shadows between the experimental TDOS and the Dynes fit. Additional pump-probe delays are shown in Fig.\,S2.
    %\textbf{d} and \textbf{h} show the difference between the experimental TDOS of panels b, c, f, and e and the corresponding Dynes function for cuts 1 and 2, respectively, in the LF (blue) and HF (red) regimes.
    }
    \label{Fig2}
\end{figure}

To study the transient evolution of the spectral weight around E$_\text{F}$ we need a spectral quantity that does not depend on the Fermi-Dirac distribution, and is robust against possible small variations of the Fermi momentum k$_\text{F}$ upon light excitation.
For this purpose, we employ the tomographic density of states (TDOS) curves, momentum integral of the DOS along the selected momentum cut (see Methods section for precise definition).
%We now discuss the transient evolution of the spectral weight around E$_\text{F}$ for cuts 1 and 2, for which the equilibrium SC gap determined from measurements at negative delays is approximately 25\,meV and 40\,meV, respectively. 
Figures\,\ref{Fig2}a and d display the $\tau <$0 ARPES (left) and the differential ARPES (right) maps for cuts 1 and 2, for which the equilibrium SC gap is approximately 25\,meV and 40\,meV, respectively.
%In an effort to disclose the transient evolution of the SC gap, and generally the spectral function, it is necessary to subtract any thermal broadening contributions to the TR-ARPES intensity. 
In Figs.\,\ref{Fig2}b,c,e,f we plot the TDOS curves for each momentum cut.
%In Figs.\,\ref{Fig2}b,c,e,f we plot the tomographic density of states (TDOS) curves (see Methods section for precise definition) for each momentum cut. TDOS curves are cleared of the contribution from the Fermi-Dirac distribution and are proportional to the DOS along the selected momentum cut, and they are robust against possible small variations of the Fermi momentum k$_\text{F}$ upon light excitation. 
%Note that the Dynes function~\cite{dynes1978direct}, which captures the DOS of 
%is even across E$_\text{F}$ (particle–hole symmetry has been experimentally verified in the long-range SC state of cuprates via ARPES \cite{yang2008emergence,matsui2003bcs}).
%and grants direct insights into transient changes in the in- and near-gap spectral weight and
%Within the BCS framework, the TDOS is well captured by the Dynes function [], which is symmetric around E$_\text{F}$.
Although our discussion below focuses on the TDOS analysis, we emphasize that similar results and conclusions are obtained by examining the EDCs at k$_\text{F}$ (see Fig.\,S3).

Figure\,\ref{Fig2}b and e display the TDOS for cuts 1 and 2 in the LF pump regime for different pump-probe delays. For $t<3.5$\,ps we observe an ultrafast filling of the SC gap (defined as the difference between the intensity at $-80$\,meV and at $E_F$) with only a minimal variation in the gap amplitude (defined as the energy of the TDOS maximum relative to $E_F$), similar to what has been previously reported upon near-IR pump excitation \cite{boschini2018collapse,smallwood2012tracking,parham2017ultrafast}.
It is important to note that all TDOS curves acquired in the LF regime (Fig.\,\ref{Fig2}b and e) are well fit by the Dynes function (red dashed curves), which is intrinsically particle-hole symmetric \cite{dynes1978direct} (see Eq.\,\ref{eq2} in the Methods). This observation is in good agreement with the experimental observation of a particle-hole symmetric equilibrium SC gap in cuprates at low temperatures \cite{yang2008emergence,matsui2003bcs}.   

%While the transient evolution of $\Delta$ and the single-particle scattering rate are closely related to T$_e(\tau)$, \textit{i.e.} can be ascribed to a pure thermal contribution, 
The light-induced disruption of the SC, which is evident as the filling of the SC gap, can be ascribed to an enhancement of the pair-breaking scattering rate \cite{kim2024tracing,armanno2025direct,boschini2018collapse}. 
%capturing the filling of the SC gap is delayed $\sim$500\,fs and is non-thermal \cite{zonno2021time,boschini2018collapse,armanno2025direct}. 
%This experimental strategy is validated by recent ultrafast spectroscopy works reporting, upon light excitation, a transient state similar to what is achieved at high magnetic fields \cite{wandel2022enhanced,jang2022characterization}, but without the actual need for an external magnetic field, which is not easy to implement in ARPES experiments \cite{huang2023angle} (certainly not with a strength high enough to disrupt the SC order).
The time dependence of the pair-breaking scattering rate (Fig.\,S4) does not follow the time dependence of the electronic temperature but is delayed by $\sim$500\,fs, which is speculated to be caused 
%It has been suggested that the delayed response of the pair-breaking scattering events comes 
by the build-up of a non-thermal bosonic population following the thermalization processes of optically excited electrons \cite{giannetti2016ultrafast}. 
%These non-thermal bosons interact with the SC condensate breaking pairs, consequently enhancing pair fluctuations and quenching the long-range phase coherence of the SC condensate \cite{giannetti2016ultrafast}.
Although the exact mechanism underlying the ultrafast enhancement of pair-breaking events has not yet been experimentally confirmed, it is well established that light excitation with photon energies exceeding the SC gap amplitude disrupts the SC order while maintaining the electronic temperature below T$_c$ and marginally affecting the SC gap amplitude \cite{zonno2021time,boschini2018collapse,armanno2025direct}. This interpretation is also validated by recent ultrafast spectroscopy works reporting, upon light excitation, a transient state similar to what is achieved at high magnetic fields \cite{wandel2022enhanced,jang2022characterization}. 
%but without the actual need for an external magnetic field, which is not easy to implement in ARPES experiments \cite{huang2023angle} (certainly not with a strength high enough to disrupt the SC order)..

\vspace{5mm}
\large\noindent\textbf{Light-induced pseudogap in the antinodal region}
\vspace{2mm}
\normalsize

We now move to the HF regime where T$_e(\tau)$ marginally exceeds T$_\text{C}$ ($\sim5\%$) for the first $\sim$0.5\,ps and the pair-breaking scattering rate reaches a value of $\sim$15-20\,meV for approximately 1-2\,ps (see Fig.\,S4). Note that while saturation of pair-breaking scattering rate at similar values has been reported by equilibrium ARPES data for temperatures $>$1.1$\cdot$T$_\text{c}$ \cite{kondo2015point}, here we observe it for electronic temperatures well below T$_\text{c}$.
%By quenching macroscopic superconductivity in a non-equilibrium fashion, we expose underlying phase coexisting with fluctuating superconductivity.
%In other words, light excitation has an effect analogous to that of an external magnetic field in quenching long-range SC, as also reported by time-resolved X-ray scattering studies \cite{wandel2022enhanced,jang2022characterization}.

By simple visual inspection of the transient evolution of the TDOS in the HF regime for both cuts 1 and 2 (see Fig.\,\ref{Fig2}c and f), we observe that the $\omega \approx -40$\,meV coherent peak remains evident at all pump-probe delays and shows no significant energy shift, as also seen in the LF data. Note, though, that for $\tau<3.5$\,ps the peak amplitude is more suppressed in the HF data than in the LF data. However, for early pump-probe delays (0.6\,ps and 1\,ps in Fig.\,\ref{Fig2}c and f) the TDOS is no longer symmetric and deviates from the Dynes function form (red dashed lines and shadows). In particular, by focusing on cut 1 in Fig.\,\ref{Fig2}c, for which we observe the $\omega > 0$ quasiparticle peak via Fermi-Dirac division, we note a marked suppression of the $\omega>0$ intensity, resulting in an asymmetric TDOS.  %Figs.\,\ref{Fig2}d and h display the difference between the experimental TDOS and the Dynes function for cuts 1 (d) and 2 (h), in the LF and HF regimes (blue and red, respectively). In the HF regime, the TDOS curves along cuts 1 and 2 exhibit statistically significant transient suppression of the $\omega > 0$ intensity.
Nevertheless, the minimum of this light-induced pseudogap state is still centered at E$_\text{F}$. 
Figure\,\ref{Fig2} provides  important,  purely experimental, evidence that for any energy range and momentum cut (either 1 or 2), all LF TDOS curves are symmetric, while HF TDOS curves are asymmetric for the first 1-2\,ps (red shadow in Fig.\,\ref{Fig2}c and f), and particle-hole symmetry is restored by 3.5\,ps.

%. I would first think you made a mistake in the experiment. I think you should clearly emphasize that this asymmetry is definetely real because in the HF data you can see that at 3.5ps the gap symmetry is restored. And over the same energy range in the LF data all the curves are Dynes symmetric. YOu provide in this and the previous section all the information to reach these important purely experimental conclusions but you didn't write it explicitly.

%that although the minimum of the gap is still centered at E$_\text{F}$, the spectral weight of the cohernet peak in the unoccupied states is quenched, resulting in an asymmetric TDOS.

\vspace{5mm}
\large\noindent\textbf{Non-thermal nature of the light-induced asymmetric pseudogap}
\vspace{2mm}
\normalsize

\begin{figure}[H]
    \centering
    \includegraphics[width=\textwidth]{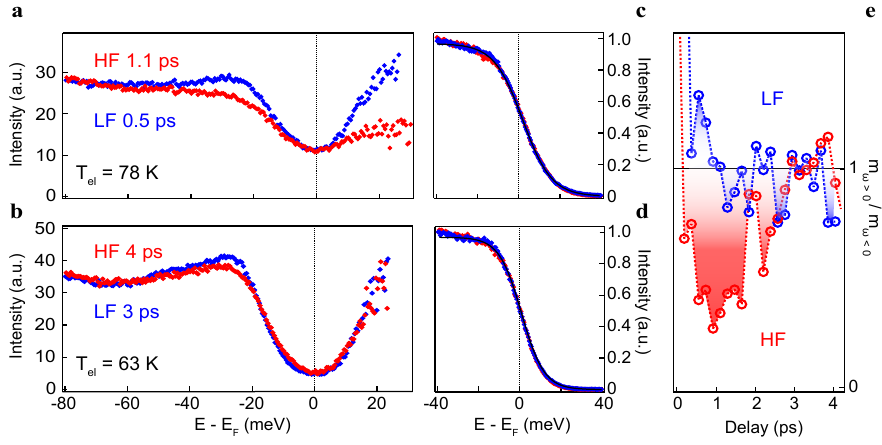}
    \caption{\textbf{Non-thermal nature of the light-induced pseudogap.}
    \textbf{a} and \textbf{b} Comparison of TDOS curves along cut 1 taken at different pump-probe delays in the LF (0.5\,ps and 3\,ps) and HF (1.1\,ps and 4\,ps) regime, but with similar $\text{T}_e$, namely $\sim$78\,K in c and $\sim$63\,K in d. The spectral weight symmetry is restored at lower $\text{T}_e$, as shown in b.
    \textbf{c} and \textbf{d} Momentum-integrated nodal EDCs, extracted from Fig.\,\ref{Fig1}b in the LF and HF cases, for the two $\text{T}_e$ discussed in a and b. In both cases, a single Fermi-Dirac fit successfully captures both curves at different pump-probe delays in the LF and HF regimes. 
    \textbf{e} Transient evolution of the linear slope ratio of the TDOS at cut 1 as defined in the main text, extracted in the range [-17;-7]\,meV and [7;17]\,meV below and above $E_\text{F}$, respectively. In the LF regime (blue), the ratio is approximately 1, evidence of particle-hole symmetry. Instead, in the HF regime (red), we report a reduction in the slope ratio, with maximal variation at around 1\,ps pump-probe delay, highlighting the transient spectral weight asymmetry. 
    }
    \label{Fig3}
\end{figure}

Before discussing the origin of this transient asymmetry in the coherent spectral weight, it is important to illustrate its non-thermal character. To do so, we compare TDOS curves acquired at different pump-probe delays in the LF and HF regimes, but sharing the same transient electronic temperature. 
Figure\,\ref{Fig3}c and d demonstrate that the momentum-integrated nodal EDC in LF at one time delay and the momentum-integrated nodal EDC in HF at a different, later time delay can be fit by the same Fermi-Dirac distribution, thus showing that the two data sets were taken at the same transient electronic temperatures. Figure\,\ref{Fig3}a and b compare the LF (blue) and HF (red) TDOS curves along cut 1 at times corresponding to the same electronic temperature, $\sim$78\,K and $\sim$63\,K respectively. The data in panel a, taken at short delays, confirm a clear suppression of the $\omega > 0$ coherent spectral weight in the HF regime that does not depend on the specific value of T$_e$ and is not found in the LF regime. At long times when, presumably, the long-range SC order is restored, the LF and HF TDOS curves are almost the same (panel b). Figure\,S6 reports similar behavior also along cut 2.

To empirically characterize the timescale of this light-induced spectral weight asymmetry, we extract the positive and negative frequency   slope $m$ of the TDOS data (cut 1) by linear fits to the data in the ranges [-17,-7]\,meV and [7,17]\,meV with respect to E$_\text{F}$. This ratio is well defined only for delays $\tau>0$. We plot the ratio of the slopes $m_{\omega>0} / m_{\omega<0}$ for cut 1 in Fig.\,\ref{Fig3}e as a function of the pump-probe delays for LF and HF regimes. Similar results are obtained along cut 2, see Fig.\,S6).  $m_{\omega>0} / m_{\omega<0}$ is expected to be equal to one if the TDOS is symmetric, and to approach zero when the $\omega>0$ coherent peak is suppressed. Indeed, while $m_{\omega>0} / m_{\omega<0}\simeq$1 for any $\tau > 0$ in the LF regime, it displays a transient evolution within the first 2\,ps in the HF regime (Fig.\,\ref{Fig3}e). Furthermore, the slope ratio in the HF regime reaches its maximum deviation from unity after approximately 0.5--1\,ps, which is reminiscent of the transient evolution of the non-thermal evolution of the pair-breaking scattering rate discussed above \cite{boschini2018collapse,armanno2025direct} (see Fig.\,S4). 
%Therefore, not only is the transient suppression of the coherent spectral weight non-thermal, but it is delayed $\sim$0.5-1\,ps, thus corroborating its direct relation to the transient suppression of the macroscopic superconducting condensate.

\vspace{5mm}
\large\noindent\textbf{Dynamic interplay between superconductivity and pseudogap}
\vspace{2mm}
\normalsize

%Above, we have demonstrated that the SC phase displays robust particle-hole symmetry when light excitation does not break the long-range phase coherence of the SC condensate (see Fig.\,\ref{Fig2}b and e). Instead, when macroscopic SC is transiently quenched, an asymmetric light-induced pseudogap emerges in the near antinodal region of optimally-doped \BSCCO (Fig.\,\ref{Fig2}c,f). 

\begin{figure}[H]
    \centering
    \includegraphics[width=0.55\textwidth]{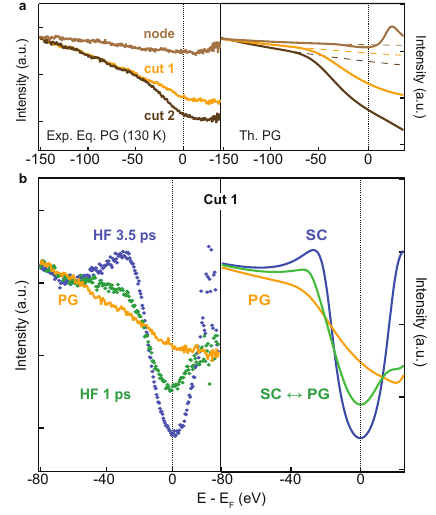}
    \caption{\textbf{Competition and coexistence between superconductivity and pseudogap.}
    \textbf{a} Experimental (top) and theoretical (bottom) TDOS in the pseudogap state. Experimental TDOS curves along the three momentum cuts of Fig.\,\ref{Fig1}a have been acquired at 130\,K, well above T$_\text{c}$. Theoretical parameters have been adjusted to generate a theoretical pseudogap curve that closely resembles the experimental data (details in the Methods and Supplementary Information).
    \textbf{b} comparison of three experimental TDOS curves along cut 1: Equilibrium pseudogap phase at 130 K (orange), light-induced pseudogap state at 1 ps in the HF regime (green), and restored superconducting particle-hole symmetric state at 3.5 ps in the HF regime (blue). 
    \textbf{c} Simulated curves using the phenomenological self-energy discussed in the Supplementary Information. When only pseudogap (orange) or SC (blue) correlations are considered, the light-induced pseudogap is not reproduced. Instead, the coexistence of SC and the pseudogap (green) provides a better fit to the experimental evidence in panel b. When long-range SC correlations are restored, the particle-hole symmetric SC contribution alone (blue) reproduces the data. 
    }
    \label{Fig4}
\end{figure}

In an effort to investigate the relationship between our findings and the possible
%better frame our findings within the context of 
competition between the SC and PG states in cuprates, we compare the light-induced pseudogap of Fig.\,\ref{Fig2}c,f and Fig.\,\ref{Fig3}a to the high-temperature equilibrium pseudogap state. First, in Fig.\,\ref{Fig4}a, left, we show the TDOS curves extracted at 130\,K along the node, as well as cuts 1 and 2, and demonstrate that only the latter two display a particle-hole asymmetric pseudogap with a minimum centered above E$_\text{F}$. Then, in Fig.\,\ref{Fig4}b, left, we provide direct comparison between the equilibrium TDOS at 130\,K (orange), the light-induced asymmetric TDOS (HF, 1\,ps, green), and the TDOS at late pump-probe delays when particle-hole symmetric SC order is restored (HF, 3.5\,ps, blue). Remarkably, we observe that the high-temperature and light-induced asymmetric TDOS curves share similar spectral weight at $\omega \sim$20\,meV, although these two curves have minima at different energy locations ($\omega \sim$20\,meV and $\omega=$0\,meV for the high-temperature and light-induced asymmetric TDOS, respectively).

The above comparison suggests that the light-induced pseudogap combines contributions from both SC correlations (minimum centered at $\omega$=0) and the pseudogap state (suppression of the spectral weight for $\omega >$0). To support this picture, we derived a phenomenological self-energy that combines both the pseudogap and superconducting physics, as captured by the effective 4x4 Hamiltonian of Eq.\,\ref{eq:interplay_hamiltonian} in the Methods and the self-energy of Eq.\,S14 in the Supplementary Information. Note that this theory captures the interplay between both phenomena, resulting in a self-energy that differs significantly from the simple independent sum of the self-energies
%simply summing the self-energies 
typically employed to simulate superconducting and pseudogap features separately (see Supplemental Information and Figs.\,S9 and S10). 
We emphasize that the pseudogap processes considered reproduce the experimental high-temperature TDOS in a satisfactory fashion (see Fig.\,\ref{Fig4}a, right, and Fig.\,S8). 
Figure\,\ref{Fig4}b, right, displays simulated TDOS, and shows that 
only a combination of both the pseudogap and SC correlations 
is able to properly describe the light-induced asymmetric pseudogap observed in the experimental data %(see also Figs.\,S8)
, thus highlighting the interplay between the two states. Figure\,S7 also shows remarkable agreement between the experimental and simulated TDOS curves along cut 2.
It is important to emphasize that it is not possible to reproduce the light-induced pseudogap state (green curve in Fig.\,\ref{Fig4}b) as a simple weighted sum (i.e., two separate self-energies) of either of the high-temperature pseudogap curves (orange curve in Fig.\,\ref{Fig4}b) and particle-hole symmetric SC state (blue curve in Fig.\,\ref{Fig4}b), as this results in a curve with a minimum that is not centered at zero energy and a more pronounced coherent peak at $\omega >$0 (see Figs.\,S9 and S10).

%it is crucial to assess whether the asymmetric DOS varies as a function of electronic temperature and how it compares to the high-temperature pseudogap phase.
%First, we select three delays in the HF regime, identified by the three green lines in Fig.\,\ref{Fig3}d, corresponding to three different electronic temperatures T$_e$. Since the long-range SC is transiently quenched at these three delays, we can track the temperature evolution of the asymmetric spectral weight as shown in Fig.\,\ref{Fig3}e. Although all three curves touch and share similar spectral weight at $\omega > 20$\,meV, the spectral weight under the remnant coherent peak at $\omega\sim-30$\,meV seems to move in-gap at $\omega=0$ as T$_e$ increases. We also include the TDOS extracted in the equilibrium pseudogap phase at 130\,K (orange line, $\varphi$=2). Note that (i) although the minimum of the HF TDOS curves is still centered at E$_\text{F}$, the minimum of the TDOS at 130\,K is at $\omega \sim$20\,meV, and (ii) the HF and 130\,K TDOS have similar spectral intensities for $\omega<$-60\,meV and $\omega>$20\,meV. Moreover, consistent with the temperature dependence shown by the HF TDOS curves, one can transition from the transient HF TDOS to the 130\,K TDOS by transferring the spectral weight from the remnant quasiparticle peak ([-50,-20]\,meV energy region) to the gap at $\omega \sim 0$, effectively filling the gap (similar experimental evidence is reported for $\varphi$=3 as shown in Fig.\,S5).

\vspace{5mm}
\large\noindent\textbf{Discussion}
\vspace{2mm}
\normalsize

%By performing the first high-resolution TR-ARPES study in the antinodal region of optimally-doped Bi2212, we unveiled an unprecedented light-induced asymmetric pseudogap. This transient state has a lifetime of $\sim$2\,ps, and appears to be related to a non-thermal enhancement of pair-breaking scattering events which, in turn, disrupt the SC order. By means of phenomenological mean-field calculations, we show that this transient spectrally asymmetric state can be reproduced only when both the SC and pseudogap phases are combined. 

These experimental and theoretical findings provide a new perspective on the interplay between superconductivity and the pseudogap in cuprates. The observation of the light-induced pseudogap state of Fig.\,\ref{Fig3}a is a direct signature of the dynamic interplay between SC and pseudogap, and offers a momentum-resolved confirmation of recent terahertz third harmonic generation data in cuprates, where the dynamic interplay between SC and pseudogap was also proposed \cite{yuan2024dynamical}. The difference between spectra obtained under low fluence, where the density of states remains particle-hole symmetric,  and high fluence, where a transient particle hole asymmetry develops, suggests that the interplay between pseudogap and superconducting order is controlled by the presence or absence of long ranged phase coherence. 
%but they coexist within the timescale of the Cooper pairs' lifetime. 
% It is important to note that it is not possible to reproduce the light-induced pseudogap state (green curve in Fig.\,\ref{Fig4}b) as a simple weighted sum of either of the experimental or theoretical  high-temperature pseudogap curves (orange curve in Fig.\,\ref{Fig4}b) and particle-hole symmetric SC state (blue curve in Fig.\,\ref{Fig4}b), as this  results in a curve with a minimum that is not centered at zero energy and a more pronounced coherent peak at $\omega >$0. The same argument also applies to the simulated TDOS curves, where simply adding the superconducting and pseudogap contributions together (i.e., two separate self-energies) does not reproduce the experimental data (see Figs.\,S8 and S9). 
Moreover, Fig.\,\ref{Fig3} clearly demonstrates that the light-induced pseudogap is not driven by the electronic temperature, and the $\sim$1\,ps response time rules out potential inhomogeneous optical excitation effects.

Another important piece of experimental evidence is the restoration of the spectral symmetry when long-range SC order is present. By visual inspection of Figs.\,\ref{Fig2}c and\,\ref{Fig4}b  for cut 1, it is possible to track the re-emergence of a symmetric spectrum for late pump-probe delays, as if the pseudogap phase was quenched by the presence of static SC order. 
%However, the potential quench of the pseudogap phase at low temperature is challenged by several experimental observations. Indeed, we note that (i) the quasiparticle peaks of the asymmetric and symmetric TDOS in Fig.\,\ref{Fig4}b fall at approximately the same energy position -- and we reiterate that
%both pseudogap and SC phases are needed to capture the asymmetric TDOS, see Fig.\,\ref{Fig4}c --, as well as (ii) the position of antinodal k$_\text{F}$ (which hits for particle-hole asymmetry of the pseudogap) does not change when cooling below T$_C$ \cite{hashimoto2010particle} and (iii) competition between the pseudogap and SC has been reported for temperatures well below T$_C$ \cite{kuspert2022pseudogap,guyard2008breakpoint,kondo2009competition}. 
In this regard, we note that the restoration of the particle-hole symmetry in the low-temperature SC state does not necessarily imply any changes in the pole responsible for the pseudogap. Instead, any reduction of the pseudogap strength may enable the transfer of enough spectral weight into the Bogoliubov coherent peaks, as corroborated by simulations in Fig.\,S9 and S10.

Our data suggest that SC correlations and the pseudogap phase do not necessarily compete when SC correlations are short-range/fluctuating, but the competition sets in only once SC becomes long-range. This is reminiscent of the competition between the charge density wave order and superconductivity in Y-based cuprates, in which the competition is only evident when SC becomes long-range (for T$<$T$_\text{c}$) \cite{ghiringhelli2012long,chang2012direct}, and, in analogy to this work, it can be softened by disrupting the SC phase \cite{wandel2022enhanced,jang2022characterization}. However, we note that charge density wave order in Bi2212 does not display any evident competition with low-temperature SC order \cite{boschini2021dynamic,da2024dynamic}, although recent work suggest an effect on the quasiparticle residue \cite{smit2025enhanced}. Given that our data clearly show a suppression of the pseudogap when long-range SC is restored (Fig.\,\ref{Fig2} and \ref{Fig4}), we deem it unlikely that charge order correlations are responsible for the pseudogap phase in cuprates.
Instead, a dynamic competition between superconductivity and spin excitations has been reported in La-based cuprates \cite{decrausaz2025dynamic}, and would support a reduction of the strength of a spin-fluctuations-induced pseudogap when long-range superconductivity is established.
Interestingly, this would imply an anti-correlation between the spectral weight of the superconducting Bogoliubov peaks and the in-gap spectral weight in the pseudogap, in agreement with previous ARPES studies \cite{kuspert2022pseudogap,kondo2009competition}. Future studies correlating the evolutions of particle-hole asymmetry, pseudogap spectral weight, and magnetic fluctuations may lay the groundwork for a comprehensive understanding of the competition and fundamental origin of unconventional superconducting and pseudogap states in cuprates.
Nevertheless, the most important finding of this work is that the comparison between the experimental data and the phenomenological theory (Fig.\,\ref{Fig4}) constitutes the first direct probe of the competition between superconductivity and pseudogap in the unoccupied states, which, in turn, dictates the low-temperature Fermiology of the cuprates.

\vspace{5mm}
\large\noindent\textbf{Methods}\label{methods}
\vspace{2mm}
\normalsize

\textbf{Experimental set-up}\\
TR-ARPES measurements have been performed at the Advanced Laser Light Source (ALLS) facility at the Institut national de la recherche scientifique (INRS), EMT centre. The experimental setup is described in detail in a previous work \cite{longa2024time}. TR-ARPES measurements used s-polarized 6\,eV (206.5\,nm) light to elicit the emission of electrons and s-polarized 300\,meV (4\,$\mu$m) pump pulses to photoexcite the sample. The samples were cleaved in an ultra-high vacuum better than 6$\cdot10^{-11}$ Torr, at a base temperature of 10\,K, and photoelectrons have been detected via a SPECS ASTRAIOS 190 hemispherical analyzer. In an effort to enhance the momentum coverage, the sample was biased to 48\,V with respect to the nose of the electron analyzer (grounded).   
The overall energy resolution was $\sim$18\,meV, and the temporal resolution was estimated to be $\sim$300\,fs by measuring ultrafast dynamics on a topological insulator (precise zero pump-probe delay was also confirmed on the topological insulator).
The pump-induced average heating of the sample was estimated by fitting the Fermi edge at negative pump-probe delays in both fluence regimes.

\textbf{Analysis of TR-ARPES data}\\
\emph{Deconvolution:} Experimental spectra have been deconvolved from the experimental energy resolution via Lucy-Richardson deconvolution algorithm. The number of iterations (15) of the algorithm has been selected to minimize the enhancement of the noise level while maintaining satisfactory matching between the raw data and the devonvoluted data convoluted by the energy resolution.
The energy resolution has been estimated via several independent methods: (i) Fermi-Dirac fit of the k-integrated nodal EDC without pump pulse on the sample; (ii) fit of the low-energy cut-off in $\Gamma$ \cite{gauthier2021expanding}; (iii) Fermi-Dirac fit of polycrystalline Au.

\emph{Data processing:} Following the deconvolution, ARPES data are divided by the transient Fermi-Dirac distribution extracted along the nodal direction (see Fig.\,\ref{Fig1}). Moreover, minor transient shifts of the electronic band structure ($<$2\,meV) are corrected following Refs.\,\cite{miller2015photoinduced,miller2017particle}. Further details are provided in the SI, Fig.\,S1.
\\

\emph{TDOS:}  
The tomographic density of states (TDOS) is obtained by dividing the momentum-integrated EDCs deconvoluted from the energy resolution (red arrows in Figs.\,\ref{Fig2}a and d indicate the integration range) by a Fermi-Dirac function evaluated at the transient electronic temperature T$_e(\tau)$ extracted along the nodal direction \cite{reber2012origin}.
Under the approximation of constant photoemission matrix elements (which is reasonable in an energy range of a few tens of meV), the photoemission intensity is $I(\textbf{k},\omega)=[f(\omega) A(\textbf{k},\omega)] \ast R(\omega)$, where $f$ is the Fermi-Dirac distribution, $A$ the one-electron removal spectral function, and $R$ the energy resolution \cite{damascelli2003angle}. Upon deconvolving the energy resolution using the Lucy-Richardson algorithm discussed above,  
\begin{equation} 
\label{eq1}
\text{TDOS}_{\varphi}(\omega)=\frac{\int_{\varphi} d\textbf{k} I(\textbf{k},\omega)}{f(\omega)}  \propto \int_{\varphi} d\textbf{k} A(\textbf{k},\omega).
\end{equation}
If the selected momentum cut $\varphi$ is perpendicular to the Fermi surface, and a well-defined superconducting gap $\Delta$ can be defined, the TDOS is well fit by the Dynes function
\begin{equation} 
\label{eq2}
\text{TDOS}_{\varphi}(\omega)=Re\frac{\omega -i\Gamma}{\sqrt{(\omega -i\Gamma)^2-\Delta^2}},
\end{equation}
which captures the DOS of a conventional BCS superconductor with $\Gamma$ being a broadening scattering term including single-particle and pair-breaking scattering contributions.
\\

\textbf{Phenomenological calculations}

%\og{My proposal for this section:}

The theoretical TDOS at the energy $\omega$ that are reported in this work were obtained by calculating 
\begin{equation}
    \text{TDOS}_\phi (\omega) = \sum_{\textbf{k} \in \phi} \frac{-\frac{1}{\pi}\text{Im}\Sigma_\textbf{k}(\omega)}{(\omega - \epsilon_\textbf{k} - \text{Re} \Sigma_\textbf{k}(\omega))^2 + (\text{Im} \Sigma_\textbf{k}( \omega))^2},
\end{equation}
where $\phi$ is a path in the Brillouin zone, $\epsilon_\textbf{k}$ is a non-interacting dispersion taken from the ARPES fits reported in Ref.~\citenum{Drozdov2018} on the two-dimensional square lattice, and $\Sigma_\textbf{k}(\omega)$ is the self-energy which we constructed phenomenologically to capture the interplay between the pseudogap and the superconducting phases.

As detailed in the Supplementary Information, the self-energy is obtain by considering that electrons and holes are coupled through a superconducting gap function $\Delta_{\textbf{k}}$, while electrons are coupled to a set of auxiliary fermions with frequency $\omega_0$ through a Higgs field $H_\textbf{k}$ that captures the pseudogap behavior. Considering scattering rates in the single-particle ($\Gamma_s$), pairing ($\Gamma_p$) and pseudogap ($\Gamma_{pg}$) channels, the effective Hamiltonian reads $\mathcal{H} = \sum_{\textbf{k}} \bm \Psi_{\textbf{k}} \boldsymbol{\mathcal{H}}_{\textbf{k}} \bm \Psi_{\textbf{k}}^\dagger$ with
\begin{equation}
    \boldsymbol{\mathcal{H}}_{\textbf{k}} = 
    \left( 
        \begin{array}{cccc}
            \epsilon_\textbf{k} & H_\textbf{k} & \Delta_\textbf{k} & 0 \\
            H_\textbf{k} & \omega_0 & 0 & \Delta_\textbf{k} \\
            \Delta_\textbf{k} & 0 & -\epsilon_{-\textbf{k}} & - H_\textbf{k} \\
            0 & \Delta_\textbf{k} & - H_\textbf{k} & - \omega_0
        \end{array}
    \right)
    \label{eq:interplay_hamiltonian}
\end{equation}
in the spinor basis
\begin{equation}
    \bm \Psi_\textbf{k} \equiv 
    \left( \begin{array}{ cccc }
        c^\dagger_{\textbf{k}\uparrow} & \bar{c}^\dagger_{\textbf{k}\uparrow} & c_{\textbf{k}\downarrow} & \bar{c}_{\textbf{k}\downarrow}
    \end{array} \right)
\end{equation}
where $c^\dagger_{\textbf{k}\sigma}$ ($c_{\textbf{k}\sigma}$) and $\bar{c}^\dagger_{\textbf{k}\sigma}$ ($c_{\textbf{k}\sigma}$) respectively create (annihilate) an electron and an auxiliary fermion with momentum \textbf{k} and spin $\sigma$.
The self-energy is extracted from the first diagonal component of the full Green's function
\begin{equation}
    \label{eq:electron_Green}
    G_\textbf{k}(\omega) \equiv \bm G_\textbf{k}(\omega)\Big\lvert_{00} = \frac{1}{\omega \bm 1 - \boldsymbol{\mathcal{H}}_{\textbf{k}} + i \bm \Gamma }\Bigg\lvert_{00}
\end{equation}
with $\bm \Gamma = \Gamma_{s} \oplus \Gamma_{pg} \oplus \Gamma_p \oplus (\Gamma_{pg} + \Gamma_p)$
by solving the Dyson equation
\begin{equation}
    \Sigma_\textbf{k}(\omega) \equiv \omega - \epsilon_\textbf{k} - G_\textbf{k}(\omega)^{-1}.
\end{equation}
The final expression can be found as Eq.~(S14) in the Supplementary Information.

\vspace{5mm}
\large\noindent\textbf{Acknowledgements}
\vspace{2mm}
\normalsize

We thank the ALLS technical team for their support in the laboratory. We thank M. Bluschke, , E. H. da Silva Neto, A. Damascelli, A. Frano, S. Smit and M. Zonno for their critical reading of the manuscript and for useful discussions, as well as Lucy Reading-Ikkanda/Simons Foundation for providing graphic design support. The work at ALLS was supported by the Canada Foundation for Innovation (CFI) -- Major Science Initiatives. We acknowledge support from the Alfred P. Sloan Foundation (F.B), the Natural Sciences and Engineering Research Council of Canada (F.B., F.L., B.J.S), the Canada Research Chairs Program (F.B.,B.J.S.), the CFI (F.B., F.L., B.J.S.); the Fonds de recherche du Qu\'{e}bec --- Nature et Technologies (F.B., F.L., B.J.S.), the Minist\`{e}re de l'\'{E}conomie, de l'Innovation et de l'\'{E}nergie --- Qu\'{e}bec (F.B., F.L., B.J.S.), PRIMA Qu\'{e}bec (F.B., F.L.), and the Gordon and Betty Moore Foundation’s EPiQS Initiative, grant GBMF12761 (F.B., F.L.). The Flatiron
Institute is a division of the Simons Foundation.
The work at BNL was supported by the US Department of Energy, office of Basic Energy Sciences, contract no. DOE-sc0012704.

\newpage

\newpage

\clearpage
\newpage

\renewcommand{\thefigure}{S\arabic{figure}}
\renewcommand{\theequation}{S\arabic{equation}}

\setcounter{figure}{0}
\setcounter{equation}{0}

\section*{Supplementary Information for:\\ \textit{Light-induced Asymmetric Pseudogap below $T_\text{c}$ in cuprates}}

\section{Data treatment and analysis}

The photoemission signal can be written as \cite{damascelli2003angle}
    \begin{equation}
    I(\boldsymbol{k},\omega) = \left[ |M|^2\cdot f(\omega) \cdot A(\boldsymbol{k},\omega)\right] \ast R(\omega),
\end{equation}
where $|M|^{2}$ is the dipole matrix element, $f(\omega)$ the electronic distribution, $A(\boldsymbol{k},\omega)$ the one-electron spectral function, and $R(\omega)$ the experimental energy resolution. The purpose of the data analysis process is to isolate the contribution of the spectral function, allowing for direct consideration of the electronic density of states (DOS).

\begin{figure}[b]
    \centering
    \includegraphics[width=0.8\textwidth]{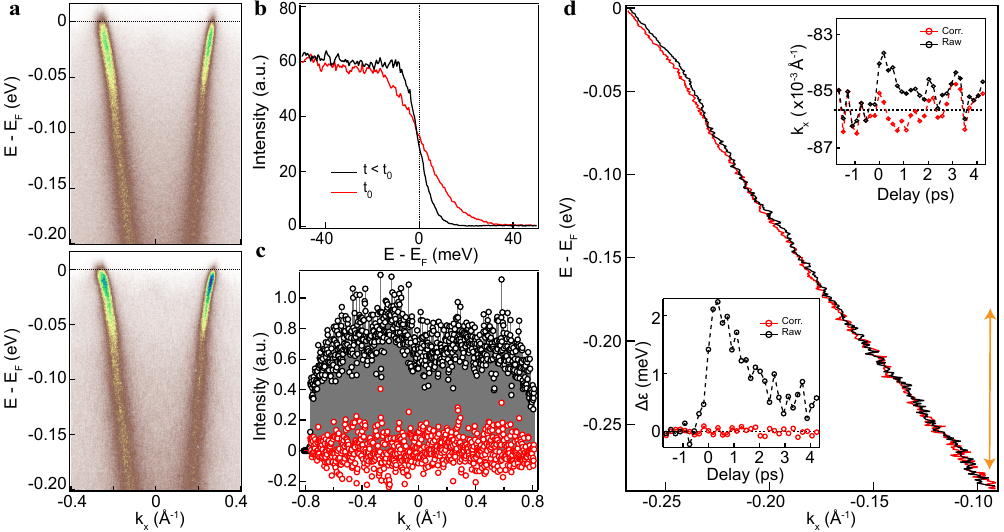}
    \caption{\textbf{Data treatment and analysis.}
    \textbf{a} Comparison between raw (top) and deconvolved (bottom) nodal ARPES spectra.
    \textbf{b} Apparent chemical potential shift upon pump excitation. 
    \textbf{c} MDCs at 100\,meV above $E_F$ (30\,meV integration window) before (black) and after (red) background subtraction. \textbf{d} Nodal electronic dispersion extracted via MDC fitting from -0.28 eV to $E_F$ before (black) and after (red) apparent chemical potential correction. Bottom inset: chemical potential shift as a function of the pump-probe delay extracted via Fermi-Dirac fit before (black) and after (red) the chemical potential shift correction. Top inset: shift of the electronic dispersion as a function of the pump-probe delay, integrated within the region indicated by the yellow arrow in d, before (black) and after (red) the chemical potential shift correction.
    }
    \label{FigS1}
\end{figure}

To achieve this, we deconvolve the energy resolution $R(\omega)$ from the photoemission intensity using the Lucy-Richardson algorithm (see Methods), as displayed in Fig.\ref{FigS1}a. %Moving on, as the matrix element is mainly dependent on the light polarization and the symmetry of the orbitals probed, it is considered constant throughout the experiment, coherently with other work on the same material [\da{review}]. 
\begin{figure}[b]
    \centering
    \includegraphics[width=0.78\textwidth]{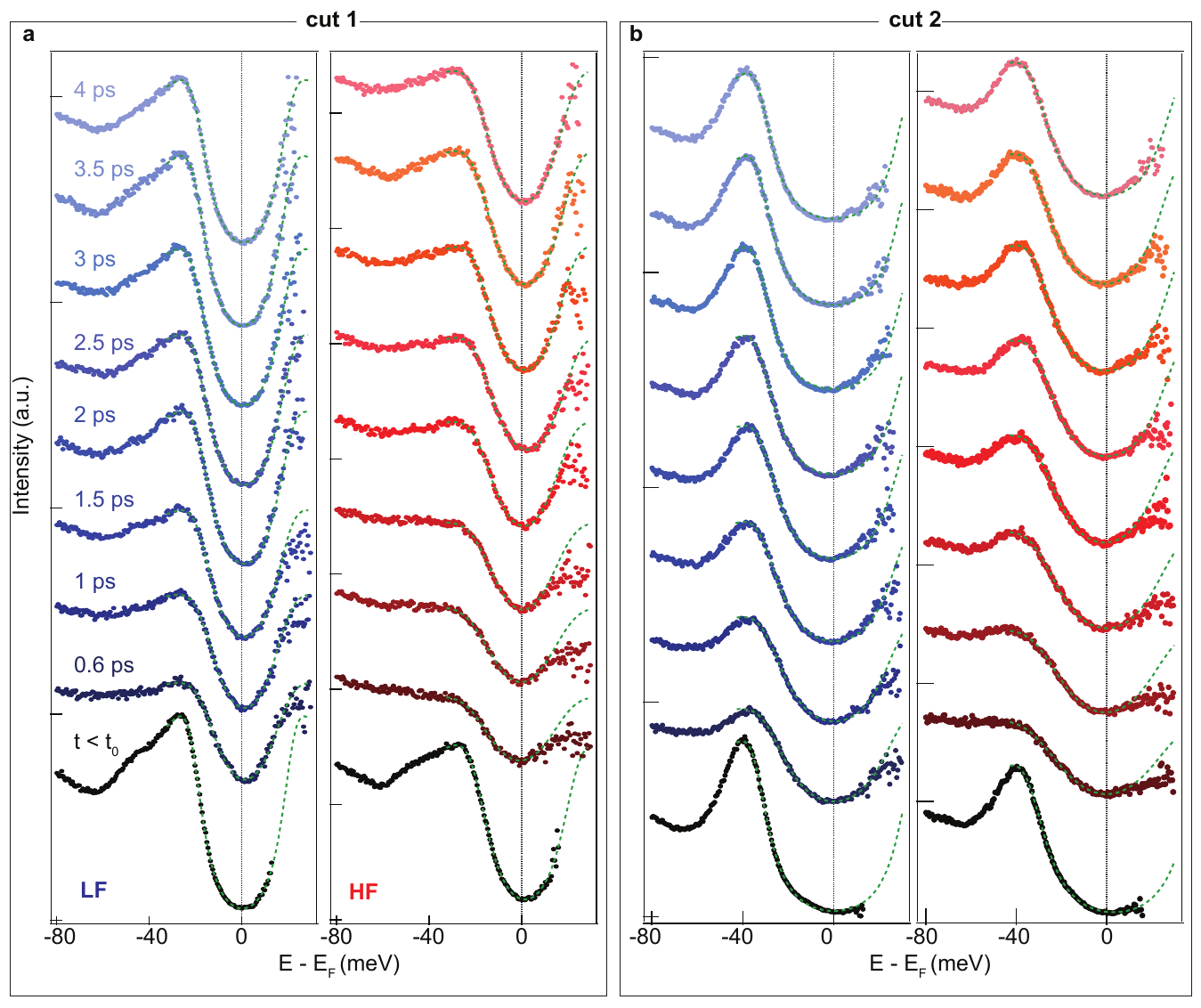}
    \caption{\textbf{Transient evolution of the TDOS curves.}
     \textbf{a} and \textbf{b}, similar to Fig.\,3 in the main text, transient evolution of the TDOS for cuts 1 and 2, respectively, in the LF (blue curves)  and HF (red curves) regimes.
    }
    \label{FigS10}
\end{figure}
Then, we remove the Fermi-Dirac distribution $f(\omega)$ contribution to access the spectral function. A transient Fermi-Dirac distribution following the electronic temperature evolution (see Fig.\,1c of the main text) can be extracted for each pump-probe delay by fitting the momentum-integrated EDC along the nodal region. Before proceeding with the Fermi-Dirac division, a constant background evaluated at $100 \pm 30$ meV above $E_{F}$ is subtracted from the whole spectrum (see Fig.\ref{FigS1}c). Moreover, in agreement with what has already been reported in TR-ARPES studies on Bi2212 \cite{miller2015photoinduced,miller2017particle}, a transient shift in the energy of the chemical potential upon pump excitation is observed and corrected for by rigidly shifting the data in energy (Fig.\,\ref{FigS1}b and bottom inset in Fig.\,\ref{FigS1}d). The fact that we can completely correct the apparent shift in the chemical potential by simply shifting the entire electronic dispersion rigidly reassures us about the absence of any photodoping effects. The rigid shift of the electronic band structure may be caused by minor changes in the local work function or excitation of incoherent phonons \cite{boschini2024time}.
%The validity of this correction is verified by reconstructing the electronic distribution of the nodal cut via MDC fit position within the energy range [-0.28;0] eV before and after the chemical potential correction (Fig. S1d). Notably, the center of mass of the band is not affected (orange arrow, top inset in Fig. S1d).

%Upon Fermi-Dirac division, the accessible energy range above $E_F$ is limited by the dynamic range of the detector

%of the nodal momentum integrated EDCs of the off-nodal momentum cuts $\varphi=1$, $\varphi=2$ and $\varphi=3$, the limit of accessible states above $E_F$ is limited to the ones within O($k_{B}T$) of $E_{F}$, hence by the dynamic range of the detector. An alternative approach to remove the contribution of $f(\omega)$ is to symmetrize the ARPES intensity about $E=E_{F}$. This approach overcomes the technical limitation of the detector, but assumes the particle-hole symmetry of the system, therefore unfeasible for this study. 

%\clearpage
%\newpage

\section{Transient evolution TDOS curves - Extended}
Figure\,\ref{FigS10} displays TDOS curves along cuts 1 and 2 in the LF (blue) and HF (red) regimes for various pump-probe delays.

%\clearpage
%\newpage
\section{Transient evolution of the spectral function at $k_{F}$}
\begin{figure}[b]
    \centering
    \includegraphics[width=0.85\textwidth]{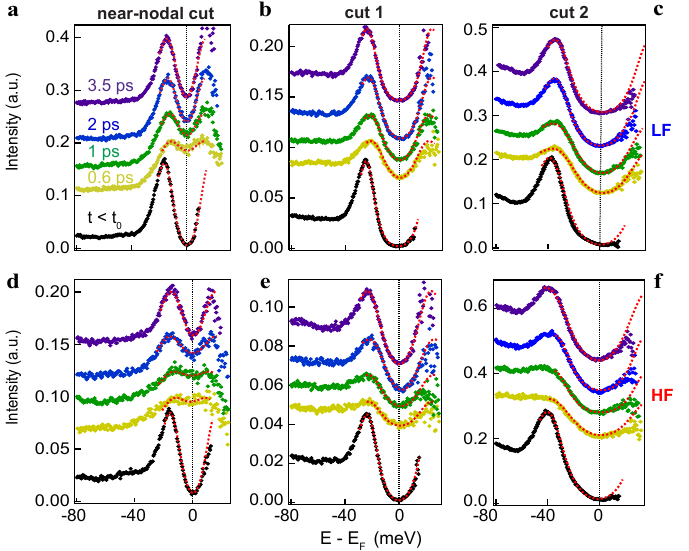}
    \caption{\textbf{Transient evolution of the off-nodal and near-antinodal spectral function at the Fermi momentum $k_F$.}
    \textbf{a}, \textbf{b} and \textbf{c} transient evolution of the spectral function at $k=k_F$ for cuts near-nodal, 1 and 2, respectively, in the LF case at various pump-probe delays. The EDCs are plotted up to the experimental limit dictated by $k_\text{B}$T$_e(\tau)$. The superimposed red dahsed line represents the Norman fit. %captures well the evolution of the SC gap and its particle-hole symmetry in excellent agreement with Fig.\,2 of the main text.
    \textbf{d}, \textbf{e} and \textbf{f} same as a,b, and c but in the HF regime. %The light excitation completely quenches the SC condensate in cut 2 and 3, while the particle-hole symmetry is preserved in cut 1, where the PG is not present.
    }
    \label{FigS2}
\end{figure}
In addition to the tomographic density of states (TDOS) analysis presented in the main text, the transient evolution of the spectral function at the Fermi wave vector $k_F$ provides essential information on the transient evolution of the superconducting gap. We look at EDCs integrated $\pm 0.01 \AA^{-1}$ around $k_{F}$ for every pump-probe delay. The EDCs along the near-nodal, 1, and 2 cuts are displayed in Fig.\,\ref{FigS2} for selected pump-probe delays. These transient EDCs are fit by Norman's model for the self energy \cite{norman1998phenomenology}: 
\begin{equation}
\label{EqNorman}
\Sigma(\omega) = -i\Gamma_s + \frac{\Delta^2}{i\Gamma_p + \omega},   
\end{equation}
where $\Delta$ is the SC gap size, $\Gamma_s$ is the single particle scattering rate and $\Gamma_p$ is the pair breaking scattering rate, i.e. the inverse of the Cooper pairs' lifetime.

In good agreement with the TDOS analysis in the main text, the particle-hole symmetric Norman fit well captures the LF data for all three cuts investigated, while a transient spectral weight asymmetry is observed at 0.6\,ps and 1\,ps pump-probe delays for cuts 1 and 2, recovering afterwards. Interestingly, we do not detect any transient asymmetry for the near-nodal cut, where the pseudogap is not present.

%\newpage
\section{Transient evolution of the pair breaking scattering rate and superconducting gap}

\begin{figure}[H]
    \centering
    \includegraphics[width=0.9\textwidth]{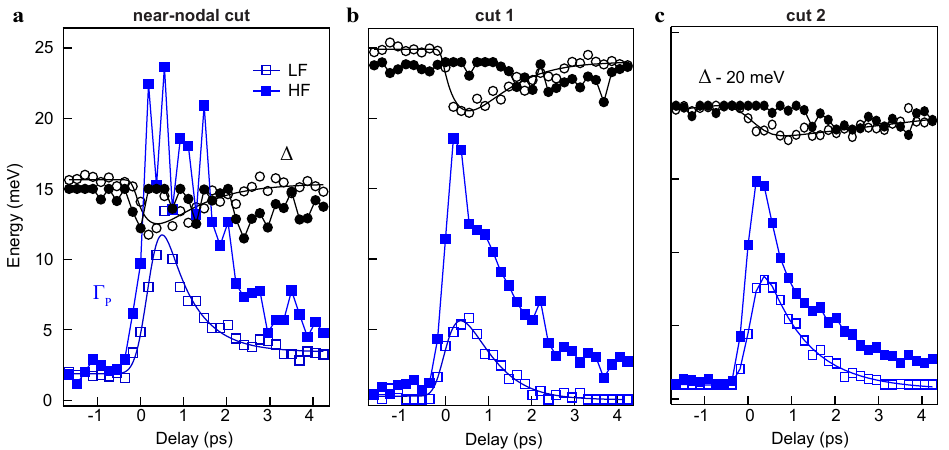}
    \caption{\textbf{Transient evolution of the pair breaking scattering rate $\Gamma_p$ and superconducting gap $\Delta$ extracted from Norman fit.}
    \textbf{a} pair breaking scattering rate $\Gamma_p$ and superconducting gap $\Delta$ for the near nodal cut, as a function of the pump-probe delay in the LF and HF regimes using Eq.\,\ref{EqNorman}. In the LF case, the evolution of both $\Gamma_p$ and $\Delta$ are captured with a single exponential fit. \textbf{b} and \textbf{c} are the same as a, but for cuts 1 and 2, respectively. %The line between points emphasises the saturation of $\Gamma_p$ observed in $\varphi=1$.
    }
    \label{FigS3}
\end{figure}

The Norman fits allow for tracking of the pair-breaking scattering rate and superconducting gap size as a function of the pump-probe delay. The single particle scattering rate ($\Gamma_S$) is extracted from the nodal cut for both LF and HF cases and fixed for the off-nodal fits, in agreement with Ref.\,\cite{boschini2018collapse}. The transient evolution of these parameters is reported in Fig.\,\ref{FigS3}. Starting from the LF case, both $\Delta$ and $\Gamma_p$ are comparable with what has been previously reported \cite{boschini2018collapse}, and their trend is captured by a single exponential decay. In the HF case, the combination of significant broadening of $\Gamma_s$ and filling of $\Delta$ makes the fit unstable. Nevertheless, we report an enhancement of $\Gamma_p$ in the HF regime, and reveal a saturation of $\Gamma_p$ at around 1\,ps for the near-nodal cut. %implying a shorter CPs lifetime at the pump-probe delay where the p-h asymmetry is observed. We speculate that a smaller gap size allowed for a more significant filling, leading to the trend observed in $\varphi=1$ and missing in the other two momentum cuts. 

%\newpage
\section{Equilibrium vs. out-of-equilibrium electronic distribution}

\begin{figure}[H]
    \centering
    \includegraphics[width=0.55\textwidth]{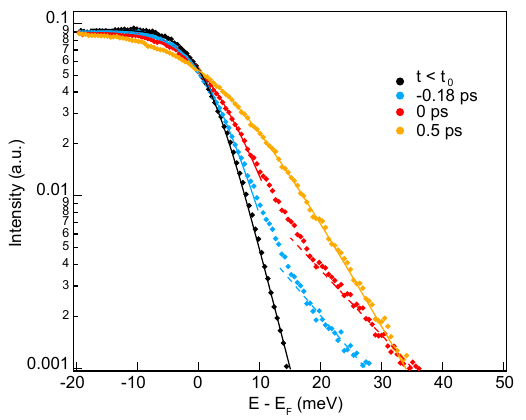}
    \caption{\textbf{Transient evolution of the nodal electronic temperature.}
    Momentum-integrated nodal EDCs at several pump-probe delays. The superimposed solid line represents Fermi-Dirac fit in the energy window [-20;50] meV for $t<t_{0}$ and 0.5 ps curves, and [-20;10] meV for the -0.18 ps and 0 ps curves. Starting from the equilibrium curve (black), the pump excitation induces a transient non-equilibrium electronic distribution at pump-probe delays around $t_{0}$, highlighted by the dashed lines. A quasi-equilibrium distribution is then reached for longer delays (200--300\,fs after pump excitation).  
    }
    \label{FigS4}
\end{figure}

To remove the thermal contribution from the photoemission intensity and track the electronic temperature throughout the experiment, it is essential to determine when an effective electronic temperature can be defined \cite{boschini2024time}. As mentioned in the main text, we extract the electronic temperature by fitting the momentum-integrated nodal EDC with a Fermi-Dirac distribution. 
%However, it is well established in the TR-ARPES community that the pump excitation induces a strong non-equilibrium electronic distribution, therefore a proper FD distribution (and consequently an electronic temperature) cannot be defined for every pump-probe delay. Here 
We use momentum-integrated nodal EDCs (Fig.\,\ref{FigS4}) to distinguish between out-of-equilibrium and quasi-equilibrium electronic distribution. Indeed, at pump-probe delays around $t_{0}$, the EDCs show an additional slope, and they are not captured by a Fermi-Dirac fit (energy window [-20; 40] meV). Instead, for $\tau >$200\,fs, the Fermi-Dirac fit is robust. 

%\newpage
\section{Non-thermal nature of the light-induced asymmetric pseudogap - extended}

\begin{figure}[H]
    \centering
    \includegraphics[width=0.67\textwidth]{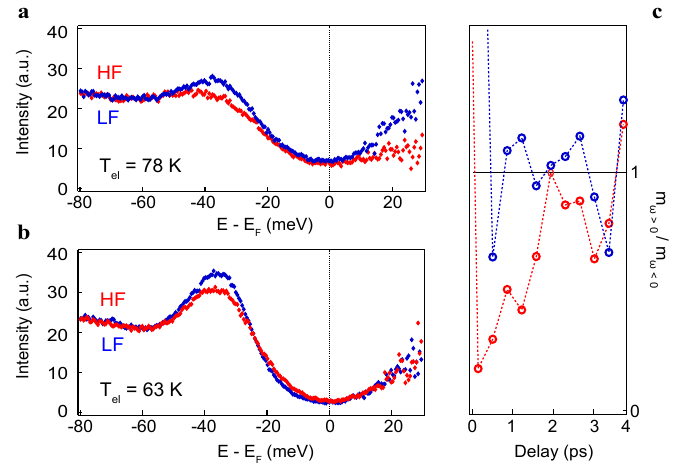}
    \caption{\textbf{Demonstration of the non-thermal character of the light-induced spectral weight asymmetry.}
    \textbf{a} and \textbf{b} Comparison of TDOS curves along cut 2 taken at different pump-probe delays in the LF (0.5\,ps and 3\,ps) and HF (1.1\,ps and 4\,ps) regime, but sharing a similar $\text{T}_e$, namely $\sim$78\,K in a and $\sim$63\,K in b. 
    \textbf{c} Transient evolution of the linear slope ratio of the TDOS along cut 2 as defined in the main text, extracted in the range [-20;-4]\,meV and [4;20]\,meV below and above $E_\text{F}$, respectively. %In the LF regime (blue) the ratio remains centered around 1 due to the symmetric nature of the SC phase. Instead, in the HF regime (red), we report a reduction in the slope ratio centered around 1\,ps pump-prove delay resulting from the spectral weight asymmetry reported previously. 
    %\textbf{d} HF TDOS at three different pump-probe delays, highlighted by the green lines in panel c (and corresponding to three different $\text{T}_e$).% We also superimpose the TDOS acquired at 130\,K in the PG phase (orange curve). As shown in the main for $\varphi=2$, one can transition from the HF to the PG TDOS (orange arrows) by transferring spectral weight from the remnant coherent peak into the gap at $\omega$=0.
    }
    \label{FigS5}
\end{figure}

To confirm the non-thermal nature of the light-induced spectral weight asymmetry, we extend Fig.\,4 of the main text for cut 2. We compare the LF and HF TDOS curves at the same electronic temperature for two different pump-probe delays (comparison of the nodal momentum-integrated EDCs at these delays can be found in Fig.\,4 of the main text). For the higher electronic temperature (78 K) we highlight the difference between the two TDOSs in Figure\,\ref{FigS5}a, while particle-hole symmetry is restored at longer delays (63 K) in Figure\,\ref{FigS5}b.

 Moreover, we repeat the slope ratio analysis to quantify the transient particle-hole asymmetry of the system in the two pumping regimes (Figure\,\ref{FigS5}c). In the LF regime (blue) the ratio remains centered around 1 due to the symmetric nature of the SC phase. Instead, in the HF regime (red), we report a transient reduction in the slope ratio.
 %centered around 1\,ps pump-prove delay resulting from the spectral weight asymmetry reported previously.  

\section{Competition and coexistence between superconductivity and pseudogap - extended}

\begin{figure}[t]
    \centering
    \includegraphics[width=0.6\textwidth]{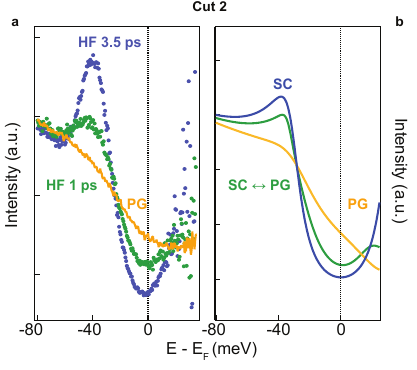}
    \caption{
    \textbf{Competition and coexistence between superconductivity and pseudogap along cut 2.}
    \textbf{a} comparison of three experimental TDOS curves along cut 2: Equilibrium pseudogap phase at 130 K (yellow), light-induced pseudogap state at 1 ps in the HF regime (green), and restored superconducting particle-hole symmetric state at 3.5 ps in the HF regime (blue). 
    \textbf{b} Simulated curves for pseudogap (orange), SC correlations (blue) and coexistence of SC and PG (green).
    %using the phenomenological self-energy discussed in the Supplementary Information. When only pseudogap (orange) and/or SC (blue) correlations are considered, the light-induced pseudogap is not reproduced. Instead, the coexistence of SC and the pseudogap (green) provides a better fit to the experimental evidence in panel b. When long-range SC correlations are restored, the particle-hole symmetric SC contribution alone (blue) reproduces the data. 
    }
    \label{FigS6}
\end{figure}

To further confirm the theoretical model presented in the main text, it is also important to extend it to cut 2. Figure\,\ref{FigS6} compares experimental (a) and theoretical (b) TDOS curves along cut 2. The experimental light-induced pseudogap is reproduced only when considering the interplay between SC and PG states.

\clearpage
\newpage

%%%%%%%%%%%%%%%%%%%%%%%%%%%%%%%%%%%%%%%%%%%%%%%%%%%%%%%%%%%%%%%%%%%%%%%%%%%%
%%%%% BEGIN THEORY SECTION %%%%%%%%%%%%%%%%%%%%%%%%%%%%%%%%%%%%%%%%%%%%%%%%%
%%%%%%%%%%%%%%%%%%%%%%%%%%%%%%%%%%%%%%%%%%%%%%%%%%%%%%%%%%%%%%%%%%%%%%%%%%%%
\section{Theory}

Throughout this work, the main quantity that we compare and study is the tomographic density of states (TDOS). At an energy $\omega$, it is obtained by calculating
    \begin{equation}
        \text{TDOS}_\phi (\omega) = \sum_{\textbf{k} \in \phi} A_\textbf{k}(\omega) \equiv \sum_{\textbf{k} \in \phi} \frac{-\frac{1}{\pi}\text{Im}\Sigma_\textbf{k}(\omega)}{(\omega - \epsilon_\textbf{k} - \text{Re} \Sigma_\textbf{k}(\omega))^2 + (\text{Im} \Sigma_\textbf{k}( \omega))^2},
    \end{equation}
    where $\phi$ is a path in the Brillouin zone, $\epsilon_\textbf{k}$ is a non-interacting dispersion of electrons in the system and $\Sigma_\textbf{k}(\omega)$ is the electronic self-energy which includes electronic correlations at the one-particle level. 
    
    In this section, we present the phenomenological models of $\epsilon_\textbf{k}$ and $\Sigma_\textbf{k}(\omega)$ used in the main text to reproduces the experimental data. 
    We start in Sec.~\ref{sec:theory_dispersion} by fixing the dispersion according to ARPES data. 
    In Sec.~\ref{sec:theory_derivation}, we derive a phenomenological self-energy that can capture the interplay between the pseudogap physics and superconductivity.
    By taking various limits of this expression, we show that it recovers the self-energies typically used in the literature when describing only the pseudogap or the superconducting state.
    We also introduce a frequency-dependent cut off to better capture the high-energy behavior of the TDOS.
    From this general framework, we study the effect of the various parameters of the model and constraint them in Sec.~\ref{sec:theory_params} by comparing calculations with experimental measurements taken in different regimes, in particular in the pseudogap and the superconducting states.
    Finally, in Sec.~\ref{sec:theory_interplay}, we investigate the interplay between the pseudogap and superconducting regimes.

\subsection{\label{sec:theory_dispersion}Non-interacting dispersion}

    To model the dispersion, we consider a two-dimensional single-orbital square lattice defined by a set of hopping amplitudes $t_{\bm \delta}$ between two sites distanced by the vector $\bm \delta$, and a chemical potential $\mu$ that fixes the number of electrons in the model. The dispersion $\epsilon_{\textbf{k}}$ at the momentum-point \textbf{k} is then given by
    \begin{equation}
        \label{eq:tight_binding} 
        \epsilon_\textbf{k}\equiv \sum_{\bm \delta} t_{\bm \delta} e^{i\textbf{k} \cdot \bm \delta} - \mu.
    \end{equation}

    For the Bi2212 material with a superconducting critical temperature $T_C \sim 91$~K (OP91), the parameters $\{t_{\bm \delta}\}$ for the first three nearest neighbors and $\mu$ were already extracted from ARPES measurements and reported in Ref.~\citenum{Drozdov2018}.
    We write them in Table~\ref{tab:tb_params} for convenience.
    In this manuscript, they also report $t_\perp = 108$~meV which characterizes the hopping between two different layers, but we do not consider it in this work.

    \begin{table}[h!]
        \centering
        
        \begin{tabular}{c|cccc}
            \hline \hline
            \quad Hamiltonian term \quad  & $\mu$ & \quad $t_{1, 0}$ \quad & \quad $t_{1, 1}$ \quad & \quad $t_{2, 0}$ \quad \\
            \hline
            Value [meV] & \quad 405 \quad & 360 & 108 & 36 \\
            \hline \hline
        \end{tabular}
        \caption{\label{tab:tb_params}Values used in  Eq.~\ref{eq:tight_binding} to model the dispersion of Bi2212-OP91.
        They were obtained from Ref.~\citenum{Drozdov2018} where they extracted them by fitting ARPES data.}
    \end{table}

    Note that these extracted values are only valid at low-energy, as this is where they were fitted. However, the effect of correlations is to narrow the quasi-particle peak and push some of its weight at high energies into the so-called Hubbard band. Therefore, this model is only good close to the Fermi energy, but is much narrower than the actual system and does not include any Hubbard band. It should thus not be trusted far from the Fermi level.

    Practically in the simulations, we use the \textsc{TBLattice} object of the \textsc{TRIQS} library to represent and manipulate this tight-binding model~\cite{PARCOLLET2015398}.

\subsection{\label{sec:theory_derivation}Phenomenological self-energy}

    While there are self-energies proposed to model the behavior of the pseudogap and of the superconducting state, there is none, to our knowledge, that capture the interplay between the two phenomena.

    \paragraph*{Superconductivity. ---} In the superconducting state, the self-energy close to the Fermi level can be modeled by~\cite{Kondo2015, PhysRevB.57.R11093} 
    \begin{equation}
        \label{eq:self_SC}
        \Sigma^{\text{SC}}_\textbf{k}(\omega) = \frac{\Delta_\textbf{k}^2}{\omega + \epsilon_{-\textbf{k}} + i\Gamma_p},
    \end{equation}
    where $\Delta_\textbf{k}$ is the superconducting gap function and, in the cuprates, it is believed to have a $d_{x^2-y^2}$ structure such as
    \begin{equation}
        \label{eq:sc_gap_func_d-wave}
        \Delta_\textbf{k} = \Delta_0\left(\cos^2 \left(\frac{ak_x}{2}\right) - \cos^2 \left( \frac{ak_y}{2} \right) \right).
    \end{equation}
    $\Gamma_p$ is the scattering rate due to pair breaking in the system.

    \paragraph*{Pseudogap. ---} The other phenomena we are interested to capture is the pseudogap regime where a \textbf{k}-dependent gap in the density of states develops. This behavior can be captured with a \textbf{k}-dependent self-energy, which has a pole at a finite frequency, which generates an asymmetric density of states around the Fermi level.
    In particular, one can derive the following general form for the pseudogap self-energy:
    \begin{equation}
        % \label{eq:self_PG}
        \Sigma^\text{PG}_{\textbf{k}}(\omega) = \frac{H_\textbf{k}\textbf{}^2}{\omega - \bar{\epsilon}_{\textbf{k}} + i\Gamma_{pg}},
    \end{equation}
    where $\bar{\epsilon}_\textbf{k}$ represents the position of the pole in frequency-space as a function of \textbf{k}, $H_\textbf{k}$ is the structure factor, and $\Gamma_{pg}$ is an associated pseudogap scattering rate.
    Some works based on high-precision numerical simulations on coarse momentum grids have suggested that the position of the pole could be dictated by $\bar{\epsilon}_\textbf{k} \rightarrow \epsilon_{\textbf{k}+(\pi, \pi)}$, the original dispersion shifted by $(\pi, \pi)$, and $H_\textbf{k} \rightarrow H_0$, a momentum-independent coupling strength~\cite{PhysRevLett.114.236402, PhysRevB.96.041105, PhysRevX.8.021048, doi:10.1073/pnas.1720580115}.
    
    We find that this parametrization does not reproduce the experimental measurements in the pseudogap regime. Instead, we need to keep the position of the pole constant in momentum-space ($\bar{\epsilon}_\textbf{k} = \omega_0$) and model the coupling strength with a $d$-wave structure $H_\textbf{k} \rightarrow H_0(\cos^2\frac{ak_x}{2} - \cos^2\frac{ak_y}{2})$, resulting in the following pseudogap self-energy:
    \begin{equation}
        \label{eq:self_PG}
        \Sigma^\text{PG}_{\textbf{k}}(\omega) = \frac{H_0\textbf{}^2(\cos^2\frac{ak_x}{2} - \cos^2\frac{ak_y}{2})^2}{\omega - \omega_0 + i\Gamma_{sf}}.
    \end{equation}

    This parametrization triggers questions about the mechanism leading to the emergence of the pseudogap; addressing these questions is an important topic for future research.

    \paragraph*{Interplay between the pseudogap and superconductivity. ---}
    To model a self-energy $\Sigma_{\textbf{k}}(\omega)$ which captures both the physics of the pseudogap and of superconductivity, we start from a phenomenological theory that includes both. We consider an Hamiltonian of the form $\mathcal{H} = \mathcal{H}_s + \mathcal{H}_{pg} + \mathcal{H}_{p}$.

    The first term describes the single-particle ($s$) dynamics of the electrons and is given by
    \begin{equation}
        \mathcal{H}_s = \sum_{\textbf{k}\sigma} \epsilon_\textbf{k} c^\dagger_{\textbf{k}\sigma} c_{\textbf{k}\sigma},
    \end{equation}
    with $c^\dagger_{\textbf{k}\sigma}$ ($c_{\textbf{k}\sigma}$) a creation (annihilation) operator of an electron with momentum $\textbf{k}$ and spin $\sigma$.

    The second term describes the coupling of the physical electrons to a set of auxiliary fermions, with the objective to simulate the processes that give rise to the pseudogap in the system. This term has the following form:
    \begin{equation}
        \mathcal{H}_{pg} = \sum_{\textbf{k}\sigma} \bar{\epsilon}_\textbf{k} \bar{c}^\dagger_{\textbf{k}\sigma} \bar{c}_{\textbf{k}\sigma} + \sum_{\textbf{k}\sigma} H_{\textbf{k}} \left[ c^\dagger_{\textbf{k}\sigma} \bar{c}_{\textbf{k}\sigma}  + \text{c.c} \right],
    \end{equation}
    where the auxiliary fermions are created (annihilated) by $\bar{c}^\dagger$ ($\bar{c}$) with a dispersion $\bar{\epsilon}_\textbf{k}$.
    In this work, we find that what reproduces best the experimental data (discussed in Sec.~\ref{sec:theory_params}, section \textit{Pseudogap state}) is to fix the dispersion of the auxiliary fermions as a simple pole at a value $\bar{\epsilon} = \omega_0$.
    The parameter $H_{\textbf{k}}$ is the strength of the coupling and measures the strength of the underlying processes. In general, it can be momentum-dependent.

    The third term describes the superconducting pairing ($p$) of the electrons and effective fermions. We consider singlet Cooper pairing with no center of mass momentum, leading to
    \begin{equation}
        \mathcal{H}_{p} = 
        \sum_{\textbf{k}\sigma} \Delta_\textbf{k} \left[ c_{\textbf{k}\sigma} c_{-\textbf{k}\bar{\sigma}} + \text{c.c.} \right] + 
        \sum_{\textbf{k}\sigma} \bar{\Delta}_\textbf{k} \left[ \bar{c}_{\textbf{k}\sigma} \bar {c}_{-\textbf{k}\bar{\sigma}} + \text{c.c.} \right],
    \end{equation}
    with $\Delta_\textbf{k}$ ($\bar{\Delta}_\textbf{k}$) the superconducting gap function (structure factor of the superconducting state) of the electrons (auxiliary fermions) and $\bar{\sigma} = - \sigma$.
    
    From this Hamiltonian, we can extract an effective self-energy affecting the physical electrons in the system. To do so, we write the Hamiltonian $\mathcal{H}$ in the Nambu basis for both the original electrons and effective fermions, that is
    \begin{equation}
        \begin{array}{l}
            \mathcal{H} = \sum_{\textbf{k}} \bm \Psi_{\textbf{k}} \boldsymbol{\mathcal{H}}_{\textbf{k}} \bm \Psi_{\textbf{k}}^\dagger  \\
            \quad \quad \text{with} \quad
        \bm \Psi_\textbf{k} \equiv 
        \left( \begin{array}{ cccc }
            c^\dagger_{\textbf{k}\uparrow} & \bar{c}^\dagger_{\textbf{k}\uparrow} & c_{\textbf{k}\downarrow} & \bar{c}_{\textbf{k}\downarrow}
        \end{array} \right)
        \end{array}
        \quad \text{and} \quad \boldsymbol{\mathcal{H}}_{\textbf{k}} = 
        \left( 
            \begin{array}{cccc}
                \epsilon_\textbf{k} & H_\textbf{k} & \Delta_\textbf{k} & 0 \\
                H_\textbf{k} & \bar{\epsilon}_\textbf{k} & 0 & \bar{\Delta}_\textbf{k} \\
                \Delta_\textbf{k} & 0 & -\epsilon_{-\textbf{k}} & - H_\textbf{k} \\
                0 & \bar{\Delta}_\textbf{k} & - H_\textbf{k} & - \bar{\epsilon}_{-\textbf{k}}
            \end{array}
        \right).
    \end{equation}
    Setting $\bar{\epsilon}_\textbf{k} = \omega_0$ leads to Eq.~4 of the main text. We then compute the associate Green's function $\bm G_\textbf{k}(\omega) \equiv \left[ \omega \bm 1 - \boldsymbol{\mathcal{H}}_\textbf{k} + i \bm \Gamma\right]^{-1}$, from which the first diagonal component is the effective Green's function of the physical electrons $G_\textbf{k}(\omega) \equiv \bm G_\textbf{k}(\omega) \big\lvert_{00}$.
    In this expression, we introduced a set of finite scattering rates $\bm \Gamma = \Gamma_{s} \oplus \Gamma_{pg} \oplus \Gamma_p \oplus (\Gamma_{pg} + \Gamma_p)$ which captures the broadening of the single-particle dispersion $\Gamma_s$, of the pseudogap processes $\Gamma_{pg}$ and of the superconducting pairs $\Gamma_p$.
    The effective self-energy is extracted using the Dyson equation
    \begin{equation}
        \Sigma_\textbf{k}(\omega) \equiv \omega - \epsilon_\textbf{k} - G_\textbf{k}(\omega)^{-1}.
    \end{equation}
    The effective self-energy can be written as
    \begin{equation}
        \label{eq:gen_self}
        \Sigma^{\text{full}}_\textbf{k}(\omega) = 
        - i \Gamma_s +
        \frac{
            H_ {\textbf{k}}^2 \cdot
            \left[\bm G^{-1}_0 \right]_{22} \cdot
            \left[\bm G^{-1}_0 \right]_{33}
            + 
            \Delta_\textbf{k}^2 \cdot
            \left[\bm G^{-1}_0 \right]_{11} \cdot
            \left[\bm G^{-1}_0 \right]_{33}
            -
            \left[ \Delta_{\textbf{k}} \bar{\Delta}_{\textbf{k}} + H_{\textbf{k}}^2 \right]^2
        }{
            \left[\bm G^{-1}_0 \right]_{11} \cdot
            \left[\bm G^{-1}_0 \right]_{22} \cdot
            \left[\bm G^{-1}_0 \right]_{33}
            - H_{\textbf{k}}^2 \cdot \left[\bm G^{-1}_0 \right]_{11}
            - \bar{\Delta}_{\textbf{k}}^2 \cdot \left[\bm G^{-1}_0 \right]_{22}
        },
    \end{equation}
    where $[\bm G^{-1}_0]_{ii} \equiv \omega - \left[\bm \epsilon_\textbf{k}\right]_{ii} + i\left[\bm \Gamma \right]_{ii}$. Taking different limits, we recover the self-energies of the individual phenomena, that is
    \begin{equation}
        \Sigma^{\text{full}}_\textbf{k}(\omega)\Big\lvert_{\Gamma_s=H_\textbf{k}=0}=\Sigma^{\text{SC}}_\textbf{k}(\omega),
        \ \
        \Sigma^{\text{full}}_\textbf{k}(\omega)\Big\lvert_{\Gamma_s=\Delta_\textbf{k}=0}=\Sigma^{\text{PG}}_\textbf{k}(\omega),
        \ \text{and} \ \
        \Sigma^{\text{full}}_\textbf{k}(\omega)\Big\lvert_{H_0=\Delta_\textbf{k}=0}=-i\Gamma_s,
    \end{equation}
    where the last one is simply an isotropic single-particle scattering.

    \paragraph*{Frequency-dependence. --- } 
    Above, the pseudogap and superconducting coupling strengths have been defined as static, but in reality, they primarily affect the low-energy spectrum close to the Fermi level. Therefore, they should be modeled with a frequency cut-off. Since a frequency dependence does not affect the results much for the superconducting gap function, we do not consider it to minimize the number of parameters in the theory. For the pseudogap coupling strength, however, it does, and we consider $H_\textbf{k}(\omega) = H_\textbf{k} f_H(\omega)$ where $f_H(\omega)$ behaves as a plateau centered around the pole $\omega_0$, with smooth edges. We employ the \textit{probability density function} implemented in the \textsc{scipy} package~\cite{2020SciPy-NMeth} and defined as
    \begin{equation}
        f_{H}(\omega) = \frac{\beta_H}{2 \Gamma(1/\beta_H)} \exp \left( - \big|(\omega - \omega_0) / \omega_{H}\big|^{\beta_H} \right),
    \end{equation}
    where $\beta_H$ is an effective temperature that controls the steepness of the function at the edges, $\omega_H$ is the width of the plateau $\omega \in [-\omega_{H},\omega_{H}]$, and $\Gamma$ is the Gamma function. We use $\beta=4$ so that the transition is smooth and $\omega_H = 50$~meV reproduces well ARPES data in the pseudogap regime (see Fig.~\ref{fig:PG_fixedPole}).

    \subsection{\label{sec:theory_params}Fixing the parameters}

    Now that we have established the parameters required to perform the calculations reported in this manuscript, we discuss the values selected. Most of them are based on ARPES results.
    
    \paragraph*{Scattering rates. ---}
    In order to select the values for the single-particle and pairing scattering rates $\Gamma_s$ and $\Gamma_p$, respectively, we compare with those extracted from ARPES and in particular the work of Ref.~\citenum{Kondo2015}.
    In the Fig.~5 of their manuscript, they report fits a self-energy of the form 
    \begin{equation}
        \Sigma_\textbf{k}(\omega=0) = \Sigma^{\text{iso}}(\omega=0) + \Sigma^{\text{SC}}_\textbf{k}(\omega=0) = -i\Gamma_s + \frac{\Delta_\textbf{k}^2}{\epsilon_{-\textbf{k}}+i\Gamma_p}
    \end{equation}
    to their data at various temperatures, at the Fermi level and along the Fermi surface. We use values similar to what they report for OD92, in particular $\Gamma_s = 7$~meV and $\Gamma_p = 10$~meV, for most calculations if not stated otherwise.

\begin{figure}[h!]
        \centering
        \includegraphics[width=\linewidth]{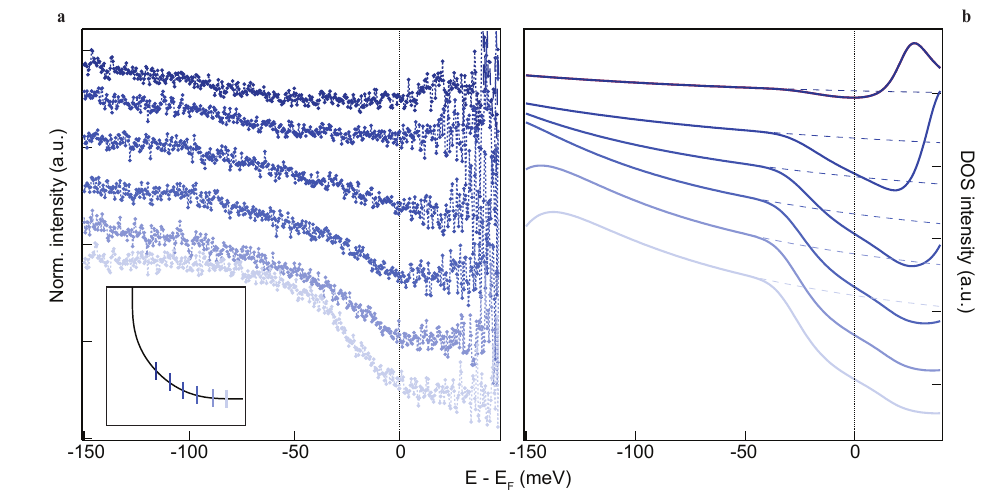}
        
        \caption{
            Comparison between the TDOS$_\phi(\omega)$ along different momentum cuts obtained from \textbf{a} ARPES and \textbf{b} using the self-energy $\Sigma^{\text{PG}}_\textbf{k}(\omega)$. The data in both panels is shifted along the $y$-axis for clarity. The parameters used for the phenomenological self-energy are detailed in the text.
        }
        \label{fig:PG_fixedPole}
    \end{figure}

    The other scattering rate $\Gamma_{pg}$ corresponding to the pseudogap process is set as $10$~meV, as this value reproduces well the experimental TDOS for various cuts in the Brillouin zone, shown in Fig.~\ref{fig:PG_fixedPole}.

    \paragraph*{Pseudogap state. ---}
    The modeling of the pseudogap state is guided by comparing calculated TDOS$_\phi(\omega)$ along different cuts $\phi_i(k_y)$ with data from ARPES inside of the pseudogap regime, higher in temperature outside the superconducting dome ($\sim 150$~K).
    In Fig.~\ref{fig:PG_fixedPole}a, the experimental data are shown whe Fig.~\ref{fig:PG_fixedPole}b the calculated TDOS.

    The self-energy used in this calculations corresponds to Eq.~(\ref{eq:self_PG}) with a pole displacement $\omega_0 = 20$~meV, the amplitude of the coupling strength $H_0 = 150$~meV and a pseudogap scattering rate $\Gamma_{pg}=10$~meV.
    We see that this self-energy reproduces well the position of the minimum of the pseudogap along different cuts, along with the filling of the gap when approaching the node at $k_x=k_y$.

    \paragraph*{Superconducting state. ---}
    The principal parameter of the superconducting part is $\Delta_0$, the amplitude of the $d$-wave gap function in Eq.~(\ref{eq:sc_gap_func_d-wave}). This parameter controls the size of the superconducting gap and is set by comparing with ARPES data at equilibrium in the superconducting phase. By comparing the TDOS obtained from our simulations with measurements presented in Fig.~\ref{FigS2}, we find that $\Delta_0 = 50$~meV is the appropriate value for this parameter.

    \paragraph*{Superconductivity of the auxiliary fermions. --- }
    The last parameter of our model is one that only appears when both superconductivity and the pseudogap physics are present ($\Delta_\textbf{k} \neq 0$ and $H_\textbf{k} \neq 0$), that is the superconductivity of the auxiliary fermions dictated by $\bar{\Delta}_\textbf{k}$.
    In this work, we simply assumed the same structure for $\bar{\Delta}_\textbf{k}$ as for $\Delta_\textbf{k}$, given by Eq.~(\ref{eq:sc_gap_func_d-wave}).

\subsection{\label{sec:theory_interplay}Interplay of SC and PG}
    
    \begin{table}[h!]
        \centering
        \begin{tabular}{
            @{\hspace{.3cm}}c@{\hspace{.3cm}}|
            @{\hspace{.3cm}}c@{\hspace{.3cm}}|
            @{\hspace{.3cm}}c@{\hspace{.3cm}}
            @{\hspace{.3cm}}c@{\hspace{.3cm}}
            @{\hspace{.3cm}}c@{\hspace{.3cm}}
            @{\hspace{.3cm}}c@{\hspace{.3cm}}|
            @{\hspace{.3cm}}c@{\hspace{.3cm}}
            @{\hspace{.3cm}}c@{\hspace{.3cm}}
            @{\hspace{.3cm}}c@{\hspace{.3cm}}|
            @{\hspace{.3cm}}c@{\hspace{.3cm}}
        }
            \hline \hline
            Self-energy parameters
                & $\Gamma_s$
                & $\omega_0$ & $H_0$ & $\beta_H$ & $\omega_H$ & $\Gamma_{pg}$ 
                & $\Delta_0$ & $\Gamma_p$
                & $\bar{\Delta}_0$ \\
            \hline
            Values [meV]
                & 7
                & 20 & 150 & 4 (unitless) & 50 & 10
                & 50 & 10 &
                50 \\
            \hline \hline
        \end{tabular}
        \caption{\label{tab:params_values}
        Values selected in Sec.~\ref{sec:theory_params} to simulate the superconducting, pseudogap and interplay states in Eq.~(\ref{eq:gen_self}). The parameters are split into sections corresponding to the single particle, pseudogap, superconducting and interplay processes, respectively.}
    \end{table}

    After deriving the self-energy Eq.~(\ref{eq:gen_self}) that captures the interplay of both superconductivity and pseudogap features in Sec.~\ref{sec:theory_derivation}, we were able to fix its parameters in Sec.~\ref{sec:theory_params} by comparing simulations with ARPES data in the low-temperature superconducting and high-temperature pseudogap phases. The parameters are summarized in Table~\ref{tab:params_values}.

    \begin{figure}
        \centering
        \includegraphics[width=\linewidth]{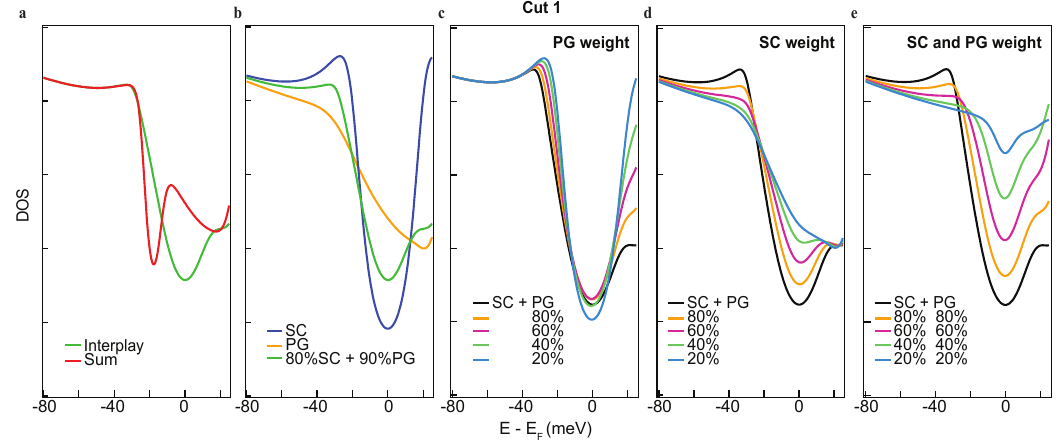}
        \caption{Interplay between the superconducting and pseudogap state along cut 1. We present the TDOS$_{\phi=1}$ for various choices of parameters. }
        \label{fig:interplay_cut1}
    \end{figure}
    \begin{figure}
        \includegraphics[width=\linewidth]{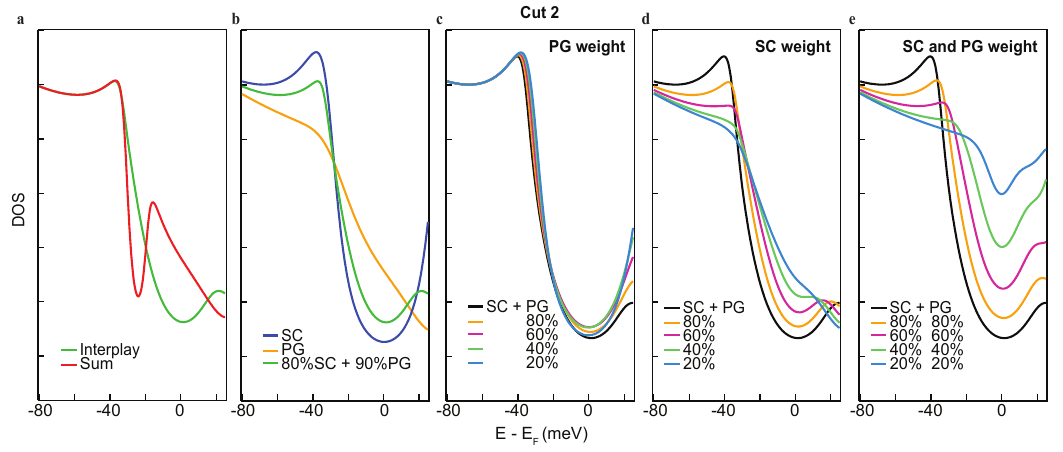}
        \caption{Interplay between the superconducting and pseudogap state along cut 2. We present the TDOS$_{\phi=2}$.}
        \label{fig:interplay_cut2}
    \end{figure}
    
    In this section, we investigate the interplay between the superconducting and pseudogap states and use it to characterize the transient state reported in the main text. In Figure~\ref{fig:interplay_cut1} (Figure~\ref{fig:interplay_cut2}), we present the TDOS$_\phi$ along cut 1 (2) for various case.
    
    In panel a, we show that summing the superconducting and pseudogap self-energies Eqs~(\ref{eq:self_SC}) and (\ref{eq:self_PG}) leads to a dramatically different TDOS that cannot reproduce the experiment ("Sum" curve in red). Instead, one really has to consider the phenomenological self-energy of Eq.~(\ref{eq:gen_self}) that includes the interplay between the two phenomena ("Interplay" curve in green).
    
    In panel b, we present the TDOS corresponding to the optimal self-energy (in green), compared with the ones corresponding to the pseudogap (yellow) and superconducting states (blue). The optimal self-energy is given by the parameters of Table~\ref{tab:params_values}, but with 80\% superconductivity ($\Delta'_0 = 0.8 \cdot \Delta_0$) and 90\% pseudogap ($H_0' = 0.9 \cdot H_0$). The pseudogap self-energy also takes the values from that table but with $\Delta_0 = 0$, while the superconducting one has $H_0 = 0$.

    In panels c-e, we present various variations of the superconducting (pseudogap) coupling strength $\Delta_0 = \bar{\Delta}_0$ ($H_0$). The need to use smaller couplings in the transient state than in the equilibrium states to better reproduce the data can be interpreted as follows: while a fraction of the superconducting pairs are melted by the light pulse, a partial pseudogap state emerges. Therefore, in the coexistence regime, the couplings are attenuated with respect to the regimes where a single state dominates.

\providecommand{\noopsort}[1]{}\providecommand{\singleletter}[1]{#1}


\begin{thebibliography}{65}%
\makeatletter
\providecommand \@ifxundefined [1]{%
 \@ifx{#1\undefined}
}%
\providecommand \@ifnum [1]{%
 \ifnum #1\expandafter \@firstoftwo
 \else \expandafter \@secondoftwo
 \fi
}%
\providecommand \@ifx [1]{%
 \ifx #1\expandafter \@firstoftwo
 \else \expandafter \@secondoftwo
 \fi
}%
\providecommand \natexlab [1]{#1}%
\providecommand \enquote  [1]{``#1''}%
\providecommand \bibnamefont  [1]{#1}%
\providecommand \bibfnamefont [1]{#1}%
\providecommand \citenamefont [1]{#1}%
\providecommand \href@noop [0]{\@secondoftwo}%
\providecommand \href [0]{\begingroup \@sanitize@url \@href}%
\providecommand \@href[1]{\@@startlink{#1}\@@href}%
\providecommand \@@href[1]{\endgroup#1\@@endlink}%
\providecommand \@sanitize@url [0]{\catcode `\\12\catcode `\$12\catcode `\&12\catcode `\#12\catcode `\^12\catcode `\_12\catcode `\%12\relax}%
\providecommand \@@startlink[1]{}%
\providecommand \@@endlink[0]{}%
\providecommand \url  [0]{\begingroup\@sanitize@url \@url }%
\providecommand \@url [1]{\endgroup\@href {#1}{\urlprefix }}%
\providecommand \urlprefix  [0]{URL }%
\providecommand \Eprint [0]{\href }%
\providecommand \doibase [0]{http://dx.doi.org/}%
\providecommand \selectlanguage [0]{\@gobble}%
\providecommand \bibinfo  [0]{\@secondoftwo}%
\providecommand \bibfield  [0]{\@secondoftwo}%
\providecommand \translation [1]{[#1]}%
\providecommand \BibitemOpen [0]{}%
\providecommand \bibitemStop [0]{}%
\providecommand \bibitemNoStop [0]{.\EOS\space}%
\providecommand \EOS [0]{\spacefactor3000\relax}%
\providecommand \BibitemShut  [1]{\csname bibitem#1\endcsname}%
\let\auto@bib@innerbib\@empty
%</preamble>
\bibitem [{\citenamefont {Keimer}\ \emph {et~al.}(2015)\citenamefont {Keimer}, \citenamefont {Kivelson}, \citenamefont {Norman}, \citenamefont {Uchida},\ and\ \citenamefont {Zaanen}}]{keimer2015quantum}%
  \BibitemOpen
  \bibfield  {author} {\bibinfo {author} {\bibfnamefont {B.}~\bibnamefont {Keimer}}, \bibinfo {author} {\bibfnamefont {S.~A.}\ \bibnamefont {Kivelson}}, \bibinfo {author} {\bibfnamefont {M.~R.}\ \bibnamefont {Norman}}, \bibinfo {author} {\bibfnamefont {S.}~\bibnamefont {Uchida}}, \ and\ \bibinfo {author} {\bibfnamefont {J.}~\bibnamefont {Zaanen}},\ }\href@noop {} {\bibfield  {journal} {\bibinfo  {journal} {Nature}\ }\textbf {\bibinfo {volume} {518}},\ \bibinfo {pages} {179} (\bibinfo {year} {2015})}\BibitemShut {NoStop}%
\bibitem [{\citenamefont {Phillips}\ \emph {et~al.}(2022)\citenamefont {Phillips}, \citenamefont {Hussey},\ and\ \citenamefont {Abbamonte}}]{phillips2022stranger}%
  \BibitemOpen
  \bibfield  {author} {\bibinfo {author} {\bibfnamefont {P.~W.}\ \bibnamefont {Phillips}}, \bibinfo {author} {\bibfnamefont {N.~E.}\ \bibnamefont {Hussey}}, \ and\ \bibinfo {author} {\bibfnamefont {P.}~\bibnamefont {Abbamonte}},\ }\href@noop {} {\bibfield  {journal} {\bibinfo  {journal} {Science}\ }\textbf {\bibinfo {volume} {377}},\ \bibinfo {pages} {eabh4273} (\bibinfo {year} {2022})}\BibitemShut {NoStop}%
\bibitem [{\citenamefont {Vishik}(2018)}]{vishik2018photoemission}%
  \BibitemOpen
  \bibfield  {author} {\bibinfo {author} {\bibfnamefont {I.}~\bibnamefont {Vishik}},\ }\href@noop {} {\bibfield  {journal} {\bibinfo  {journal} {Reports on Progress in Physics}\ }\textbf {\bibinfo {volume} {81}},\ \bibinfo {pages} {062501} (\bibinfo {year} {2018})}\BibitemShut {NoStop}%
\bibitem [{\citenamefont {Kordyuk}(2015)}]{kordyuk2015pseudogap}%
  \BibitemOpen
  \bibfield  {author} {\bibinfo {author} {\bibfnamefont {A.}~\bibnamefont {Kordyuk}},\ }\href@noop {} {\bibfield  {journal} {\bibinfo  {journal} {Low Temperature Physics}\ }\textbf {\bibinfo {volume} {41}},\ \bibinfo {pages} {319} (\bibinfo {year} {2015})}\BibitemShut {NoStop}%
\bibitem [{\citenamefont {Yang}\ \emph {et~al.}(2008)\citenamefont {Yang}, \citenamefont {Rameau}, \citenamefont {Johnson}, \citenamefont {Valla}, \citenamefont {Tsvelik},\ and\ \citenamefont {Gu}}]{yang2008emergence}%
  \BibitemOpen
  \bibfield  {author} {\bibinfo {author} {\bibfnamefont {H.-B.}\ \bibnamefont {Yang}}, \bibinfo {author} {\bibfnamefont {J.}~\bibnamefont {Rameau}}, \bibinfo {author} {\bibfnamefont {P.}~\bibnamefont {Johnson}}, \bibinfo {author} {\bibfnamefont {T.}~\bibnamefont {Valla}}, \bibinfo {author} {\bibfnamefont {A.}~\bibnamefont {Tsvelik}}, \ and\ \bibinfo {author} {\bibfnamefont {G.}~\bibnamefont {Gu}},\ }\href@noop {} {\bibfield  {journal} {\bibinfo  {journal} {Nature}\ }\textbf {\bibinfo {volume} {456}},\ \bibinfo {pages} {77} (\bibinfo {year} {2008})}\BibitemShut {NoStop}%
\bibitem [{\citenamefont {Chen}\ \emph {et~al.}(2019)\citenamefont {Chen}, \citenamefont {Hashimoto}, \citenamefont {He}, \citenamefont {Song}, \citenamefont {Xu}, \citenamefont {He}, \citenamefont {Devereaux}, \citenamefont {Eisaki}, \citenamefont {Lu}, \citenamefont {Zaanen} \emph {et~al.}}]{chen2019incoherent}%
  \BibitemOpen
  \bibfield  {author} {\bibinfo {author} {\bibfnamefont {S.-D.}\ \bibnamefont {Chen}}, \bibinfo {author} {\bibfnamefont {M.}~\bibnamefont {Hashimoto}}, \bibinfo {author} {\bibfnamefont {Y.}~\bibnamefont {He}}, \bibinfo {author} {\bibfnamefont {D.}~\bibnamefont {Song}}, \bibinfo {author} {\bibfnamefont {K.-J.}\ \bibnamefont {Xu}}, \bibinfo {author} {\bibfnamefont {J.-F.}\ \bibnamefont {He}}, \bibinfo {author} {\bibfnamefont {T.~P.}\ \bibnamefont {Devereaux}}, \bibinfo {author} {\bibfnamefont {H.}~\bibnamefont {Eisaki}}, \bibinfo {author} {\bibfnamefont {D.-H.}\ \bibnamefont {Lu}}, \bibinfo {author} {\bibfnamefont {J.}~\bibnamefont {Zaanen}},  \emph {et~al.},\ }\href@noop {} {\bibfield  {journal} {\bibinfo  {journal} {Science}\ }\textbf {\bibinfo {volume} {366}},\ \bibinfo {pages} {1099} (\bibinfo {year} {2019})}\BibitemShut {NoStop}%
\bibitem [{\citenamefont {Hashimoto}\ \emph {et~al.}(2015)\citenamefont {Hashimoto}, \citenamefont {Nowadnick}, \citenamefont {He}, \citenamefont {Vishik}, \citenamefont {Moritz}, \citenamefont {He}, \citenamefont {Tanaka}, \citenamefont {Moore}, \citenamefont {Lu}, \citenamefont {Yoshida} \emph {et~al.}}]{hashimoto2015direct}%
  \BibitemOpen
  \bibfield  {author} {\bibinfo {author} {\bibfnamefont {M.}~\bibnamefont {Hashimoto}}, \bibinfo {author} {\bibfnamefont {E.~A.}\ \bibnamefont {Nowadnick}}, \bibinfo {author} {\bibfnamefont {R.-H.}\ \bibnamefont {He}}, \bibinfo {author} {\bibfnamefont {I.~M.}\ \bibnamefont {Vishik}}, \bibinfo {author} {\bibfnamefont {B.}~\bibnamefont {Moritz}}, \bibinfo {author} {\bibfnamefont {Y.}~\bibnamefont {He}}, \bibinfo {author} {\bibfnamefont {K.}~\bibnamefont {Tanaka}}, \bibinfo {author} {\bibfnamefont {R.~G.}\ \bibnamefont {Moore}}, \bibinfo {author} {\bibfnamefont {D.}~\bibnamefont {Lu}}, \bibinfo {author} {\bibfnamefont {Y.}~\bibnamefont {Yoshida}},  \emph {et~al.},\ }\href@noop {} {\bibfield  {journal} {\bibinfo  {journal} {Nature materials}\ }\textbf {\bibinfo {volume} {14}},\ \bibinfo {pages} {37} (\bibinfo {year} {2015})}\BibitemShut {NoStop}%
\bibitem [{\citenamefont {He}\ \emph {et~al.}(2011)\citenamefont {He}, \citenamefont {Hashimoto}, \citenamefont {Karapetyan}, \citenamefont {Koralek}, \citenamefont {Hinton}, \citenamefont {Testaud}, \citenamefont {Nathan}, \citenamefont {Yoshida}, \citenamefont {Yao}, \citenamefont {Tanaka} \emph {et~al.}}]{he2011single}%
  \BibitemOpen
  \bibfield  {author} {\bibinfo {author} {\bibfnamefont {R.-H.}\ \bibnamefont {He}}, \bibinfo {author} {\bibfnamefont {M.}~\bibnamefont {Hashimoto}}, \bibinfo {author} {\bibfnamefont {H.}~\bibnamefont {Karapetyan}}, \bibinfo {author} {\bibfnamefont {J.}~\bibnamefont {Koralek}}, \bibinfo {author} {\bibfnamefont {J.}~\bibnamefont {Hinton}}, \bibinfo {author} {\bibfnamefont {J.}~\bibnamefont {Testaud}}, \bibinfo {author} {\bibfnamefont {V.}~\bibnamefont {Nathan}}, \bibinfo {author} {\bibfnamefont {Y.}~\bibnamefont {Yoshida}}, \bibinfo {author} {\bibfnamefont {H.}~\bibnamefont {Yao}}, \bibinfo {author} {\bibfnamefont {K.}~\bibnamefont {Tanaka}},  \emph {et~al.},\ }\href@noop {} {\bibfield  {journal} {\bibinfo  {journal} {Science}\ }\textbf {\bibinfo {volume} {331}},\ \bibinfo {pages} {1579} (\bibinfo {year} {2011})}\BibitemShut {NoStop}%
\bibitem [{\citenamefont {Matsui}\ \emph {et~al.}(2003)\citenamefont {Matsui}, \citenamefont {Sato}, \citenamefont {Takahashi}, \citenamefont {Wang}, \citenamefont {Yang}, \citenamefont {Ding}, \citenamefont {Fujii}, \citenamefont {Watanabe},\ and\ \citenamefont {Matsuda}}]{matsui2003bcs}%
  \BibitemOpen
  \bibfield  {author} {\bibinfo {author} {\bibfnamefont {H.}~\bibnamefont {Matsui}}, \bibinfo {author} {\bibfnamefont {T.}~\bibnamefont {Sato}}, \bibinfo {author} {\bibfnamefont {T.}~\bibnamefont {Takahashi}}, \bibinfo {author} {\bibfnamefont {S.-C.}\ \bibnamefont {Wang}}, \bibinfo {author} {\bibfnamefont {H.-B.}\ \bibnamefont {Yang}}, \bibinfo {author} {\bibfnamefont {H.}~\bibnamefont {Ding}}, \bibinfo {author} {\bibfnamefont {T.}~\bibnamefont {Fujii}}, \bibinfo {author} {\bibfnamefont {T.}~\bibnamefont {Watanabe}}, \ and\ \bibinfo {author} {\bibfnamefont {A.}~\bibnamefont {Matsuda}},\ }\href@noop {} {\bibfield  {journal} {\bibinfo  {journal} {Physical review letters}\ }\textbf {\bibinfo {volume} {90}},\ \bibinfo {pages} {217002} (\bibinfo {year} {2003})}\BibitemShut {NoStop}%
\bibitem [{\citenamefont {Lee}\ \emph {et~al.}(2007)\citenamefont {Lee}, \citenamefont {Vishik}, \citenamefont {Tanaka}, \citenamefont {Lu}, \citenamefont {Sasagawa}, \citenamefont {Nagaosa}, \citenamefont {Devereaux}, \citenamefont {Hussain},\ and\ \citenamefont {Shen}}]{lee2007abrupt}%
  \BibitemOpen
  \bibfield  {author} {\bibinfo {author} {\bibfnamefont {W.-S.}\ \bibnamefont {Lee}}, \bibinfo {author} {\bibfnamefont {I.}~\bibnamefont {Vishik}}, \bibinfo {author} {\bibfnamefont {K.}~\bibnamefont {Tanaka}}, \bibinfo {author} {\bibfnamefont {D.}~\bibnamefont {Lu}}, \bibinfo {author} {\bibfnamefont {T.}~\bibnamefont {Sasagawa}}, \bibinfo {author} {\bibfnamefont {N.}~\bibnamefont {Nagaosa}}, \bibinfo {author} {\bibfnamefont {T.}~\bibnamefont {Devereaux}}, \bibinfo {author} {\bibfnamefont {Z.}~\bibnamefont {Hussain}}, \ and\ \bibinfo {author} {\bibfnamefont {Z.-X.}\ \bibnamefont {Shen}},\ }\href@noop {} {\bibfield  {journal} {\bibinfo  {journal} {Nature}\ }\textbf {\bibinfo {volume} {450}},\ \bibinfo {pages} {81} (\bibinfo {year} {2007})}\BibitemShut {NoStop}%
\bibitem [{\citenamefont {Kondo}\ \emph {et~al.}(2013)\citenamefont {Kondo}, \citenamefont {Palczewski}, \citenamefont {Hamaya}, \citenamefont {Takeuchi}, \citenamefont {Wen}, \citenamefont {Xu}, \citenamefont {Gu},\ and\ \citenamefont {Kaminski}}]{kondo2013formation}%
  \BibitemOpen
  \bibfield  {author} {\bibinfo {author} {\bibfnamefont {T.}~\bibnamefont {Kondo}}, \bibinfo {author} {\bibfnamefont {A.~D.}\ \bibnamefont {Palczewski}}, \bibinfo {author} {\bibfnamefont {Y.}~\bibnamefont {Hamaya}}, \bibinfo {author} {\bibfnamefont {T.}~\bibnamefont {Takeuchi}}, \bibinfo {author} {\bibfnamefont {J.}~\bibnamefont {Wen}}, \bibinfo {author} {\bibfnamefont {Z.}~\bibnamefont {Xu}}, \bibinfo {author} {\bibfnamefont {G.}~\bibnamefont {Gu}}, \ and\ \bibinfo {author} {\bibfnamefont {A.}~\bibnamefont {Kaminski}},\ }\href@noop {} {\bibfield  {journal} {\bibinfo  {journal} {Physical review letters}\ }\textbf {\bibinfo {volume} {111}},\ \bibinfo {pages} {157003} (\bibinfo {year} {2013})}\BibitemShut {NoStop}%
\bibitem [{\citenamefont {Hashimoto}\ \emph {et~al.}(2010)\citenamefont {Hashimoto}, \citenamefont {He}, \citenamefont {Tanaka}, \citenamefont {Testaud}, \citenamefont {Meevasana}, \citenamefont {Moore}, \citenamefont {Lu}, \citenamefont {Yao}, \citenamefont {Yoshida}, \citenamefont {Eisaki} \emph {et~al.}}]{hashimoto2010particle}%
  \BibitemOpen
  \bibfield  {author} {\bibinfo {author} {\bibfnamefont {M.}~\bibnamefont {Hashimoto}}, \bibinfo {author} {\bibfnamefont {R.-H.}\ \bibnamefont {He}}, \bibinfo {author} {\bibfnamefont {K.}~\bibnamefont {Tanaka}}, \bibinfo {author} {\bibfnamefont {J.-P.}\ \bibnamefont {Testaud}}, \bibinfo {author} {\bibfnamefont {W.}~\bibnamefont {Meevasana}}, \bibinfo {author} {\bibfnamefont {R.~G.}\ \bibnamefont {Moore}}, \bibinfo {author} {\bibfnamefont {D.}~\bibnamefont {Lu}}, \bibinfo {author} {\bibfnamefont {H.}~\bibnamefont {Yao}}, \bibinfo {author} {\bibfnamefont {Y.}~\bibnamefont {Yoshida}}, \bibinfo {author} {\bibfnamefont {H.}~\bibnamefont {Eisaki}},  \emph {et~al.},\ }\href@noop {} {\bibfield  {journal} {\bibinfo  {journal} {Nature Physics}\ }\textbf {\bibinfo {volume} {6}},\ \bibinfo {pages} {414} (\bibinfo {year} {2010})}\BibitemShut {NoStop}%
\bibitem [{\citenamefont {Yang}\ \emph {et~al.}(2011)\citenamefont {Yang}, \citenamefont {Rameau}, \citenamefont {Pan}, \citenamefont {Gu}, \citenamefont {Johnson}, \citenamefont {Claus}, \citenamefont {Hinks},\ and\ \citenamefont {Kidd}}]{yang2011reconstructed}%
  \BibitemOpen
  \bibfield  {author} {\bibinfo {author} {\bibfnamefont {H.-B.}\ \bibnamefont {Yang}}, \bibinfo {author} {\bibfnamefont {J.}~\bibnamefont {Rameau}}, \bibinfo {author} {\bibfnamefont {Z.-H.}\ \bibnamefont {Pan}}, \bibinfo {author} {\bibfnamefont {G.}~\bibnamefont {Gu}}, \bibinfo {author} {\bibfnamefont {P.}~\bibnamefont {Johnson}}, \bibinfo {author} {\bibfnamefont {H.}~\bibnamefont {Claus}}, \bibinfo {author} {\bibfnamefont {D.}~\bibnamefont {Hinks}}, \ and\ \bibinfo {author} {\bibfnamefont {T.}~\bibnamefont {Kidd}},\ }\href@noop {} {\bibfield  {journal} {\bibinfo  {journal} {Physical review letters}\ }\textbf {\bibinfo {volume} {107}},\ \bibinfo {pages} {047003} (\bibinfo {year} {2011})}\BibitemShut {NoStop}%
\bibitem [{\citenamefont {H{\"u}fner}\ \emph {et~al.}(2008)\citenamefont {H{\"u}fner}, \citenamefont {Hossain}, \citenamefont {Damascelli},\ and\ \citenamefont {Sawatzky}}]{hufner2008two}%
  \BibitemOpen
  \bibfield  {author} {\bibinfo {author} {\bibfnamefont {S.}~\bibnamefont {H{\"u}fner}}, \bibinfo {author} {\bibfnamefont {M.}~\bibnamefont {Hossain}}, \bibinfo {author} {\bibfnamefont {A.}~\bibnamefont {Damascelli}}, \ and\ \bibinfo {author} {\bibfnamefont {G.}~\bibnamefont {Sawatzky}},\ }\href@noop {} {\bibfield  {journal} {\bibinfo  {journal} {Reports on Progress in Physics}\ }\textbf {\bibinfo {volume} {71}},\ \bibinfo {pages} {062501} (\bibinfo {year} {2008})}\BibitemShut {NoStop}%
\bibitem [{\citenamefont {Guyard}\ \emph {et~al.}(2008)\citenamefont {Guyard}, \citenamefont {Le~Tacon}, \citenamefont {Cazayous}, \citenamefont {Sacuto}, \citenamefont {Georges}, \citenamefont {Colson},\ and\ \citenamefont {Forget}}]{guyard2008breakpoint}%
  \BibitemOpen
  \bibfield  {author} {\bibinfo {author} {\bibfnamefont {W.}~\bibnamefont {Guyard}}, \bibinfo {author} {\bibfnamefont {M.}~\bibnamefont {Le~Tacon}}, \bibinfo {author} {\bibfnamefont {M.}~\bibnamefont {Cazayous}}, \bibinfo {author} {\bibfnamefont {A.}~\bibnamefont {Sacuto}}, \bibinfo {author} {\bibfnamefont {A.}~\bibnamefont {Georges}}, \bibinfo {author} {\bibfnamefont {D.}~\bibnamefont {Colson}}, \ and\ \bibinfo {author} {\bibfnamefont {A.}~\bibnamefont {Forget}},\ }\href@noop {} {\bibfield  {journal} {\bibinfo  {journal} {Physical Review B—Condensed Matter and Materials Physics}\ }\textbf {\bibinfo {volume} {77}},\ \bibinfo {pages} {024524} (\bibinfo {year} {2008})}\BibitemShut {NoStop}%
\bibitem [{\citenamefont {Sakai}\ \emph {et~al.}(2013)\citenamefont {Sakai}, \citenamefont {Blanc}, \citenamefont {Civelli}, \citenamefont {Gallais}, \citenamefont {Cazayous}, \citenamefont {Méasson}, \citenamefont {Wen}, \citenamefont {Xu}, \citenamefont {Gu}, \citenamefont {Sangiovanni}, \citenamefont {Motome}, \citenamefont {Held}, \citenamefont {Sacuto}, \citenamefont {Georges},\ and\ \citenamefont {Imada}}]{sakai2013raman}%
  \BibitemOpen
  \bibfield  {author} {\bibinfo {author} {\bibfnamefont {S.}~\bibnamefont {Sakai}}, \bibinfo {author} {\bibfnamefont {S.}~\bibnamefont {Blanc}}, \bibinfo {author} {\bibfnamefont {M.}~\bibnamefont {Civelli}}, \bibinfo {author} {\bibfnamefont {Y.}~\bibnamefont {Gallais}}, \bibinfo {author} {\bibfnamefont {M.}~\bibnamefont {Cazayous}}, \bibinfo {author} {\bibfnamefont {M.-A.}\ \bibnamefont {Méasson}}, \bibinfo {author} {\bibfnamefont {J.~S.}\ \bibnamefont {Wen}}, \bibinfo {author} {\bibfnamefont {Z.~J.}\ \bibnamefont {Xu}}, \bibinfo {author} {\bibfnamefont {G.~D.}\ \bibnamefont {Gu}}, \bibinfo {author} {\bibfnamefont {G.}~\bibnamefont {Sangiovanni}}, \bibinfo {author} {\bibfnamefont {Y.}~\bibnamefont {Motome}}, \bibinfo {author} {\bibfnamefont {K.}~\bibnamefont {Held}}, \bibinfo {author} {\bibfnamefont {A.}~\bibnamefont {Sacuto}}, \bibinfo {author} {\bibfnamefont {A.}~\bibnamefont {Georges}}, \ and\ \bibinfo {author} {\bibfnamefont {M.}~\bibnamefont {Imada}},\ }\href@noop {} {\bibfield  {journal} {\bibinfo
  {journal} {Physical review letters}\ }\textbf {\bibinfo {volume} {111}},\ \bibinfo {pages} {107001} (\bibinfo {year} {2013})}\BibitemShut {NoStop}%
\bibitem [{\citenamefont {Kondo}\ \emph {et~al.}(2009)\citenamefont {Kondo}, \citenamefont {Khasanov}, \citenamefont {Takeuchi}, \citenamefont {Schmalian},\ and\ \citenamefont {Kaminski}}]{kondo2009competition}%
  \BibitemOpen
  \bibfield  {author} {\bibinfo {author} {\bibfnamefont {T.}~\bibnamefont {Kondo}}, \bibinfo {author} {\bibfnamefont {R.}~\bibnamefont {Khasanov}}, \bibinfo {author} {\bibfnamefont {T.}~\bibnamefont {Takeuchi}}, \bibinfo {author} {\bibfnamefont {J.}~\bibnamefont {Schmalian}}, \ and\ \bibinfo {author} {\bibfnamefont {A.}~\bibnamefont {Kaminski}},\ }\href@noop {} {\bibfield  {journal} {\bibinfo  {journal} {Nature}\ }\textbf {\bibinfo {volume} {457}},\ \bibinfo {pages} {296} (\bibinfo {year} {2009})}\BibitemShut {NoStop}%
\bibitem [{\citenamefont {Huecker}\ \emph {et~al.}(2014)\citenamefont {Huecker}, \citenamefont {Christensen}, \citenamefont {Holmes}, \citenamefont {Blackburn}, \citenamefont {Forgan}, \citenamefont {Liang}, \citenamefont {Bonn}, \citenamefont {Hardy}, \citenamefont {Gutowski}, \citenamefont {Zimmermann} \emph {et~al.}}]{huecker2014competing}%
  \BibitemOpen
  \bibfield  {author} {\bibinfo {author} {\bibfnamefont {M.}~\bibnamefont {Huecker}}, \bibinfo {author} {\bibfnamefont {N.~B.}\ \bibnamefont {Christensen}}, \bibinfo {author} {\bibfnamefont {A.}~\bibnamefont {Holmes}}, \bibinfo {author} {\bibfnamefont {E.}~\bibnamefont {Blackburn}}, \bibinfo {author} {\bibfnamefont {E.~M.}\ \bibnamefont {Forgan}}, \bibinfo {author} {\bibfnamefont {R.}~\bibnamefont {Liang}}, \bibinfo {author} {\bibfnamefont {D.}~\bibnamefont {Bonn}}, \bibinfo {author} {\bibfnamefont {W.}~\bibnamefont {Hardy}}, \bibinfo {author} {\bibfnamefont {O.}~\bibnamefont {Gutowski}}, \bibinfo {author} {\bibfnamefont {M.~v.}\ \bibnamefont {Zimmermann}},  \emph {et~al.},\ }\href@noop {} {\bibfield  {journal} {\bibinfo  {journal} {Physical Review B}\ }\textbf {\bibinfo {volume} {90}},\ \bibinfo {pages} {054514} (\bibinfo {year} {2014})}\BibitemShut {NoStop}%
\bibitem [{\citenamefont {Sobota}\ \emph {et~al.}(2021)\citenamefont {Sobota}, \citenamefont {He},\ and\ \citenamefont {Shen}}]{sobota2021angle}%
  \BibitemOpen
  \bibfield  {author} {\bibinfo {author} {\bibfnamefont {J.~A.}\ \bibnamefont {Sobota}}, \bibinfo {author} {\bibfnamefont {Y.}~\bibnamefont {He}}, \ and\ \bibinfo {author} {\bibfnamefont {Z.-X.}\ \bibnamefont {Shen}},\ }\href@noop {} {\bibfield  {journal} {\bibinfo  {journal} {Reviews of Modern Physics}\ }\textbf {\bibinfo {volume} {93}},\ \bibinfo {pages} {025006} (\bibinfo {year} {2021})}\BibitemShut {NoStop}%
\bibitem [{\citenamefont {Damascelli}\ \emph {et~al.}(2003)\citenamefont {Damascelli}, \citenamefont {Hussain},\ and\ \citenamefont {Shen}}]{damascelli2003angle}%
  \BibitemOpen
  \bibfield  {author} {\bibinfo {author} {\bibfnamefont {A.}~\bibnamefont {Damascelli}}, \bibinfo {author} {\bibfnamefont {Z.}~\bibnamefont {Hussain}}, \ and\ \bibinfo {author} {\bibfnamefont {Z.-X.}\ \bibnamefont {Shen}},\ }\href@noop {} {\bibfield  {journal} {\bibinfo  {journal} {Reviews of modern physics}\ }\textbf {\bibinfo {volume} {75}},\ \bibinfo {pages} {473} (\bibinfo {year} {2003})}\BibitemShut {NoStop}%
\bibitem [{\citenamefont {Kondo}\ \emph {et~al.}(2011)\citenamefont {Kondo}, \citenamefont {Hamaya}, \citenamefont {Palczewski}, \citenamefont {Takeuchi}, \citenamefont {Wen}, \citenamefont {Xu}, \citenamefont {Gu}, \citenamefont {Schmalian},\ and\ \citenamefont {Kaminski}}]{kondo2011disentangling}%
  \BibitemOpen
  \bibfield  {author} {\bibinfo {author} {\bibfnamefont {T.}~\bibnamefont {Kondo}}, \bibinfo {author} {\bibfnamefont {Y.}~\bibnamefont {Hamaya}}, \bibinfo {author} {\bibfnamefont {A.~D.}\ \bibnamefont {Palczewski}}, \bibinfo {author} {\bibfnamefont {T.}~\bibnamefont {Takeuchi}}, \bibinfo {author} {\bibfnamefont {J.}~\bibnamefont {Wen}}, \bibinfo {author} {\bibfnamefont {Z.}~\bibnamefont {Xu}}, \bibinfo {author} {\bibfnamefont {G.}~\bibnamefont {Gu}}, \bibinfo {author} {\bibfnamefont {J.}~\bibnamefont {Schmalian}}, \ and\ \bibinfo {author} {\bibfnamefont {A.}~\bibnamefont {Kaminski}},\ }\href@noop {} {\bibfield  {journal} {\bibinfo  {journal} {Nature Physics}\ }\textbf {\bibinfo {volume} {7}},\ \bibinfo {pages} {21} (\bibinfo {year} {2011})}\BibitemShut {NoStop}%
\bibitem [{\citenamefont {K{\"u}spert}\ \emph {et~al.}(2022)\citenamefont {K{\"u}spert}, \citenamefont {Cohn~Wagner}, \citenamefont {Lin}, \citenamefont {Von~Arx}, \citenamefont {Wang}, \citenamefont {Kramer}, \citenamefont {Pudelko}, \citenamefont {Plumb}, \citenamefont {Matt}, \citenamefont {Fatuzzo} \emph {et~al.}}]{kuspert2022pseudogap}%
  \BibitemOpen
  \bibfield  {author} {\bibinfo {author} {\bibfnamefont {J.}~\bibnamefont {K{\"u}spert}}, \bibinfo {author} {\bibfnamefont {R.}~\bibnamefont {Cohn~Wagner}}, \bibinfo {author} {\bibfnamefont {C.}~\bibnamefont {Lin}}, \bibinfo {author} {\bibfnamefont {K.}~\bibnamefont {Von~Arx}}, \bibinfo {author} {\bibfnamefont {Q.}~\bibnamefont {Wang}}, \bibinfo {author} {\bibfnamefont {K.}~\bibnamefont {Kramer}}, \bibinfo {author} {\bibfnamefont {W.}~\bibnamefont {Pudelko}}, \bibinfo {author} {\bibfnamefont {N.}~\bibnamefont {Plumb}}, \bibinfo {author} {\bibfnamefont {C.}~\bibnamefont {Matt}}, \bibinfo {author} {\bibfnamefont {C.}~\bibnamefont {Fatuzzo}},  \emph {et~al.},\ }\href@noop {} {\bibfield  {journal} {\bibinfo  {journal} {Physical review research}\ }\textbf {\bibinfo {volume} {4}},\ \bibinfo {pages} {043015} (\bibinfo {year} {2022})}\BibitemShut {NoStop}%
\bibitem [{\citenamefont {Pushp}\ \emph {et~al.}(2009)\citenamefont {Pushp}, \citenamefont {Parker}, \citenamefont {Pasupathy}, \citenamefont {Gomes}, \citenamefont {Ono}, \citenamefont {Wen}, \citenamefont {Xu}, \citenamefont {Gu},\ and\ \citenamefont {Yazdani}}]{pushp2009extending}%
  \BibitemOpen
  \bibfield  {author} {\bibinfo {author} {\bibfnamefont {A.}~\bibnamefont {Pushp}}, \bibinfo {author} {\bibfnamefont {C.~V.}\ \bibnamefont {Parker}}, \bibinfo {author} {\bibfnamefont {A.~N.}\ \bibnamefont {Pasupathy}}, \bibinfo {author} {\bibfnamefont {K.~K.}\ \bibnamefont {Gomes}}, \bibinfo {author} {\bibfnamefont {S.}~\bibnamefont {Ono}}, \bibinfo {author} {\bibfnamefont {J.}~\bibnamefont {Wen}}, \bibinfo {author} {\bibfnamefont {Z.}~\bibnamefont {Xu}}, \bibinfo {author} {\bibfnamefont {G.}~\bibnamefont {Gu}}, \ and\ \bibinfo {author} {\bibfnamefont {A.}~\bibnamefont {Yazdani}},\ }\href@noop {} {\bibfield  {journal} {\bibinfo  {journal} {Science}\ }\textbf {\bibinfo {volume} {324}},\ \bibinfo {pages} {1689} (\bibinfo {year} {2009})}\BibitemShut {NoStop}%
\bibitem [{\citenamefont {Gull}\ \emph {et~al.}(2013)\citenamefont {Gull}, \citenamefont {Parcollet},\ and\ \citenamefont {Millis}}]{gull2013superconductivity}%
  \BibitemOpen
  \bibfield  {author} {\bibinfo {author} {\bibfnamefont {E.}~\bibnamefont {Gull}}, \bibinfo {author} {\bibfnamefont {O.}~\bibnamefont {Parcollet}}, \ and\ \bibinfo {author} {\bibfnamefont {A.~J.}\ \bibnamefont {Millis}},\ }\href@noop {} {\bibfield  {journal} {\bibinfo  {journal} {Physical review letters}\ }\textbf {\bibinfo {volume} {110}},\ \bibinfo {pages} {216405} (\bibinfo {year} {2013})}\BibitemShut {NoStop}%
\bibitem [{\citenamefont {Gull}\ and\ \citenamefont {Millis}(2015)}]{gull2015quasiparticle}%
  \BibitemOpen
  \bibfield  {author} {\bibinfo {author} {\bibfnamefont {E.}~\bibnamefont {Gull}}\ and\ \bibinfo {author} {\bibfnamefont {A.~J.}\ \bibnamefont {Millis}},\ }\href@noop {} {\bibfield  {journal} {\bibinfo  {journal} {Physical Review B}\ }\textbf {\bibinfo {volume} {91}},\ \bibinfo {pages} {085116} (\bibinfo {year} {2015})}\BibitemShut {NoStop}%
\bibitem [{\citenamefont {Civelli}\ \emph {et~al.}(2008)\citenamefont {Civelli}, \citenamefont {Capone}, \citenamefont {Georges}, \citenamefont {Haule}, \citenamefont {Parcollet}, \citenamefont {Stanescu},\ and\ \citenamefont {Kotliar}}]{civelli2008}%
  \BibitemOpen
  \bibfield  {author} {\bibinfo {author} {\bibfnamefont {M.}~\bibnamefont {Civelli}}, \bibinfo {author} {\bibfnamefont {M.}~\bibnamefont {Capone}}, \bibinfo {author} {\bibfnamefont {A.}~\bibnamefont {Georges}}, \bibinfo {author} {\bibfnamefont {K.}~\bibnamefont {Haule}}, \bibinfo {author} {\bibfnamefont {O.}~\bibnamefont {Parcollet}}, \bibinfo {author} {\bibfnamefont {T.}~\bibnamefont {Stanescu}}, \ and\ \bibinfo {author} {\bibfnamefont {G.}~\bibnamefont {Kotliar}},\ }\href@noop {} {\bibfield  {journal} {\bibinfo  {journal} {Physical review letters}\ }\textbf {\bibinfo {volume} {100}},\ \bibinfo {pages} {046402} (\bibinfo {year} {2008})}\BibitemShut {NoStop}%
\bibitem [{\citenamefont {Giannetti}\ \emph {et~al.}(2016)\citenamefont {Giannetti}, \citenamefont {Capone}, \citenamefont {Fausti}, \citenamefont {Fabrizio}, \citenamefont {Parmigiani},\ and\ \citenamefont {Mihailovic}}]{giannetti2016ultrafast}%
  \BibitemOpen
  \bibfield  {author} {\bibinfo {author} {\bibfnamefont {C.}~\bibnamefont {Giannetti}}, \bibinfo {author} {\bibfnamefont {M.}~\bibnamefont {Capone}}, \bibinfo {author} {\bibfnamefont {D.}~\bibnamefont {Fausti}}, \bibinfo {author} {\bibfnamefont {M.}~\bibnamefont {Fabrizio}}, \bibinfo {author} {\bibfnamefont {F.}~\bibnamefont {Parmigiani}}, \ and\ \bibinfo {author} {\bibfnamefont {D.}~\bibnamefont {Mihailovic}},\ }\href@noop {} {\bibfield  {journal} {\bibinfo  {journal} {Advances in Physics}\ }\textbf {\bibinfo {volume} {65}},\ \bibinfo {pages} {58} (\bibinfo {year} {2016})}\BibitemShut {NoStop}%
\bibitem [{\citenamefont {Boschini}\ \emph {et~al.}(2024)\citenamefont {Boschini}, \citenamefont {Zonno},\ and\ \citenamefont {Damascelli}}]{boschini2024time}%
  \BibitemOpen
  \bibfield  {author} {\bibinfo {author} {\bibfnamefont {F.}~\bibnamefont {Boschini}}, \bibinfo {author} {\bibfnamefont {M.}~\bibnamefont {Zonno}}, \ and\ \bibinfo {author} {\bibfnamefont {A.}~\bibnamefont {Damascelli}},\ }\href@noop {} {\bibfield  {journal} {\bibinfo  {journal} {Reviews of Modern Physics}\ }\textbf {\bibinfo {volume} {96}},\ \bibinfo {pages} {015003} (\bibinfo {year} {2024})}\BibitemShut {NoStop}%
\bibitem [{\citenamefont {Na}\ \emph {et~al.}(2023)\citenamefont {Na}, \citenamefont {Mills},\ and\ \citenamefont {Jones}}]{na2023advancing}%
  \BibitemOpen
  \bibfield  {author} {\bibinfo {author} {\bibfnamefont {M.}~\bibnamefont {Na}}, \bibinfo {author} {\bibfnamefont {A.~K.}\ \bibnamefont {Mills}}, \ and\ \bibinfo {author} {\bibfnamefont {D.~J.}\ \bibnamefont {Jones}},\ }\href@noop {} {\bibfield  {journal} {\bibinfo  {journal} {Physics Reports}\ }\textbf {\bibinfo {volume} {1036}},\ \bibinfo {pages} {1} (\bibinfo {year} {2023})}\BibitemShut {NoStop}%
\bibitem [{\citenamefont {Cilento}\ \emph {et~al.}(2016)\citenamefont {Cilento}, \citenamefont {Crepaldi}, \citenamefont {Manzoni}, \citenamefont {Sterzi}, \citenamefont {Zacchigna}, \citenamefont {Bugnon}, \citenamefont {Berger},\ and\ \citenamefont {Parmigiani}}]{cilento2016advancing}%
  \BibitemOpen
  \bibfield  {author} {\bibinfo {author} {\bibfnamefont {F.}~\bibnamefont {Cilento}}, \bibinfo {author} {\bibfnamefont {A.}~\bibnamefont {Crepaldi}}, \bibinfo {author} {\bibfnamefont {G.}~\bibnamefont {Manzoni}}, \bibinfo {author} {\bibfnamefont {A.}~\bibnamefont {Sterzi}}, \bibinfo {author} {\bibfnamefont {M.}~\bibnamefont {Zacchigna}}, \bibinfo {author} {\bibfnamefont {P.}~\bibnamefont {Bugnon}}, \bibinfo {author} {\bibfnamefont {H.}~\bibnamefont {Berger}}, \ and\ \bibinfo {author} {\bibfnamefont {F.}~\bibnamefont {Parmigiani}},\ }\href@noop {} {\bibfield  {journal} {\bibinfo  {journal} {Journal of Electron Spectroscopy and Related Phenomena}\ }\textbf {\bibinfo {volume} {207}},\ \bibinfo {pages} {7} (\bibinfo {year} {2016})}\BibitemShut {NoStop}%
\bibitem [{\citenamefont {Gauthier}\ \emph {et~al.}(2021)\citenamefont {Gauthier}, \citenamefont {Sobota}, \citenamefont {Pfau}, \citenamefont {Gauthier}, \citenamefont {Soifer}, \citenamefont {Bachmann}, \citenamefont {Fisher}, \citenamefont {Shen},\ and\ \citenamefont {Kirchmann}}]{gauthier2021expanding}%
  \BibitemOpen
  \bibfield  {author} {\bibinfo {author} {\bibfnamefont {N.}~\bibnamefont {Gauthier}}, \bibinfo {author} {\bibfnamefont {J.~A.}\ \bibnamefont {Sobota}}, \bibinfo {author} {\bibfnamefont {H.}~\bibnamefont {Pfau}}, \bibinfo {author} {\bibfnamefont {A.}~\bibnamefont {Gauthier}}, \bibinfo {author} {\bibfnamefont {H.}~\bibnamefont {Soifer}}, \bibinfo {author} {\bibfnamefont {M.~D.}\ \bibnamefont {Bachmann}}, \bibinfo {author} {\bibfnamefont {I.~R.}\ \bibnamefont {Fisher}}, \bibinfo {author} {\bibfnamefont {Z.-X.}\ \bibnamefont {Shen}}, \ and\ \bibinfo {author} {\bibfnamefont {P.~S.}\ \bibnamefont {Kirchmann}},\ }\href@noop {} {\bibfield  {journal} {\bibinfo  {journal} {Review of Scientific Instruments}\ }\textbf {\bibinfo {volume} {92}} (\bibinfo {year} {2021})}\BibitemShut {NoStop}%
\bibitem [{\citenamefont {Longa}\ \emph {et~al.}(2024)\citenamefont {Longa}, \citenamefont {Parent}, \citenamefont {Frimpong}, \citenamefont {Armanno}, \citenamefont {Gauthier}, \citenamefont {L{\'e}gar{\'e}}, \citenamefont {Boschini},\ and\ \citenamefont {Jargot}}]{longa2024time}%
  \BibitemOpen
  \bibfield  {author} {\bibinfo {author} {\bibfnamefont {A.}~\bibnamefont {Longa}}, \bibinfo {author} {\bibfnamefont {J.-M.}\ \bibnamefont {Parent}}, \bibinfo {author} {\bibfnamefont {B.~K.}\ \bibnamefont {Frimpong}}, \bibinfo {author} {\bibfnamefont {D.}~\bibnamefont {Armanno}}, \bibinfo {author} {\bibfnamefont {N.}~\bibnamefont {Gauthier}}, \bibinfo {author} {\bibfnamefont {F.}~\bibnamefont {L{\'e}gar{\'e}}}, \bibinfo {author} {\bibfnamefont {F.}~\bibnamefont {Boschini}}, \ and\ \bibinfo {author} {\bibfnamefont {G.}~\bibnamefont {Jargot}},\ }\href@noop {} {\bibfield  {journal} {\bibinfo  {journal} {Optics Express}\ }\textbf {\bibinfo {volume} {32}},\ \bibinfo {pages} {29549} (\bibinfo {year} {2024})}\BibitemShut {NoStop}%
\bibitem [{\citenamefont {King}\ \emph {et~al.}(2011)\citenamefont {King}, \citenamefont {Rosen}, \citenamefont {Meevasana}, \citenamefont {Tamai}, \citenamefont {Rozbicki}, \citenamefont {Comin}, \citenamefont {Levy}, \citenamefont {Fournier}, \citenamefont {Yoshida}, \citenamefont {Eisaki} \emph {et~al.}}]{king2011structural}%
  \BibitemOpen
  \bibfield  {author} {\bibinfo {author} {\bibfnamefont {P.~D.}\ \bibnamefont {King}}, \bibinfo {author} {\bibfnamefont {J.}~\bibnamefont {Rosen}}, \bibinfo {author} {\bibfnamefont {W.}~\bibnamefont {Meevasana}}, \bibinfo {author} {\bibfnamefont {A.}~\bibnamefont {Tamai}}, \bibinfo {author} {\bibfnamefont {E.}~\bibnamefont {Rozbicki}}, \bibinfo {author} {\bibfnamefont {R.}~\bibnamefont {Comin}}, \bibinfo {author} {\bibfnamefont {G.}~\bibnamefont {Levy}}, \bibinfo {author} {\bibfnamefont {D.}~\bibnamefont {Fournier}}, \bibinfo {author} {\bibfnamefont {Y.}~\bibnamefont {Yoshida}}, \bibinfo {author} {\bibfnamefont {H.}~\bibnamefont {Eisaki}},  \emph {et~al.},\ }\href@noop {} {\bibfield  {journal} {\bibinfo  {journal} {Physical review letters}\ }\textbf {\bibinfo {volume} {106}},\ \bibinfo {pages} {127005} (\bibinfo {year} {2011})}\BibitemShut {NoStop}%
\bibitem [{\citenamefont {Baldini}\ \emph {et~al.}(2020)\citenamefont {Baldini}, \citenamefont {Sentef}, \citenamefont {Acharya}, \citenamefont {Brumme}, \citenamefont {Sheveleva}, \citenamefont {Lyzwa}, \citenamefont {Pomjakushina}, \citenamefont {Bernhard}, \citenamefont {Van~Schilfgaarde}, \citenamefont {Carbone} \emph {et~al.}}]{baldini2020electron}%
  \BibitemOpen
  \bibfield  {author} {\bibinfo {author} {\bibfnamefont {E.}~\bibnamefont {Baldini}}, \bibinfo {author} {\bibfnamefont {M.~A.}\ \bibnamefont {Sentef}}, \bibinfo {author} {\bibfnamefont {S.}~\bibnamefont {Acharya}}, \bibinfo {author} {\bibfnamefont {T.}~\bibnamefont {Brumme}}, \bibinfo {author} {\bibfnamefont {E.}~\bibnamefont {Sheveleva}}, \bibinfo {author} {\bibfnamefont {F.}~\bibnamefont {Lyzwa}}, \bibinfo {author} {\bibfnamefont {E.}~\bibnamefont {Pomjakushina}}, \bibinfo {author} {\bibfnamefont {C.}~\bibnamefont {Bernhard}}, \bibinfo {author} {\bibfnamefont {M.}~\bibnamefont {Van~Schilfgaarde}}, \bibinfo {author} {\bibfnamefont {F.}~\bibnamefont {Carbone}},  \emph {et~al.},\ }\href@noop {} {\bibfield  {journal} {\bibinfo  {journal} {Proceedings of the National Academy of Sciences}\ }\textbf {\bibinfo {volume} {117}},\ \bibinfo {pages} {6409} (\bibinfo {year} {2020})}\BibitemShut {NoStop}%
\bibitem [{\citenamefont {Perfetti}\ \emph {et~al.}(2007)\citenamefont {Perfetti}, \citenamefont {Loukakos}, \citenamefont {Lisowski}, \citenamefont {Bovensiepen}, \citenamefont {Eisaki},\ and\ \citenamefont {Wolf}}]{perfetti2007ultrafast}%
  \BibitemOpen
  \bibfield  {author} {\bibinfo {author} {\bibfnamefont {L.}~\bibnamefont {Perfetti}}, \bibinfo {author} {\bibfnamefont {P.}~\bibnamefont {Loukakos}}, \bibinfo {author} {\bibfnamefont {M.}~\bibnamefont {Lisowski}}, \bibinfo {author} {\bibfnamefont {U.}~\bibnamefont {Bovensiepen}}, \bibinfo {author} {\bibfnamefont {H.}~\bibnamefont {Eisaki}}, \ and\ \bibinfo {author} {\bibfnamefont {M.}~\bibnamefont {Wolf}},\ }\href@noop {} {\bibfield  {journal} {\bibinfo  {journal} {Physical review letters}\ }\textbf {\bibinfo {volume} {99}},\ \bibinfo {pages} {197001} (\bibinfo {year} {2007})}\BibitemShut {NoStop}%
\bibitem [{\citenamefont {Boschini}\ \emph {et~al.}(2018)\citenamefont {Boschini}, \citenamefont {da~Silva~Neto}, \citenamefont {Razzoli}, \citenamefont {Zonno}, \citenamefont {Peli}, \citenamefont {Day}, \citenamefont {Michiardi}, \citenamefont {Schneider}, \citenamefont {Zwartsenberg}, \citenamefont {Nigge} \emph {et~al.}}]{boschini2018collapse}%
  \BibitemOpen
  \bibfield  {author} {\bibinfo {author} {\bibfnamefont {F.}~\bibnamefont {Boschini}}, \bibinfo {author} {\bibfnamefont {E.}~\bibnamefont {da~Silva~Neto}}, \bibinfo {author} {\bibfnamefont {E.}~\bibnamefont {Razzoli}}, \bibinfo {author} {\bibfnamefont {M.}~\bibnamefont {Zonno}}, \bibinfo {author} {\bibfnamefont {S.}~\bibnamefont {Peli}}, \bibinfo {author} {\bibfnamefont {R.}~\bibnamefont {Day}}, \bibinfo {author} {\bibfnamefont {M.}~\bibnamefont {Michiardi}}, \bibinfo {author} {\bibfnamefont {M.}~\bibnamefont {Schneider}}, \bibinfo {author} {\bibfnamefont {B.}~\bibnamefont {Zwartsenberg}}, \bibinfo {author} {\bibfnamefont {P.}~\bibnamefont {Nigge}},  \emph {et~al.},\ }\href@noop {} {\bibfield  {journal} {\bibinfo  {journal} {Nature materials}\ }\textbf {\bibinfo {volume} {17}},\ \bibinfo {pages} {416} (\bibinfo {year} {2018})}\BibitemShut {NoStop}%
\bibitem [{\citenamefont {Smallwood}\ \emph {et~al.}(2012)\citenamefont {Smallwood}, \citenamefont {Hinton}, \citenamefont {Jozwiak}, \citenamefont {Zhang}, \citenamefont {Koralek}, \citenamefont {Eisaki}, \citenamefont {Lee}, \citenamefont {Orenstein},\ and\ \citenamefont {Lanzara}}]{smallwood2012tracking}%
  \BibitemOpen
  \bibfield  {author} {\bibinfo {author} {\bibfnamefont {C.~L.}\ \bibnamefont {Smallwood}}, \bibinfo {author} {\bibfnamefont {J.~P.}\ \bibnamefont {Hinton}}, \bibinfo {author} {\bibfnamefont {C.}~\bibnamefont {Jozwiak}}, \bibinfo {author} {\bibfnamefont {W.}~\bibnamefont {Zhang}}, \bibinfo {author} {\bibfnamefont {J.~D.}\ \bibnamefont {Koralek}}, \bibinfo {author} {\bibfnamefont {H.}~\bibnamefont {Eisaki}}, \bibinfo {author} {\bibfnamefont {D.-H.}\ \bibnamefont {Lee}}, \bibinfo {author} {\bibfnamefont {J.}~\bibnamefont {Orenstein}}, \ and\ \bibinfo {author} {\bibfnamefont {A.}~\bibnamefont {Lanzara}},\ }\href@noop {} {\bibfield  {journal} {\bibinfo  {journal} {Science}\ }\textbf {\bibinfo {volume} {336}},\ \bibinfo {pages} {1137} (\bibinfo {year} {2012})}\BibitemShut {NoStop}%
\bibitem [{\citenamefont {Parham}\ \emph {et~al.}(2017)\citenamefont {Parham}, \citenamefont {Li}, \citenamefont {Nummy}, \citenamefont {Waugh}, \citenamefont {Zhou}, \citenamefont {Griffith}, \citenamefont {Schneeloch}, \citenamefont {Zhong}, \citenamefont {Gu},\ and\ \citenamefont {Dessau}}]{parham2017ultrafast}%
  \BibitemOpen
  \bibfield  {author} {\bibinfo {author} {\bibfnamefont {S.}~\bibnamefont {Parham}}, \bibinfo {author} {\bibfnamefont {H.}~\bibnamefont {Li}}, \bibinfo {author} {\bibfnamefont {T.}~\bibnamefont {Nummy}}, \bibinfo {author} {\bibfnamefont {J.~A.}\ \bibnamefont {Waugh}}, \bibinfo {author} {\bibfnamefont {X.}~\bibnamefont {Zhou}}, \bibinfo {author} {\bibfnamefont {J.}~\bibnamefont {Griffith}}, \bibinfo {author} {\bibfnamefont {J.}~\bibnamefont {Schneeloch}}, \bibinfo {author} {\bibfnamefont {R.~D.}\ \bibnamefont {Zhong}}, \bibinfo {author} {\bibfnamefont {G.}~\bibnamefont {Gu}}, \ and\ \bibinfo {author} {\bibfnamefont {D.~S.}\ \bibnamefont {Dessau}},\ }\href@noop {} {\bibfield  {journal} {\bibinfo  {journal} {Physical Review X}\ }\textbf {\bibinfo {volume} {7}},\ \bibinfo {pages} {041013} (\bibinfo {year} {2017})}\BibitemShut {NoStop}%
\bibitem [{\citenamefont {Dynes}\ \emph {et~al.}(1978)\citenamefont {Dynes}, \citenamefont {Narayanamurti},\ and\ \citenamefont {Garno}}]{dynes1978direct}%
  \BibitemOpen
  \bibfield  {author} {\bibinfo {author} {\bibfnamefont {R.~C.}\ \bibnamefont {Dynes}}, \bibinfo {author} {\bibfnamefont {V.}~\bibnamefont {Narayanamurti}}, \ and\ \bibinfo {author} {\bibfnamefont {J.~P.}\ \bibnamefont {Garno}},\ }\href@noop {} {\bibfield  {journal} {\bibinfo  {journal} {Physical Review Letters}\ }\textbf {\bibinfo {volume} {41}},\ \bibinfo {pages} {1509} (\bibinfo {year} {1978})}\BibitemShut {NoStop}%
\bibitem [{\citenamefont {Kim}\ \emph {et~al.}(2024)\citenamefont {Kim}, \citenamefont {Kovalev}, \citenamefont {Udina}, \citenamefont {Haenel}, \citenamefont {Kim}, \citenamefont {Puviani}, \citenamefont {Cristiani}, \citenamefont {Ilyakov}, \citenamefont {de~Oliveira}, \citenamefont {Ponomaryov} \emph {et~al.}}]{kim2024tracing}%
  \BibitemOpen
  \bibfield  {author} {\bibinfo {author} {\bibfnamefont {M.-J.}\ \bibnamefont {Kim}}, \bibinfo {author} {\bibfnamefont {S.}~\bibnamefont {Kovalev}}, \bibinfo {author} {\bibfnamefont {M.}~\bibnamefont {Udina}}, \bibinfo {author} {\bibfnamefont {R.}~\bibnamefont {Haenel}}, \bibinfo {author} {\bibfnamefont {G.}~\bibnamefont {Kim}}, \bibinfo {author} {\bibfnamefont {M.}~\bibnamefont {Puviani}}, \bibinfo {author} {\bibfnamefont {G.}~\bibnamefont {Cristiani}}, \bibinfo {author} {\bibfnamefont {I.}~\bibnamefont {Ilyakov}}, \bibinfo {author} {\bibfnamefont {T.~V.}\ \bibnamefont {de~Oliveira}}, \bibinfo {author} {\bibfnamefont {A.}~\bibnamefont {Ponomaryov}},  \emph {et~al.},\ }\href@noop {} {\bibfield  {journal} {\bibinfo  {journal} {Science Advances}\ }\textbf {\bibinfo {volume} {10}},\ \bibinfo {pages} {eadi7598} (\bibinfo {year} {2024})}\BibitemShut {NoStop}%
\bibitem [{\citenamefont {Armanno}\ \emph {et~al.}(2025)\citenamefont {Armanno}, \citenamefont {Goto}, \citenamefont {Parent}, \citenamefont {Lapointe}, \citenamefont {Longa}, \citenamefont {Zhong}, \citenamefont {Schneeloch}, \citenamefont {Gu}, \citenamefont {Jargot}, \citenamefont {Ibrahim} \emph {et~al.}}]{armanno2025direct}%
  \BibitemOpen
  \bibfield  {author} {\bibinfo {author} {\bibfnamefont {D.}~\bibnamefont {Armanno}}, \bibinfo {author} {\bibfnamefont {F.}~\bibnamefont {Goto}}, \bibinfo {author} {\bibfnamefont {J.-M.}\ \bibnamefont {Parent}}, \bibinfo {author} {\bibfnamefont {S.}~\bibnamefont {Lapointe}}, \bibinfo {author} {\bibfnamefont {A.}~\bibnamefont {Longa}}, \bibinfo {author} {\bibfnamefont {R.}~\bibnamefont {Zhong}}, \bibinfo {author} {\bibfnamefont {J.}~\bibnamefont {Schneeloch}}, \bibinfo {author} {\bibfnamefont {G.}~\bibnamefont {Gu}}, \bibinfo {author} {\bibfnamefont {G.}~\bibnamefont {Jargot}}, \bibinfo {author} {\bibfnamefont {H.}~\bibnamefont {Ibrahim}},  \emph {et~al.},\ }\href@noop {} {\bibfield  {journal} {\bibinfo  {journal} {arXiv preprint arXiv:2505.03900}\ } (\bibinfo {year} {2025})}\BibitemShut {NoStop}%
\bibitem [{\citenamefont {Zonno}\ \emph {et~al.}(2021)\citenamefont {Zonno}, \citenamefont {Boschini},\ and\ \citenamefont {Damascelli}}]{zonno2021time}%
  \BibitemOpen
  \bibfield  {author} {\bibinfo {author} {\bibfnamefont {M.}~\bibnamefont {Zonno}}, \bibinfo {author} {\bibfnamefont {F.}~\bibnamefont {Boschini}}, \ and\ \bibinfo {author} {\bibfnamefont {A.}~\bibnamefont {Damascelli}},\ }\href@noop {} {\bibfield  {journal} {\bibinfo  {journal} {Journal of Electron Spectroscopy and Related Phenomena}\ }\textbf {\bibinfo {volume} {251}},\ \bibinfo {pages} {147091} (\bibinfo {year} {2021})}\BibitemShut {NoStop}%
\bibitem [{\citenamefont {Wandel}\ \emph {et~al.}(2022)\citenamefont {Wandel}, \citenamefont {Boschini}, \citenamefont {da~Silva~Neto}, \citenamefont {Shen}, \citenamefont {Na}, \citenamefont {Zohar}, \citenamefont {Wang}, \citenamefont {Welch}, \citenamefont {Seaberg}, \citenamefont {Koralek} \emph {et~al.}}]{wandel2022enhanced}%
  \BibitemOpen
  \bibfield  {author} {\bibinfo {author} {\bibfnamefont {S.}~\bibnamefont {Wandel}}, \bibinfo {author} {\bibfnamefont {F.}~\bibnamefont {Boschini}}, \bibinfo {author} {\bibfnamefont {E.}~\bibnamefont {da~Silva~Neto}}, \bibinfo {author} {\bibfnamefont {L.}~\bibnamefont {Shen}}, \bibinfo {author} {\bibfnamefont {M.}~\bibnamefont {Na}}, \bibinfo {author} {\bibfnamefont {S.}~\bibnamefont {Zohar}}, \bibinfo {author} {\bibfnamefont {Y.}~\bibnamefont {Wang}}, \bibinfo {author} {\bibfnamefont {S.}~\bibnamefont {Welch}}, \bibinfo {author} {\bibfnamefont {M.}~\bibnamefont {Seaberg}}, \bibinfo {author} {\bibfnamefont {J.}~\bibnamefont {Koralek}},  \emph {et~al.},\ }\href@noop {} {\bibfield  {journal} {\bibinfo  {journal} {Science}\ }\textbf {\bibinfo {volume} {376}},\ \bibinfo {pages} {860} (\bibinfo {year} {2022})}\BibitemShut {NoStop}%
\bibitem [{\citenamefont {Jang}\ \emph {et~al.}(2022)\citenamefont {Jang}, \citenamefont {Song}, \citenamefont {Kihara}, \citenamefont {Liu}, \citenamefont {Lee}, \citenamefont {Park}, \citenamefont {Kim}, \citenamefont {Kim}, \citenamefont {Coslovich}, \citenamefont {Nakata} \emph {et~al.}}]{jang2022characterization}%
  \BibitemOpen
  \bibfield  {author} {\bibinfo {author} {\bibfnamefont {H.}~\bibnamefont {Jang}}, \bibinfo {author} {\bibfnamefont {S.}~\bibnamefont {Song}}, \bibinfo {author} {\bibfnamefont {T.}~\bibnamefont {Kihara}}, \bibinfo {author} {\bibfnamefont {Y.}~\bibnamefont {Liu}}, \bibinfo {author} {\bibfnamefont {S.-J.}\ \bibnamefont {Lee}}, \bibinfo {author} {\bibfnamefont {S.-Y.}\ \bibnamefont {Park}}, \bibinfo {author} {\bibfnamefont {M.}~\bibnamefont {Kim}}, \bibinfo {author} {\bibfnamefont {H.-D.}\ \bibnamefont {Kim}}, \bibinfo {author} {\bibfnamefont {G.}~\bibnamefont {Coslovich}}, \bibinfo {author} {\bibfnamefont {S.}~\bibnamefont {Nakata}},  \emph {et~al.},\ }\href@noop {} {\bibfield  {journal} {\bibinfo  {journal} {Science advances}\ }\textbf {\bibinfo {volume} {8}},\ \bibinfo {pages} {eabk0832} (\bibinfo {year} {2022})}\BibitemShut {NoStop}%
\bibitem [{\citenamefont {Kondo}\ \emph {et~al.}(2015{\natexlab{a}})\citenamefont {Kondo}, \citenamefont {Malaeb}, \citenamefont {Ishida}, \citenamefont {Sasagawa}, \citenamefont {Sakamoto}, \citenamefont {Takeuchi}, \citenamefont {Tohyama},\ and\ \citenamefont {Shin}}]{kondo2015point}%
  \BibitemOpen
  \bibfield  {author} {\bibinfo {author} {\bibfnamefont {T.}~\bibnamefont {Kondo}}, \bibinfo {author} {\bibfnamefont {W.}~\bibnamefont {Malaeb}}, \bibinfo {author} {\bibfnamefont {Y.}~\bibnamefont {Ishida}}, \bibinfo {author} {\bibfnamefont {T.}~\bibnamefont {Sasagawa}}, \bibinfo {author} {\bibfnamefont {H.}~\bibnamefont {Sakamoto}}, \bibinfo {author} {\bibfnamefont {T.}~\bibnamefont {Takeuchi}}, \bibinfo {author} {\bibfnamefont {T.}~\bibnamefont {Tohyama}}, \ and\ \bibinfo {author} {\bibfnamefont {S.}~\bibnamefont {Shin}},\ }\href@noop {} {\bibfield  {journal} {\bibinfo  {journal} {Nature communications}\ }\textbf {\bibinfo {volume} {6}},\ \bibinfo {pages} {7699} (\bibinfo {year} {2015}{\natexlab{a}})}\BibitemShut {NoStop}%
\bibitem [{\citenamefont {Yuan}\ \emph {et~al.}(2024)\citenamefont {Yuan}, \citenamefont {Shi}, \citenamefont {Yue}, \citenamefont {Li}, \citenamefont {Wang}, \citenamefont {Xu}, \citenamefont {Xu}, \citenamefont {Wang}, \citenamefont {Gan}, \citenamefont {Chen} \emph {et~al.}}]{yuan2024dynamical}%
  \BibitemOpen
  \bibfield  {author} {\bibinfo {author} {\bibfnamefont {J.}~\bibnamefont {Yuan}}, \bibinfo {author} {\bibfnamefont {L.}~\bibnamefont {Shi}}, \bibinfo {author} {\bibfnamefont {L.}~\bibnamefont {Yue}}, \bibinfo {author} {\bibfnamefont {B.}~\bibnamefont {Li}}, \bibinfo {author} {\bibfnamefont {Z.}~\bibnamefont {Wang}}, \bibinfo {author} {\bibfnamefont {S.}~\bibnamefont {Xu}}, \bibinfo {author} {\bibfnamefont {T.}~\bibnamefont {Xu}}, \bibinfo {author} {\bibfnamefont {Y.}~\bibnamefont {Wang}}, \bibinfo {author} {\bibfnamefont {Z.}~\bibnamefont {Gan}}, \bibinfo {author} {\bibfnamefont {F.}~\bibnamefont {Chen}},  \emph {et~al.},\ }\href@noop {} {\bibfield  {journal} {\bibinfo  {journal} {Science Advances}\ }\textbf {\bibinfo {volume} {10}},\ \bibinfo {pages} {eadg9211} (\bibinfo {year} {2024})}\BibitemShut {NoStop}%
\bibitem [{\citenamefont {Ghiringhelli}\ \emph {et~al.}(2012)\citenamefont {Ghiringhelli}, \citenamefont {Le~Tacon}, \citenamefont {Minola}, \citenamefont {Blanco-Canosa}, \citenamefont {Mazzoli}, \citenamefont {Brookes}, \citenamefont {De~Luca}, \citenamefont {Frano}, \citenamefont {Hawthorn}, \citenamefont {He} \emph {et~al.}}]{ghiringhelli2012long}%
  \BibitemOpen
  \bibfield  {author} {\bibinfo {author} {\bibfnamefont {G.}~\bibnamefont {Ghiringhelli}}, \bibinfo {author} {\bibfnamefont {M.}~\bibnamefont {Le~Tacon}}, \bibinfo {author} {\bibfnamefont {M.}~\bibnamefont {Minola}}, \bibinfo {author} {\bibfnamefont {S.}~\bibnamefont {Blanco-Canosa}}, \bibinfo {author} {\bibfnamefont {C.}~\bibnamefont {Mazzoli}}, \bibinfo {author} {\bibfnamefont {N.}~\bibnamefont {Brookes}}, \bibinfo {author} {\bibfnamefont {G.}~\bibnamefont {De~Luca}}, \bibinfo {author} {\bibfnamefont {A.}~\bibnamefont {Frano}}, \bibinfo {author} {\bibfnamefont {D.}~\bibnamefont {Hawthorn}}, \bibinfo {author} {\bibfnamefont {F.}~\bibnamefont {He}},  \emph {et~al.},\ }\href@noop {} {\bibfield  {journal} {\bibinfo  {journal} {Science}\ }\textbf {\bibinfo {volume} {337}},\ \bibinfo {pages} {821} (\bibinfo {year} {2012})}\BibitemShut {NoStop}%
\bibitem [{\citenamefont {Chang}\ \emph {et~al.}(2012)\citenamefont {Chang}, \citenamefont {Blackburn}, \citenamefont {Holmes}, \citenamefont {Christensen}, \citenamefont {Larsen}, \citenamefont {Mesot}, \citenamefont {Liang}, \citenamefont {Bonn}, \citenamefont {Hardy}, \citenamefont {Watenphul} \emph {et~al.}}]{chang2012direct}%
  \BibitemOpen
  \bibfield  {author} {\bibinfo {author} {\bibfnamefont {J.}~\bibnamefont {Chang}}, \bibinfo {author} {\bibfnamefont {E.}~\bibnamefont {Blackburn}}, \bibinfo {author} {\bibfnamefont {A.}~\bibnamefont {Holmes}}, \bibinfo {author} {\bibfnamefont {N.~B.}\ \bibnamefont {Christensen}}, \bibinfo {author} {\bibfnamefont {J.}~\bibnamefont {Larsen}}, \bibinfo {author} {\bibfnamefont {J.}~\bibnamefont {Mesot}}, \bibinfo {author} {\bibfnamefont {R.}~\bibnamefont {Liang}}, \bibinfo {author} {\bibfnamefont {D.}~\bibnamefont {Bonn}}, \bibinfo {author} {\bibfnamefont {W.}~\bibnamefont {Hardy}}, \bibinfo {author} {\bibfnamefont {A.}~\bibnamefont {Watenphul}},  \emph {et~al.},\ }\href@noop {} {\bibfield  {journal} {\bibinfo  {journal} {Nature Physics}\ }\textbf {\bibinfo {volume} {8}},\ \bibinfo {pages} {871} (\bibinfo {year} {2012})}\BibitemShut {NoStop}%
\bibitem [{\citenamefont {Boschini}\ \emph {et~al.}(2021)\citenamefont {Boschini}, \citenamefont {Minola}, \citenamefont {Sutarto}, \citenamefont {Schierle}, \citenamefont {Bluschke}, \citenamefont {Das}, \citenamefont {Yang}, \citenamefont {Michiardi}, \citenamefont {Shao}, \citenamefont {Feng} \emph {et~al.}}]{boschini2021dynamic}%
  \BibitemOpen
  \bibfield  {author} {\bibinfo {author} {\bibfnamefont {F.}~\bibnamefont {Boschini}}, \bibinfo {author} {\bibfnamefont {M.}~\bibnamefont {Minola}}, \bibinfo {author} {\bibfnamefont {R.}~\bibnamefont {Sutarto}}, \bibinfo {author} {\bibfnamefont {E.}~\bibnamefont {Schierle}}, \bibinfo {author} {\bibfnamefont {M.}~\bibnamefont {Bluschke}}, \bibinfo {author} {\bibfnamefont {S.}~\bibnamefont {Das}}, \bibinfo {author} {\bibfnamefont {Y.}~\bibnamefont {Yang}}, \bibinfo {author} {\bibfnamefont {M.}~\bibnamefont {Michiardi}}, \bibinfo {author} {\bibfnamefont {Y.}~\bibnamefont {Shao}}, \bibinfo {author} {\bibfnamefont {X.}~\bibnamefont {Feng}},  \emph {et~al.},\ }\href@noop {} {\bibfield  {journal} {\bibinfo  {journal} {Nature communications}\ }\textbf {\bibinfo {volume} {12}},\ \bibinfo {pages} {597} (\bibinfo {year} {2021})}\BibitemShut {NoStop}%
\bibitem [{\citenamefont {da~Silva~Neto}\ \emph {et~al.}(2024)\citenamefont {da~Silva~Neto}, \citenamefont {Frano},\ and\ \citenamefont {Boschini}}]{da2024dynamic}%
  \BibitemOpen
  \bibfield  {author} {\bibinfo {author} {\bibfnamefont {E.~H.}\ \bibnamefont {da~Silva~Neto}}, \bibinfo {author} {\bibfnamefont {A.}~\bibnamefont {Frano}}, \ and\ \bibinfo {author} {\bibfnamefont {F.}~\bibnamefont {Boschini}},\ }\href@noop {} {\bibfield  {journal} {\bibinfo  {journal} {Frontiers in Electronic Materials}\ }\textbf {\bibinfo {volume} {4}},\ \bibinfo {pages} {1473324} (\bibinfo {year} {2024})}\BibitemShut {NoStop}%
\bibitem [{\citenamefont {Smit}\ \emph {et~al.}(2025)\citenamefont {Smit}, \citenamefont {Bluschke}, \citenamefont {Moen}, \citenamefont {Heinsdorf}, \citenamefont {Zavatti}, \citenamefont {Bellomia}, \citenamefont {Giuli}, \citenamefont {Dufresne}, \citenamefont {Suen}, \citenamefont {Zimmermann} \emph {et~al.}}]{smit2025enhanced}%
  \BibitemOpen
  \bibfield  {author} {\bibinfo {author} {\bibfnamefont {S.}~\bibnamefont {Smit}}, \bibinfo {author} {\bibfnamefont {M.}~\bibnamefont {Bluschke}}, \bibinfo {author} {\bibfnamefont {P.}~\bibnamefont {Moen}}, \bibinfo {author} {\bibfnamefont {N.}~\bibnamefont {Heinsdorf}}, \bibinfo {author} {\bibfnamefont {E.}~\bibnamefont {Zavatti}}, \bibinfo {author} {\bibfnamefont {G.}~\bibnamefont {Bellomia}}, \bibinfo {author} {\bibfnamefont {S.}~\bibnamefont {Giuli}}, \bibinfo {author} {\bibfnamefont {S.}~\bibnamefont {Dufresne}}, \bibinfo {author} {\bibfnamefont {C.}~\bibnamefont {Suen}}, \bibinfo {author} {\bibfnamefont {V.}~\bibnamefont {Zimmermann}},  \emph {et~al.},\ }\href@noop {} {\bibfield  {journal} {\bibinfo  {journal} {arXiv preprint arXiv:2506.01448}\ } (\bibinfo {year} {2025})}\BibitemShut {NoStop}%
\bibitem [{\citenamefont {Decrausaz}\ \emph {et~al.}(2025)\citenamefont {Decrausaz}, \citenamefont {Pikulski}, \citenamefont {Ivashko}, \citenamefont {Christensen}, \citenamefont {Choi}, \citenamefont {Udby}, \citenamefont {Niedermayer}, \citenamefont {Lefmann}, \citenamefont {R{\o}nnow}, \citenamefont {Mesot} \emph {et~al.}}]{decrausaz2025dynamic}%
  \BibitemOpen
  \bibfield  {author} {\bibinfo {author} {\bibfnamefont {B.}~\bibnamefont {Decrausaz}}, \bibinfo {author} {\bibfnamefont {M.}~\bibnamefont {Pikulski}}, \bibinfo {author} {\bibfnamefont {O.}~\bibnamefont {Ivashko}}, \bibinfo {author} {\bibfnamefont {N.}~\bibnamefont {Christensen}}, \bibinfo {author} {\bibfnamefont {J.}~\bibnamefont {Choi}}, \bibinfo {author} {\bibfnamefont {L.}~\bibnamefont {Udby}}, \bibinfo {author} {\bibfnamefont {C.}~\bibnamefont {Niedermayer}}, \bibinfo {author} {\bibfnamefont {K.}~\bibnamefont {Lefmann}}, \bibinfo {author} {\bibfnamefont {H.}~\bibnamefont {R{\o}nnow}}, \bibinfo {author} {\bibfnamefont {J.}~\bibnamefont {Mesot}},  \emph {et~al.},\ }\href@noop {} {\bibfield  {journal} {\bibinfo  {journal} {Physical Review Research}\ }\textbf {\bibinfo {volume} {7}},\ \bibinfo {pages} {023131} (\bibinfo {year} {2025})}\BibitemShut {NoStop}%
\bibitem [{\citenamefont {Miller}\ \emph {et~al.}(2015)\citenamefont {Miller}, \citenamefont {Smallwood}, \citenamefont {Zhang}, \citenamefont {Eisaki}, \citenamefont {Orenstein},\ and\ \citenamefont {Lanzara}}]{miller2015photoinduced}%
  \BibitemOpen
  \bibfield  {author} {\bibinfo {author} {\bibfnamefont {T.~L.}\ \bibnamefont {Miller}}, \bibinfo {author} {\bibfnamefont {C.~L.}\ \bibnamefont {Smallwood}}, \bibinfo {author} {\bibfnamefont {W.}~\bibnamefont {Zhang}}, \bibinfo {author} {\bibfnamefont {H.}~\bibnamefont {Eisaki}}, \bibinfo {author} {\bibfnamefont {J.}~\bibnamefont {Orenstein}}, \ and\ \bibinfo {author} {\bibfnamefont {A.}~\bibnamefont {Lanzara}},\ }\href@noop {} {\bibfield  {journal} {\bibinfo  {journal} {Physical Review B}\ }\textbf {\bibinfo {volume} {92}},\ \bibinfo {pages} {144506} (\bibinfo {year} {2015})}\BibitemShut {NoStop}%
\bibitem [{\citenamefont {Miller}\ \emph {et~al.}(2017)\citenamefont {Miller}, \citenamefont {Zhang}, \citenamefont {Eisaki},\ and\ \citenamefont {Lanzara}}]{miller2017particle}%
  \BibitemOpen
  \bibfield  {author} {\bibinfo {author} {\bibfnamefont {T.~L.}\ \bibnamefont {Miller}}, \bibinfo {author} {\bibfnamefont {W.}~\bibnamefont {Zhang}}, \bibinfo {author} {\bibfnamefont {H.}~\bibnamefont {Eisaki}}, \ and\ \bibinfo {author} {\bibfnamefont {A.}~\bibnamefont {Lanzara}},\ }\href@noop {} {\bibfield  {journal} {\bibinfo  {journal} {Physical Review Letters}\ }\textbf {\bibinfo {volume} {118}},\ \bibinfo {pages} {097001} (\bibinfo {year} {2017})}\BibitemShut {NoStop}%
\bibitem [{\citenamefont {Reber}\ \emph {et~al.}(2012)\citenamefont {Reber}, \citenamefont {Plumb}, \citenamefont {Sun}, \citenamefont {Cao}, \citenamefont {Wang}, \citenamefont {McElroy}, \citenamefont {Iwasawa}, \citenamefont {Arita}, \citenamefont {Wen}, \citenamefont {Xu} \emph {et~al.}}]{reber2012origin}%
  \BibitemOpen
  \bibfield  {author} {\bibinfo {author} {\bibfnamefont {T.}~\bibnamefont {Reber}}, \bibinfo {author} {\bibfnamefont {N.}~\bibnamefont {Plumb}}, \bibinfo {author} {\bibfnamefont {Z.}~\bibnamefont {Sun}}, \bibinfo {author} {\bibfnamefont {Y.}~\bibnamefont {Cao}}, \bibinfo {author} {\bibfnamefont {Q.}~\bibnamefont {Wang}}, \bibinfo {author} {\bibfnamefont {K.}~\bibnamefont {McElroy}}, \bibinfo {author} {\bibfnamefont {H.}~\bibnamefont {Iwasawa}}, \bibinfo {author} {\bibfnamefont {M.}~\bibnamefont {Arita}}, \bibinfo {author} {\bibfnamefont {J.}~\bibnamefont {Wen}}, \bibinfo {author} {\bibfnamefont {Z.}~\bibnamefont {Xu}},  \emph {et~al.},\ }\href@noop {} {\bibfield  {journal} {\bibinfo  {journal} {Nature Physics}\ }\textbf {\bibinfo {volume} {8}},\ \bibinfo {pages} {606} (\bibinfo {year} {2012})}\BibitemShut {NoStop}%
\bibitem [{\citenamefont {Drozdov}\ \emph {et~al.}(2018)\citenamefont {Drozdov}, \citenamefont {Pletikosi{\'c}}, \citenamefont {Kim}, \citenamefont {Fujita}, \citenamefont {Gu}, \citenamefont {Davis}, \citenamefont {Johnson}, \citenamefont {Bo{\v{z}}ovi{\'c}},\ and\ \citenamefont {Valla}}]{Drozdov2018}%
  \BibitemOpen
  \bibfield  {author} {\bibinfo {author} {\bibfnamefont {I.~K.}\ \bibnamefont {Drozdov}}, \bibinfo {author} {\bibfnamefont {I.}~\bibnamefont {Pletikosi{\'c}}}, \bibinfo {author} {\bibfnamefont {C.-K.}\ \bibnamefont {Kim}}, \bibinfo {author} {\bibfnamefont {K.}~\bibnamefont {Fujita}}, \bibinfo {author} {\bibfnamefont {G.}~\bibnamefont {Gu}}, \bibinfo {author} {\bibfnamefont {J.~S.}\ \bibnamefont {Davis}}, \bibinfo {author} {\bibfnamefont {P.}~\bibnamefont {Johnson}}, \bibinfo {author} {\bibfnamefont {I.}~\bibnamefont {Bo{\v{z}}ovi{\'c}}}, \ and\ \bibinfo {author} {\bibfnamefont {T.}~\bibnamefont {Valla}},\ }\href@noop {} {\bibfield  {journal} {\bibinfo  {journal} {Nature communications}\ }\textbf {\bibinfo {volume} {9}},\ \bibinfo {pages} {5210} (\bibinfo {year} {2018})}\BibitemShut {NoStop}%
\bibitem [{\citenamefont {Norman}\ \emph {et~al.}(1998{\natexlab{a}})\citenamefont {Norman}, \citenamefont {Randeria}, \citenamefont {Ding},\ and\ \citenamefont {Campuzano}}]{norman1998phenomenology}%
  \BibitemOpen
  \bibfield  {author} {\bibinfo {author} {\bibfnamefont {M.~R.}\ \bibnamefont {Norman}}, \bibinfo {author} {\bibfnamefont {M.}~\bibnamefont {Randeria}}, \bibinfo {author} {\bibfnamefont {H.}~\bibnamefont {Ding}}, \ and\ \bibinfo {author} {\bibfnamefont {J.}~\bibnamefont {Campuzano}},\ }\href@noop {} {\bibfield  {journal} {\bibinfo  {journal} {Physical Review B}\ }\textbf {\bibinfo {volume} {57}},\ \bibinfo {pages} {R11093} (\bibinfo {year} {1998}{\natexlab{a}})}\BibitemShut {NoStop}%
\bibitem [{\citenamefont {Parcollet}\ \emph {et~al.}(2015)\citenamefont {Parcollet}, \citenamefont {Ferrero}, \citenamefont {Ayral}, \citenamefont {Hafermann}, \citenamefont {Krivenko}, \citenamefont {Messio},\ and\ \citenamefont {Seth}}]{PARCOLLET2015398}%
  \BibitemOpen
  \bibfield  {author} {\bibinfo {author} {\bibfnamefont {O.}~\bibnamefont {Parcollet}}, \bibinfo {author} {\bibfnamefont {M.}~\bibnamefont {Ferrero}}, \bibinfo {author} {\bibfnamefont {T.}~\bibnamefont {Ayral}}, \bibinfo {author} {\bibfnamefont {H.}~\bibnamefont {Hafermann}}, \bibinfo {author} {\bibfnamefont {I.}~\bibnamefont {Krivenko}}, \bibinfo {author} {\bibfnamefont {L.}~\bibnamefont {Messio}}, \ and\ \bibinfo {author} {\bibfnamefont {P.}~\bibnamefont {Seth}},\ }\href {\doibase https://doi.org/10.1016/j.cpc.2015.04.023} {\bibfield  {journal} {\bibinfo  {journal} {Computer Physics Communications}\ }\textbf {\bibinfo {volume} {196}},\ \bibinfo {pages} {398} (\bibinfo {year} {2015})}\BibitemShut {NoStop}%
\bibitem [{\citenamefont {Kondo}\ \emph {et~al.}(2015{\natexlab{b}})\citenamefont {Kondo}, \citenamefont {Malaeb}, \citenamefont {Ishida}, \citenamefont {Sasagawa}, \citenamefont {Sakamoto}, \citenamefont {Takeuchi}, \citenamefont {Tohyama},\ and\ \citenamefont {Shin}}]{Kondo2015}%
  \BibitemOpen
  \bibfield  {author} {\bibinfo {author} {\bibfnamefont {T.}~\bibnamefont {Kondo}}, \bibinfo {author} {\bibfnamefont {W.}~\bibnamefont {Malaeb}}, \bibinfo {author} {\bibfnamefont {Y.}~\bibnamefont {Ishida}}, \bibinfo {author} {\bibfnamefont {T.}~\bibnamefont {Sasagawa}}, \bibinfo {author} {\bibfnamefont {H.}~\bibnamefont {Sakamoto}}, \bibinfo {author} {\bibfnamefont {T.}~\bibnamefont {Takeuchi}}, \bibinfo {author} {\bibfnamefont {T.}~\bibnamefont {Tohyama}}, \ and\ \bibinfo {author} {\bibfnamefont {S.}~\bibnamefont {Shin}},\ }\href {\doibase 10.1038/ncomms8699} {\bibfield  {journal} {\bibinfo  {journal} {Nature Communications}\ }\textbf {\bibinfo {volume} {6}},\ \bibinfo {pages} {7699} (\bibinfo {year} {2015}{\natexlab{b}})}\BibitemShut {NoStop}%
\bibitem [{\citenamefont {Norman}\ \emph {et~al.}(1998{\natexlab{b}})\citenamefont {Norman}, \citenamefont {Randeria}, \citenamefont {Ding},\ and\ \citenamefont {Campuzano}}]{PhysRevB.57.R11093}%
  \BibitemOpen
  \bibfield  {author} {\bibinfo {author} {\bibfnamefont {M.~R.}\ \bibnamefont {Norman}}, \bibinfo {author} {\bibfnamefont {M.}~\bibnamefont {Randeria}}, \bibinfo {author} {\bibfnamefont {H.}~\bibnamefont {Ding}}, \ and\ \bibinfo {author} {\bibfnamefont {J.~C.}\ \bibnamefont {Campuzano}},\ }\href {\doibase 10.1103/PhysRevB.57.R11093} {\bibfield  {journal} {\bibinfo  {journal} {Phys. Rev. B}\ }\textbf {\bibinfo {volume} {57}},\ \bibinfo {pages} {R11093} (\bibinfo {year} {1998}{\natexlab{b}})}\BibitemShut {NoStop}%
\bibitem [{\citenamefont {Gunnarsson}\ \emph {et~al.}(2015)\citenamefont {Gunnarsson}, \citenamefont {Sch\"afer}, \citenamefont {LeBlanc}, \citenamefont {Gull}, \citenamefont {Merino}, \citenamefont {Sangiovanni}, \citenamefont {Rohringer},\ and\ \citenamefont {Toschi}}]{PhysRevLett.114.236402}%
  \BibitemOpen
  \bibfield  {author} {\bibinfo {author} {\bibfnamefont {O.}~\bibnamefont {Gunnarsson}}, \bibinfo {author} {\bibfnamefont {T.}~\bibnamefont {Sch\"afer}}, \bibinfo {author} {\bibfnamefont {J.~P.~F.}\ \bibnamefont {LeBlanc}}, \bibinfo {author} {\bibfnamefont {E.}~\bibnamefont {Gull}}, \bibinfo {author} {\bibfnamefont {J.}~\bibnamefont {Merino}}, \bibinfo {author} {\bibfnamefont {G.}~\bibnamefont {Sangiovanni}}, \bibinfo {author} {\bibfnamefont {G.}~\bibnamefont {Rohringer}}, \ and\ \bibinfo {author} {\bibfnamefont {A.}~\bibnamefont {Toschi}},\ }\href {\doibase 10.1103/PhysRevLett.114.236402} {\bibfield  {journal} {\bibinfo  {journal} {Phys. Rev. Lett.}\ }\textbf {\bibinfo {volume} {114}},\ \bibinfo {pages} {236402} (\bibinfo {year} {2015})}\BibitemShut {NoStop}%
\bibitem [{\citenamefont {Wu}\ \emph {et~al.}(2017)\citenamefont {Wu}, \citenamefont {Ferrero}, \citenamefont {Georges},\ and\ \citenamefont {Kozik}}]{PhysRevB.96.041105}%
  \BibitemOpen
  \bibfield  {author} {\bibinfo {author} {\bibfnamefont {W.}~\bibnamefont {Wu}}, \bibinfo {author} {\bibfnamefont {M.}~\bibnamefont {Ferrero}}, \bibinfo {author} {\bibfnamefont {A.}~\bibnamefont {Georges}}, \ and\ \bibinfo {author} {\bibfnamefont {E.}~\bibnamefont {Kozik}},\ }\href {\doibase 10.1103/PhysRevB.96.041105} {\bibfield  {journal} {\bibinfo  {journal} {Phys. Rev. B}\ }\textbf {\bibinfo {volume} {96}},\ \bibinfo {pages} {041105} (\bibinfo {year} {2017})}\BibitemShut {NoStop}%
\bibitem [{\citenamefont {Wu}\ \emph {et~al.}(2018)\citenamefont {Wu}, \citenamefont {Scheurer}, \citenamefont {Chatterjee}, \citenamefont {Sachdev}, \citenamefont {Georges},\ and\ \citenamefont {Ferrero}}]{PhysRevX.8.021048}%
  \BibitemOpen
  \bibfield  {author} {\bibinfo {author} {\bibfnamefont {W.}~\bibnamefont {Wu}}, \bibinfo {author} {\bibfnamefont {M.~S.}\ \bibnamefont {Scheurer}}, \bibinfo {author} {\bibfnamefont {S.}~\bibnamefont {Chatterjee}}, \bibinfo {author} {\bibfnamefont {S.}~\bibnamefont {Sachdev}}, \bibinfo {author} {\bibfnamefont {A.}~\bibnamefont {Georges}}, \ and\ \bibinfo {author} {\bibfnamefont {M.}~\bibnamefont {Ferrero}},\ }\href {\doibase 10.1103/PhysRevX.8.021048} {\bibfield  {journal} {\bibinfo  {journal} {Phys. Rev. X}\ }\textbf {\bibinfo {volume} {8}},\ \bibinfo {pages} {021048} (\bibinfo {year} {2018})}\BibitemShut {NoStop}%
\bibitem [{\citenamefont {Scheurer}\ \emph {et~al.}(2018)\citenamefont {Scheurer}, \citenamefont {Chatterjee}, \citenamefont {Wu}, \citenamefont {Ferrero}, \citenamefont {Georges},\ and\ \citenamefont {Sachdev}}]{doi:10.1073/pnas.1720580115}%
  \BibitemOpen
  \bibfield  {author} {\bibinfo {author} {\bibfnamefont {M.~S.}\ \bibnamefont {Scheurer}}, \bibinfo {author} {\bibfnamefont {S.}~\bibnamefont {Chatterjee}}, \bibinfo {author} {\bibfnamefont {W.}~\bibnamefont {Wu}}, \bibinfo {author} {\bibfnamefont {M.}~\bibnamefont {Ferrero}}, \bibinfo {author} {\bibfnamefont {A.}~\bibnamefont {Georges}}, \ and\ \bibinfo {author} {\bibfnamefont {S.}~\bibnamefont {Sachdev}},\ }\href {\doibase 10.1073/pnas.1720580115} {\bibfield  {journal} {\bibinfo  {journal} {Proceedings of the National Academy of Sciences}\ }\textbf {\bibinfo {volume} {115}},\ \bibinfo {pages} {E3665} (\bibinfo {year} {2018})},\ \Eprint {http://arxiv.org/abs/https://www.pnas.org/doi/pdf/10.1073/pnas.1720580115} {https://www.pnas.org/doi/pdf/10.1073/pnas.1720580115} \BibitemShut {NoStop}%
\bibitem [{\citenamefont {Virtanen}\ \emph {et~al.}(2020)\citenamefont {Virtanen}, \citenamefont {Gommers}, \citenamefont {Oliphant}, \citenamefont {Haberland}, \citenamefont {Reddy}, \citenamefont {Cournapeau}, \citenamefont {Burovski}, \citenamefont {Peterson}, \citenamefont {Weckesser}, \citenamefont {Bright}, \citenamefont {{van der Walt}}, \citenamefont {Brett}, \citenamefont {Wilson}, \citenamefont {Millman}, \citenamefont {Mayorov}, \citenamefont {Nelson}, \citenamefont {Jones}, \citenamefont {Kern}, \citenamefont {Larson}, \citenamefont {Carey}, \citenamefont {Polat}, \citenamefont {Feng}, \citenamefont {Moore}, \citenamefont {{VanderPlas}}, \citenamefont {Laxalde}, \citenamefont {Perktold}, \citenamefont {Cimrman}, \citenamefont {Henriksen}, \citenamefont {Quintero}, \citenamefont {Harris}, \citenamefont {Archibald}, \citenamefont {Ribeiro}, \citenamefont {Pedregosa}, \citenamefont {{van Mulbregt}},\ and\ \citenamefont {{SciPy 1.0 Contributors}}}]{2020SciPy-NMeth}%
  \BibitemOpen
  \bibfield  {author} {\bibinfo {author} {\bibfnamefont {P.}~\bibnamefont {Virtanen}}, \bibinfo {author} {\bibfnamefont {R.}~\bibnamefont {Gommers}}, \bibinfo {author} {\bibfnamefont {T.~E.}\ \bibnamefont {Oliphant}}, \bibinfo {author} {\bibfnamefont {M.}~\bibnamefont {Haberland}}, \bibinfo {author} {\bibfnamefont {T.}~\bibnamefont {Reddy}}, \bibinfo {author} {\bibfnamefont {D.}~\bibnamefont {Cournapeau}}, \bibinfo {author} {\bibfnamefont {E.}~\bibnamefont {Burovski}}, \bibinfo {author} {\bibfnamefont {P.}~\bibnamefont {Peterson}}, \bibinfo {author} {\bibfnamefont {W.}~\bibnamefont {Weckesser}}, \bibinfo {author} {\bibfnamefont {J.}~\bibnamefont {Bright}}, \bibinfo {author} {\bibfnamefont {S.~J.}\ \bibnamefont {{van der Walt}}}, \bibinfo {author} {\bibfnamefont {M.}~\bibnamefont {Brett}}, \bibinfo {author} {\bibfnamefont {J.}~\bibnamefont {Wilson}}, \bibinfo {author} {\bibfnamefont {K.~J.}\ \bibnamefont {Millman}}, \bibinfo {author} {\bibfnamefont {N.}~\bibnamefont {Mayorov}}, \bibinfo {author} {\bibfnamefont
  {A.~R.~J.}\ \bibnamefont {Nelson}}, \bibinfo {author} {\bibfnamefont {E.}~\bibnamefont {Jones}}, \bibinfo {author} {\bibfnamefont {R.}~\bibnamefont {Kern}}, \bibinfo {author} {\bibfnamefont {E.}~\bibnamefont {Larson}}, \bibinfo {author} {\bibfnamefont {C.~J.}\ \bibnamefont {Carey}}, \bibinfo {author} {\bibfnamefont {{\.I}.}~\bibnamefont {Polat}}, \bibinfo {author} {\bibfnamefont {Y.}~\bibnamefont {Feng}}, \bibinfo {author} {\bibfnamefont {E.~W.}\ \bibnamefont {Moore}}, \bibinfo {author} {\bibfnamefont {J.}~\bibnamefont {{VanderPlas}}}, \bibinfo {author} {\bibfnamefont {D.}~\bibnamefont {Laxalde}}, \bibinfo {author} {\bibfnamefont {J.}~\bibnamefont {Perktold}}, \bibinfo {author} {\bibfnamefont {R.}~\bibnamefont {Cimrman}}, \bibinfo {author} {\bibfnamefont {I.}~\bibnamefont {Henriksen}}, \bibinfo {author} {\bibfnamefont {E.~A.}\ \bibnamefont {Quintero}}, \bibinfo {author} {\bibfnamefont {C.~R.}\ \bibnamefont {Harris}}, \bibinfo {author} {\bibfnamefont {A.~M.}\ \bibnamefont {Archibald}}, \bibinfo {author}
  {\bibfnamefont {A.~H.}\ \bibnamefont {Ribeiro}}, \bibinfo {author} {\bibfnamefont {F.}~\bibnamefont {Pedregosa}}, \bibinfo {author} {\bibfnamefont {P.}~\bibnamefont {{van Mulbregt}}}, \ and\ \bibinfo {author} {\bibnamefont {{SciPy 1.0 Contributors}}},\ }\href {\doibase 10.1038/s41592-019-0686-2} {\bibfield  {journal} {\bibinfo  {journal} {Nature Methods}\ }\textbf {\bibinfo {volume} {17}},\ \bibinfo {pages} {261} (\bibinfo {year} {2020})}\BibitemShut {NoStop}%
\end{thebibliography}
\end{document}